\documentclass[aps, preprintnumbers, superscriptaddress, nofootinbibt, twocolumn]{revtex4-2}

\usepackage{makeidx}
\usepackage{amsfonts}
\usepackage{amsmath}
\usepackage{amssymb}
\usepackage{eurosym}
\usepackage{graphicx}
\usepackage{tensor}
\usepackage[dvipsnames]{xcolor}
\usepackage[colorlinks, allcolors=Blue]{hyperref}
\usepackage[capitalize]{cleveref}
\usepackage{color}
\usepackage{revsymb}

\def\be{\begin{equation}}
\def\ee{\end{equation}}
\def\bea{\begin{eqnarray}}
\def\eea{\end{eqnarray}}

\begin{document}

\title{Weyl geometric effects on the propagation of light in gravitational fields}

\author{Marius A. Oancea}
\email{marius.oancea@univie.ac.at}
\affiliation{University of Vienna, Faculty of Physics, Boltzmanngasse 5, 1090 Vienna,
Austria}

\author{Tiberiu Harko}
\email{tiberiu.harko@aira.astro.ro}
\affiliation{Department of Theoretical Physics, National Institute of Physics and Nuclear Engineering (IFIN-HH), Bucharest, 077125 Romania,}
\affiliation{Department of Physics, Babes-Bolyai University, Kogalniceanu Street, Cluj-Napoca 400084, Romania,}
\affiliation{Astronomical Observatory, 19 Ciresilor Street, Cluj-Napoca 400487, Romania}

\preprint{\href{https://doi.org/10.1103/PhysRevD.109.064020}{DOI: 10.1103/PhysRevD.109.064020}}

\begin{abstract}
We consider the effects of Weyl geometry on the propagation of electromagnetic wave packets and on the gravitational spin Hall effect of light. It is usually assumed that in vacuum the electromagnetic waves propagate along null geodesics, a result which follows from the geometrical optics approximation. However, this model is valid only in the limit of infinitely high frequencies. At large but finite frequencies, the ray dynamics is affected by the wave polarization. Therefore, the propagation of the electromagnetic waves can deviate from null geodesics, and this phenomenon is known as the gravitational spin Hall effect of light. On the other hand, Maxwell's equations have the remarkable property of conformal invariance. This property is a cornerstone of Weyl geometry and the corresponding gravitational theories. As a first step in our study, we obtain the polarization-dependent ray equations in Weyl geometry, describing the gravitational spin Hall effect of localized electromagnetic wave packets in the presence of nonmetricity. As a specific example of the spin Hall effect of light in Weyl geometry, we consider the case of the simplest conformally invariant action, constructed from the square of the Weyl scalar, and the strength of the Weyl vector only. The action is linearized in the Weyl scalar by introducing an auxiliary scalar field. In static spherical symmetry, this theory admits an exact black hole solution, which generalizes the standard Schwarzschild solution through the presence of two new terms in the metric, having a linear and a quadratic dependence on the radial coordinate. We numerically study the polarization-dependent propagation of light rays in this exact Weyl geometric metric, and the effects of the presence of the Weyl vector on the magnitude of the spin Hall effect are estimated.
\end{abstract}

\maketitle

\section{Introduction}

Maxwell's equations have a natural invariance with respect to the group of conformal transformations $g_{\mu\nu}\rightarrow \hat{g}_{\mu \nu}=\Sigma ^2 g_{\mu \nu}$, where the dimensionless conformal factor $\Sigma(x)$ is a smooth, strictly positive function of the spacetime coordinates $x^\mu$ \cite{Wald,Far}. The conformal invariance of Maxwell's equations has fundamental physical consequences and implies that because the photon is massless, no specific mass or length scale is associated with the electromagnetic field \cite{Far}. In the geometrical optics limit of Maxwell's equations, which is based on the assumption that the wavelengths of the electromagnetic waves are negligible as compared to the radius of curvature of spacetime, light travels along null geodesics, which are left invariant by the conformal transformations, except for a change of parametrization \cite{Wald}.

On the other hand, conformal invariance is the cornerstone of Weyl geometry \cite{Weyl}, which is an important extension of Riemann geometry. The starting point for building up Weyl geometry is the replacement of the metric compatibility condition of Riemann geometry with a more general condition. Assume that the covariant derivative of the metric tensor does not vanish identically but is given by $\tilde{\nabla}_\alpha g_{\mu \nu} = Q_{\alpha \mu \nu}$, where $Q_{\alpha \mu \nu}$ is called the nonmetricity tensor. Initially, Weyl adopted for the nonmetricity tensor the particular form $Q_{\alpha \mu \nu} = \omega _{\alpha}g_{\mu \nu}$, where $\omega_{\alpha}$ is the Weyl vector. For a detailed discussion of Weyl geometry, as well as of its historical development, see Ref. \cite{Scholz}. Weyl also introduced important ideas regarding the necessity of conformal invariance of physical laws, and he also proposed reformulating Einstein's gravitational theory as a conformally invariant physical theory. Conformally invariant theories of gravity, as well as of elementary particle physics, were recently considered in Refs. \cite{C1,C2,C3,C4,C5,C6,C7}. In a general sense, the action that describes an arbitrary physical system is conformally invariant if the variation of the action $S\left[ g_{\mu \nu},\phi \right]$ with respect to the group of conformal transformations vanishes, $\delta _cS\left[g_{\mu \nu},\phi\right]=\int{d^n x\left(\delta L/\delta \phi\right)\delta _c\phi}=0$ \cite{Kaz}.

We can also consider the Weyl rescaling, representing the simultaneous transformations of the physical fields and of the metric, given by $\hat{g}_{\mu \nu}(x)=e^{2\sigma (x)}g_{\mu \nu}$, and $\hat{\phi}=e^{-\Delta \sigma(x)}\phi(x)$ \cite{Kaz}, under which the action transforms as $\delta _\sigma S\left[g_{\mu \nu},\phi\right]=\int{d^n x\sigma \left[2\left(\delta L/\delta g_{\mu \nu}\right)g_{\mu \nu}-\Delta_n\left(\delta L/\delta \phi\right)\phi\right]}$. See Ref. \cite{Kaz} for an in-depth discussion of conformal invariance and Weyl invariance.

An interesting extension of Weyl's theory was proposed by Dirac \cite{D1,D2}, which is based on the introduction of a new geometric quantity, the Dirac gauge function $\beta$, which describes the geometric and physical properties of the spacetime manifold, together with the symmetric metric tensor $g_{\mu \nu}$ and the geometric Weyl vector $\omega _\mu$. For physical and cosmological applications of the Weyl-Dirac theory, see Refs. \cite{R1,R2}. The Weyl geometric theory and its possible physical implications were also investigated in Refs. \cite{U1,U2,U3}. In particular, a new scalar field, called a measure field, was introduced in the theory. The measure field plays the role of a measure associated with each world point. If all the physical quantities are measured with a standard given by the measure field, then it is possible to formulate all the field equations in a manifestly gauge-invariant way.

The simplest possible gravitational action with conformal symmetry in a purely Riemann geometry, implemented locally, is the conformal gravity model, in which the Lagrangian density is given by $C_{\alpha \beta \gamma \delta}^2$ \cite{M1,M2,M3,M4,M5,M6}, where $C_{\alpha \beta \gamma \delta}$ is the Weyl tensor defined in Riemann geometry. Conformal Weyl gravity is purely geometric and does not contain the Weyl gauge field $\omega_\mu$ or a scalar field. In four dimensions, the theory has the important property of invariance under local Weyl gauge transformations.

An attractive approach to Weyl's theory and to its physical applications was considered in Refs. \cite{Gh1,Gh2,Gh3,Gh4,Gh5,Gh6,Gh7,Gh8} by adopting a viewpoint from elementary particle physics. The key idea of this approach is the linearization in the gravitational action of the Weyl quadratic term $\tilde{R}^2$  by introducing an auxiliary scalar field $\phi_0$ \cite{Gh1}. Thus, in the linearized version of Weyl quadratic theory, a spontaneous symmetry breaking of the $D(1)$ group can be implemented through a geometric Stueckelberg-type mechanism \cite{Gh2,Gh3,Gh4}. Consequently, the Weyl gauge field becomes massive, with the mass term originating from the spin-zero mode of the geometric $\tilde{R}^2$ term in the total gravitational action.

From a technical point of view, the Stueckelberg mechanism is introduced by replacing the scalar field $\phi_0$ with a constant value, that is, its vacuum expectation value, $\phi _0\rightarrow \langle \phi _0\rangle$. Once the Weyl vector field has become massive, it includes the auxiliary scalar field $\phi_0$, which no longer appears in the scalar-vector-tensor formulation of the theory, and we recover the initial tensor-vector theory, as proposed by Weyl. However, the Einstein-Proca action, which arises in the broken phase, can be obtained directly from the Weyl action, by eliminating the auxiliary scalar field $\phi_0$ \cite{Gh5,Gh6,Gh7}.

Moreover, the scalar mode also leads to the existence of the Planck scale and of the cosmological constant. In Weyl geometric gravity, the Planck scale, all mass scales, and the cosmological constant, originate from geometry \cite{Gh8}. The Higgs field, playing a fundamental role in the standard model of elementary particle physics, is generated by the Weyl boson fusion in the early Universe.

Black hole solutions in the linearized Weyl geometric gravity were considered in Ref. \cite{Yang}. Although generally the vacuum field equations of Weyl geometric gravity cannot be solved exactly, an exact solution, corresponding to a Weyl-type black hole, can be obtained in the particular case in which the Weyl vector has only a radial component. This solution represents an extension of the Schwarzschild line element, with two extra terms appearing in the metric.

The behavior of the galactic rotation curves in the exact solution of Weyl geometric gravity was considered in Ref. \cite{Pi1}, where it was shown that a dark matter density profile and an effective geometric mass can also be introduced. Three particular cases corresponding to some specific functional forms of the Weyl vector were also studied. The predictions of the Weyl geometric theoretical model were compared with a selected sample of galactic rotation curves, by also introducing an explicit breaking of the conformal invariance, which allows one to fix the numerical values of the free parameters of the model. The obtained results did show that Weyl geometric models can be considered as viable theoretical alternatives to the dark matter paradigm.

Another interesting and universal phenomenon in physics, generally encountered for waves with internal structure propagating in inhomogeneous media, is the spin-orbit coupling between the internal (spin) and the external (average position and momentum) degrees of freedom of a wave packet \cite{SHE_review, SHE_review1, opt1}. The universality of spin-orbit coupling is a consequence of the conservation of angular momentum \cite{rev}. The spin has an important effect on dynamics and can generally be seen as particles following spin-dependent trajectories. For example, this leads to the spin Hall effect of electrons propagating in condensed matter systems \cite{originalSHE1,originalSHE2,originalSHE3,originalSHE4,SHE_review,SHE_review1}.

Spin-orbit interactions are also present in optics \cite{opt1, opt2}, where electromagnetic waves can generally be seen to propagate in a polarization-dependent way. In this case, the spin internal degree of freedom is now represented by the state of polarization of the wave packet. This leads to the spin Hall effect of light, represented by the polarization-dependent spatial splitting of light that occurs during refraction and reflection at an optical interface \cite{opt5,B1}. In this case, linearly polarized light, representing the superposition of left- and right-circularly polarized components, experiences polarization-dependent separation at the optical interface. The presence of this effect is a direct consequence of the spin-dependent correction terms that appear when considering the boundary conditions of a wave packet, rather than using the single-ray approximation \cite{opt3}. The spin Hall effect of light was proposed in Ref. \cite{opt4} by analogy with the standard Hall effect, with the spin-1 photons playing the role of the spin-1/2 charges and a refractive index gradient representing the electric potential gradient. The spin-dependent displacement, which is perpendicular to the refractive index gradient, was detected for photons passing through an air-glass interface in Ref. \cite{opt5} (see also Ref. \cite{B1}, where a similar experimental result was reported), indicating the universality of the spin Hall effect for wave packets of different nature.

Similar effects are also present for electromagnetic waves propagating in curved spacetimes. The scattering of electromagnetic plane waves in gravitational fields has been considered in Refs. \cite{Plebansky-Maxwell,Mashoon1, Mashoon2, Mashoon3, Mashoon4, Mashoon5,leite2017absorption,leite2018absorption}. In these works, the main result is that electromagnetic plane waves are scattered in a polarization-dependent way by rotating gravitational objects, such as Kerr black holes, and there is no polarization-dependent effect when considering static spherically symmetric gravitational objects, such as Schwarzschild black holes (similar results were also obtained in Refs. \cite{Frolov1, Frolov2,covariantSpinoptics}). However, in this paper we are mainly interested in the propagation of localized electromagnetic wave packets. In this case, the spin Hall effect of light in the presence of a gravitational field was investigated in detail in Refs. \cite{O1,O2} (similar results were also obtained in Refs. \cite{G1,Frolov2020} and other higher-order geometrical optics treatments can be found in Refs. \cite{spinorSpinoptics2,Dolan,Harte2018,HarteOptics2}). One important difference between localized wave packets and plane waves is that they behave differently in static spherically symmetric spacetimes. Although polarization-dependent scattering is absent for plane waves in this case, localized wave packets experience a nontrivial effect, following frequency- and polarization-dependent trajectories \cite{O1}. Similar effects have also been shown to affect the propagation of other types of fields in curved spacetime, such as linearized gravitational waves \cite{GSHE_GW,GSHE_lensing,GSHE_lensing_letter,SHE_GW}, as well as massive and massless Dirac fields \cite{audretsch,rudiger,GSHE_Dirac}. More exotic forms of spin Hall effects in gravitational fields have also been investigated for massless particles with anyonic spin \cite{marsot2022,gray2022,marsot2022hall,bicak2023}. For a review of the work done in this direction, see Refs. \cite{GSHE_rev,GSHE_review}. From a theoretical point of view, it is usually assumed in many investigations that in vacuum electromagnetic waves travel along null geodesics. This description corresponds to the geometrical optics approximation of Maxwell's equations. However, it is important to note that this approach to wave propagation is rigorously valid only in the limit of infinitely high frequencies of light. At high but finite frequencies, diffraction effects can still be neglected, but a spin-orbit coupling appears, and ray propagation is influenced by the wave polarization \cite{O1}. Therefore, the path of the electromagnetic waves can depart from the null geodesics. This is called the gravitational spin Hall effect of light \cite{O1}.

A fully covariant Wentzel-Kramers-Brillouin (WKB) approach for the propagation of electromagnetic wave packets in arbitrary curved spacetimes was introduced and developed in Ref. \cite{O1}. The ray equations, depending on the polarization of light, which describe the gravitational spin Hall effect were obtained, and the role of the curvature of spacetime was pointed out. The polarization-dependent ray dynamics in the Schwarzschild spacetime was also investigated numerically, and the magnitude and importance of the effect were briefly assessed in an astrophysical context. It is important to note that the gravitational spin Hall effect is analogous to the spin Hall effect of light in inhomogeneous media, an effect whose existence was confirmed experimentally \cite{opt5,B1}.

The main goal of the present manuscript is to extend the investigation of the gravitational spin Hall effect to more general geometries than the Riemannian one considered in Refs. \cite{O1,O2}, and to investigate the non-Riemannian effects induced on the propagation of light by the change of the geometrical structure of the base spacetime manifold. In this study we will concentrate on the ray propagation in conformally invariant Weyl geometry \cite{Weyl, Scholz}, which has the important property that the form of Maxwell's equations coincides with their Riemannian counterparts. In the present investigation, we consider first the polarization-dependent ray equations in Weyl geometry, obtained by using the covariant WKB approach from first principles. These equations describe the gravitational spin Hall effect of light in the presence of nonmetricity, generated by the presence of the Weyl vector. We also consider a specific example of the spin Hall effect of light in Weyl geometry, by considering a conformally invariant, Weyl-type gravitational model, constructed from the square of the Weyl scalar and from the strength of the Weyl vector only. This gravitational action can be reformulated as a scalar-vector-tensor theory by linearizing it in the Weyl scalar via the introduction of an auxiliary scalar field. In static spherical symmetry, and with the Weyl vector assumed to have only a radial component, this conformally invariant geometric theory has an exact black hole solution, which generalizes the standard Schwarzschild solution through the presence of two new terms in the metric, having a linear and a quadratic dependence on the radial coordinate. We numerically study the polarization-dependent propagation of light rays in this exact Weyl geometric metric, and the effects of the presence of the Weyl vector on the magnitude of the spin Hall effect are presented and discussed in detail.

The present paper is organized as follows. We introduce the basics of Weyl geometry, gravitational action, and the black hole solution of Weyl geometric gravity in \cref{sect1}. The conformal invariance of Maxwell's equations is also discussed. The propagation of high-frequency electromagnetic waves in Weyl geometry is discussed in \cref{sect2}, where the equations of the spin Hall effect for light are also obtained. We investigate the spin Hall effect in the background geometry of the exact black hole solution of Weyl geometric gravity in \cref{sect3}. Finally, we discuss and conclude our work in \cref{sect4}. Complete details of the computation of the curvature tensor of the Weyl geometry are presented in \cref{app}.

\section{From Weyl geometry to Weyl geometric gravity}\label{sect1}

In this section, we briefly review the basics of Weyl geometry, discuss the conformal invariance of Maxwell's equations, and, after introducing the action and the field equations of Weyl geometric gravity, we present a spherically symmetric solution of the vacuum equations of the theory.

\subsection{Weyl geometry}

One of the basic properties of Weyl geometry is the variation of the length of a vector under parallel transport. If a vector with initial $\ell = g_{\mu \nu} v^\mu v^\nu$ is parallel transported between the infinitesimally close points $x^{\mu }$ and $x^{\mu }+\delta x^{\mu }$, then in Weyl geometry the length of the vector will change according to
\begin{equation}\label{eq1}
\delta \ell =- \ell \alpha \omega _{\mu }\delta x^{\mu },
\end{equation}
where $\omega _{\mu }$ is the Weyl vector field, and $\alpha $ is a constant, called the Weyl gauge coupling constant. In the above equation, the change in length $\delta \ell$ is proportional to the initial length $\ell$ of the vector. This means that null vectors, for which $\ell = 0$, do not change in length under parallel transport in Weyl geometry. On the other hand, the lengths of timelike and spacelike vectors will generally change under parallel transport in Weyl geometry.

The second important characteristic of Weyl geometry is the abandonment of the metric compatibility condition $\nabla _{\alpha }g_{\mu \nu }=0$ of Riemann geometry. The nonmetricity $Q_{\lambda \mu \nu }$ can be defined through the covariant derivative of the metric tensor, which in Weyl geometry takes the form of
\begin{equation}  \label{nmetr}
\tilde{\nabla}_{\lambda }g_{\mu \nu }=-\alpha \omega _{\lambda }g_{\mu \nu
}\equiv Q_{\lambda \mu \nu }, \quad \tilde{\nabla}_{\lambda }g^{\mu \nu }=\alpha
\omega _{\lambda }g^{\mu \nu },
\end{equation}
where $\alpha $ denotes the Weyl gauge coupling constant. From the nonmetricity condition in \cref{nmetr}, we immediately obtain the connection $\tilde{\Gamma}_{\mu \nu }^{\lambda }$ of Weyl geometry as
\begin{equation}  \label{Con}
\tilde{\Gamma}_{\mu \nu }^{\lambda }=\Gamma _{\mu \nu }^{\lambda }+\frac{1}{2%
}\alpha \Big[\delta _{\mu }^{\lambda }\omega _{\nu }+\delta _{\nu }^{\lambda
}\omega _{\mu }-g_{\mu \nu }\omega ^{\lambda }\Big]=\Gamma _{\mu \nu
}^{\lambda }+\Psi _{\mu \nu }^{\lambda },
\end{equation}
where $\Gamma_{\mu \nu }^{\lambda }$ is the standard Levi-Civita connection associated with the metric $g_{\mu \nu }$:
\begin{equation}
\Gamma _{\mu \nu }^{\lambda }=\frac{1}{2}g^{\lambda \sigma}\left(\partial
_\nu g_{\sigma \mu}+\partial_\mu g_{\sigma \nu}-\partial _\sigma g_{\mu
\nu}\right).
\end{equation}
In the following, we denote the physical and geometric quantities in Weyl geometry by a tilde. With the help of the Weyl connection, we define the Weyl covariant derivative of a vector $v_{\mu }$ as
\begin{subequations}
\begin{align}
    \tilde{\nabla}_{\lambda }v_{\mu } &= \partial _{\lambda }v_{\mu }-\tilde{\Gamma}_{\lambda \mu }^{\nu }v_{\nu}=\nabla _\lambda v_\mu-\Psi_{\lambda \mu }^{\nu }v_{\nu }, \\
    \tilde{\nabla}_{\lambda }v^{\mu } &= \partial_{\lambda }v^{\mu }+\tilde{\Gamma}_{\lambda \nu }^{\mu }v^{\nu}=\nabla_{\lambda }v^{\mu }+\Psi_{\lambda \nu }^{\mu }v^{\nu },
\end{align}
\end{subequations}
where $\nabla_\lambda$ denotes the covariant derivative constructed with the Levi-Civita connection. 

Generally, the length of a vector $v^\mu$ is defined as the square root of plus/minus (depending on the signature and the spacelike/timelike vector type) $\ell =g_{\mu \nu} v^\mu v^\nu$. Under parallel transport (with respect to the connection $\tilde{\Gamma}^{\lambda}_{\mu\nu}$) along a curve $\gamma(\tau)$, the length of $v^\mu$ will change according to the equation
\begin{equation}
    \dot{\ell} = \dot{\gamma}^{\nu} \tilde{\nabla}_\nu \ell = - \ell \alpha \omega_\mu \dot{\gamma}^\mu,
\end{equation}
which is equivalent to Eq.~(\ref{eq1}). This equation can be integrated, and we obtain
\begin{equation}
    \ell(\tau) = \ell(0) e^{-\alpha \int_0^\tau \omega_\mu \dot{\gamma}^\mu(\tau') d \tau' }.
\end{equation}
Note that, on the basis of the above result, vectors cannot change their causal character after finite parallel transport in Weyl geometry.  

By contracting \cref{Con}, we obtain
\begin{equation}
    \omega_{\mu } = \frac{1}{2\alpha }\left( \tilde{\Gamma}^{\lambda}_{\lambda \mu}-\Gamma^{\lambda}_{\lambda \mu} \right) .
\end{equation}
Furthermore, using the relation $\Gamma^{\lambda}_{\lambda \mu}=\partial_\mu \ln \sqrt{-g}$, we obtain
\begin{equation}
    \tilde{\Gamma}^{\lambda}_{\lambda \mu} =\partial_\mu \ln \sqrt{-g}+2\alpha \omega _\mu.
\end{equation}
In Weyl geometry, the determinant of the metric is a scalar density of weight 2, satisfying the relations
\begin{subequations}
    \begin{align}
    \tilde{\nabla}_\mu g &= \partial_\mu g - 2\tilde{\Gamma}^{\lambda}_{\lambda \mu} g, \\
    \tilde{\nabla}_\mu \ln \sqrt{-g} &= \partial _\mu \ln \sqrt{-g}-\tilde{\Gamma}^{\lambda}_{\lambda \mu}.
    \end{align}
\end{subequations}
Another important geometrical and physical quantity, the field strength $\tilde{W}_{\mu\nu}$ of the Weyl vector $\omega_\mu$, is defined as
\begin{equation}\label{W}
\tilde{W}_{\mu\nu} = \tilde{\nabla}_{\mu} \omega_{\nu} - \tilde{\nabla}_{\nu} \omega_{\mu} = \partial _{\mu}\omega
_\nu-\partial _\nu \omega _\mu.
\end{equation}
We can also express the action of covariant derivative commutators on vectors and covectors as
\begin{subequations}
\begin{align}
\left( \tilde{\nabla}_{\mu }\tilde{\nabla}_{\nu }-\tilde{\nabla}_{\nu }%
\tilde{\nabla}_{\mu }\right) v^{\sigma }& =\tilde{R}\indices{^\sigma_{\rho
\mu \nu}}v^{\rho }, \\
\left( \tilde{\nabla}_{\mu }\tilde{\nabla}_{\nu }-\tilde{\nabla}_{\nu }%
\tilde{\nabla}_{\mu }\right) v_{\sigma }& =-\tilde{R}\indices{^\rho_{\sigma
\mu \nu}}v_{\rho },
\end{align}
where the curvature tensor $\tilde{R}\indices{^{\lambda }_{\mu \nu \sigma }}$ of Weyl geometry is defined according to the standard definition:
\end{subequations}
\begin{equation}
\tilde{R}\indices{^{\lambda }_{\mu \nu \sigma }} =\partial _{\nu }\tilde{\Gamma}_{\mu
\sigma }^{\lambda }-\partial _{\sigma }\tilde{\Gamma}_{\mu \nu }^{\lambda }+%
\tilde{\Gamma}_{\rho \nu }^{\lambda }\tilde{\Gamma}_{\mu \sigma }^{\rho }-%
\tilde{\Gamma}_{\rho \sigma }^{\lambda }\tilde{\Gamma}_{\mu \nu }^{\rho },
\end{equation}
If we consider a general symmetric connection of the form $\tilde{\Gamma}_{\mu \nu }^{\lambda }=\Gamma _{\mu \nu }^{\lambda }+\Psi_{\mu \nu }^{\lambda }$, the curvature tensor can be expanded as
\bea
\tilde{R}\indices{^{\lambda }_{\mu \nu \sigma }} &=& R\indices{^{\lambda }_{\mu \nu \sigma }} + \nabla
_{\nu }\Psi _{\mu \sigma }^{\lambda } - \nabla _{\sigma }\Psi _{\mu \nu
}^{\lambda } \nonumber\\
    && +\Psi _{\rho \nu }^{\lambda }\Psi _{\mu \sigma }^{\rho }-\Psi
_{\rho \sigma }^{\lambda }\Psi _{\mu \nu }^{\rho },
\eea
where $R\indices{^{\lambda }_{\mu \nu \sigma }}$ is the curvature tensor of Riemann geometry. Taking into account that $\Psi _{\mu \nu }^{\lambda }$ is given by \cref{Con}, we obtain
\begin{eqnarray}
&&\tilde{R}\indices{^{\lambda }_{\mu \nu \sigma }} = R\indices{^{\lambda }_{\mu \nu \sigma }} +\frac{1}{2}\alpha \Bigg[ \tilde{W}_{\nu \sigma }\delta _{\mu }^{\lambda
}+\left( \delta _{\sigma }^{\lambda }\nabla _{\nu }-\delta _{\nu }^{\lambda
}\nabla _{\sigma }\right) \omega _{\mu }\nonumber\\
&&+\left( g_{\mu \nu }\nabla _{\sigma
}-g_{\mu \sigma }\nabla _{\nu }\right) \omega ^{\lambda }\Bigg]
+\frac{1}{4}\alpha ^{2}\Bigg[ \left( \omega ^{2}g_{\mu \nu }-\omega _{\mu
}\omega _{\nu }\right) \delta _{\sigma }^{\lambda }\nonumber\\
&&-\left( \omega ^{2}g_{\mu
\sigma }-\omega _{\mu }\omega _{\sigma }\right) \delta _{\nu }^{\lambda
}+\left( g_{\mu \sigma }\omega _{\nu }-g_{\mu \nu }\omega _{\sigma }\right)
\omega ^{\lambda }\Bigg],
\end{eqnarray}
where $\omega ^2=\omega_\rho \omega ^\rho$ (see \cref{app} for the derivation of the above relations). It follows from straightforward calculations that the curvature tensor in Weyl geometry satisfies the following symmetries:
\begin{subequations}\label{eq:R_sym}
\begin{align}
& \tilde{R}_{\mu \nu \rho \sigma }=-\tilde{R}_{\mu \nu \sigma \rho }, \\
& \tilde{R}_{\mu \nu \rho \sigma }=-\tilde{R}_{\nu \mu \rho \sigma }+\alpha
g_{\mu \nu }\tilde{W}_{\rho \sigma }, \\
& \tilde{R}_{\mu \nu \rho \sigma }=\tilde{R}_{\rho \sigma \mu \nu }+\frac{%
\alpha }{2}\Big(g_{\mu \nu }\tilde{W}_{\rho \sigma }-g_{\rho \sigma }\tilde{W}_{\mu \nu }
\notag \\
& \quad +g_{\nu \sigma }\tilde{W}_{\mu \rho }-g_{\nu \rho }\tilde{W}_{\mu \sigma }+g_{\mu
\rho }\tilde{W}_{\nu \sigma }-g_{\mu \sigma }\tilde{W}_{\nu \rho }\Big), \\
& \tilde{R}_{\mu \nu }=\tilde{R}_{\nu \mu }+2\alpha \tilde{W}_{\mu \nu }.
\end{align}
\end{subequations}

Note that the curvature tensor does not satisfy the same symmetries as in Riemann geometry, unless $\tilde{W}_{\mu \nu }=0$. The contractions of the Weyl curvature tensor are defined as
\begin{equation}
\tilde{R}_{\mu \nu }=\tilde{R}\indices{^{\lambda }_{\mu \lambda \nu }}, \qquad\tilde{R}%
=g^{\mu \sigma }\tilde{R}_{\mu \sigma }.
\end{equation}
The Weyl scalar takes the form
\begin{equation}
\tilde{R}=R-3\alpha \nabla _{\mu }\omega ^{\mu }-\frac{3}{2} \alpha
 ^{2}\omega _{\mu }\omega ^{\mu },  \label{R}
\end{equation}
where $R$ is the Ricci scalar defined in Riemann geometry.

Under a conformal transformation with a conformal factor $\Sigma$, the
metric tensor, of the Weyl field, and of a scalar field $\phi$ transform as
\begin{subequations} \label{a2}
    \begin{align}
        \hat g_{\mu\nu} &= \Sigma ^2g_{\mu\nu},\\
        \hat \omega _\mu &= \omega_\mu - \frac{2}{\alpha} \partial_\mu \ln\Sigma, \\
        \hat \phi &= \Sigma^{-1} \phi.
    \end{align}
\end{subequations}
\subsection{Conformal invariance of Maxwell's equations}

In Riemann geometry, electromagnetic fields are described by the potential $A_\mu$ and the field strength $F_{\mu \nu}$, defined as the antisymmetrized derivative of the potential:
\begin{equation}\label{F}
F_{\mu \nu}=\nabla_\mu A_\nu-\nabla _\nu A_\mu.
\end{equation}
Note the formal analogy between the above equation and \cref{W}, which was used by Weyl \cite{Weyl} to propose a unified theory of gravitation and electromagnetism. However, in the present work, we consider $\tilde{W}_{\mu\nu}$ as a purely geometric quantity that has no direct physical or geometric relation to the electromagnetic potential. However, as we shall see, the presence of a Weyl geometric structure on the spacetime manifold may have important implications on the behavior of the electromagnetic fields.

Maxwell's equations in Riemann geometry can be derived from the action \cite{Far}
\begin{equation}
S_{(em)}=\int{\left(-\frac{1}{4}F_{\mu \nu}F^{\mu \nu}+4\pi A_\mu j^{\mu}\right)%
\sqrt{-g} \, d^4x},
\end{equation}
where $j^{\mu}$ is the $4$-current. The equations take the covariant form
\begin{equation}
\nabla _\nu F^{\mu \nu}=-4\pi j^{\mu},\;\;\;\varepsilon^{\alpha \beta \mu \nu}\nabla _\beta
F_{\mu \nu}=0,
\end{equation}
where $\varepsilon^{\alpha \beta \mu \nu}$ is the Levi-Civita tensor. The potential satisfies the equation
\begin{equation} \label{weq}
\left( \nabla^\nu \nabla_\mu - \delta^\nu_\mu \nabla^\sigma \nabla_\sigma \right) A_\nu = 0.
\end{equation}
Since the term $\Psi ^{\lambda}_{\mu \nu}$ in the Weyl connection is symmetric, it immediately follows that the definition of the electromagnetic field tensor takes the same form in Weyl geometry:
\begin{equation}
\tilde{F}_{\mu \nu}=F_{\mu \nu}=\tilde{\nabla}_\mu A_\nu-\tilde{\nabla} _\nu A_\mu=\nabla_\mu A_\nu-\nabla _\nu A_\mu.
\end{equation}
Thus, the relation between the field tensor and the potentials is the same in both Riemann and Weyl geometries. On the other hand, the contravariant and mixed components of the electromagnetic field tensor have the transformation rules,
\begin{equation}
\tilde{F}^{\mu \nu}=\Sigma ^{-4}F^{\mu \nu}, \qquad \tilde{F}\indices{_\nu^\mu}=\Sigma ^{-2}F\indices{_\nu^\mu}.
\end{equation}
Therefore, the conformally rescaled Maxwell's equations take the form
\begin{equation}
\tilde{g}^{\mu \nu}\tilde{\nabla}_\nu\tilde{F}_{\mu \sigma}=-4\pi \tilde{j}_\sigma, \qquad \varepsilon^{\alpha \beta \mu \nu}\tilde{\nabla} _\beta
\tilde{F}_{\mu \nu}=0,
\end{equation}
with the current $j_\sigma$ having the transformation law $\tilde{j}_\sigma =\Sigma ^{-2}j_\sigma$ \cite{Far}. 

One can also show that the wave equation \cref{weq} is also conformally invariant \cite{Far}. Moreover, the action of the electromagnetic field is also invariant with respect to the conformal transformations, and in the Weyl geometry takes the form
\begin{equation}
\tilde{S}_{(em)}=\int{\left(-\frac{1}{4}\tilde{F}_{\mu \nu}\tilde{F}^{\mu \nu}+4\pi \tilde{A}_\mu \tilde{j}^{\mu}\right)%
\sqrt{-\tilde{g}} \, d^4x}.
\end{equation}
In \cref{sect2}, we will also show that the null geodesic equations are invariant under conformal transformations. This result follows from the conformal invariance of Maxwell's equations, together with the geometrical optics approximation.

\subsection{Weyl geometric gravity: Action, field equations, and black hole solutions}

The simplest gravitational Lagrangian density, which is conformally invariant, can be introduced in Weyl geometry according to the definition \cite{Weyl,Gh3,Gh4,Gh5,Gh6,Gh7}
\begin{equation}  \label{inA}
L_W=\left( \frac{1}{4! \, \xi^2} \tilde R^2 - \frac{1}{4} \tilde{W}_{\mu\nu} \tilde{W}^{\mu\nu} \right) \sqrt{-\tilde{g}},
\end{equation}
where we have denoted by $\xi < 1$ the parameter of the perturbative coupling. The Lagrangian $L_W$ can be linearized by the replacement $\tilde{R}^2\rightarrow 2 \phi_0^2\,\tilde R-\phi_0^4$, where $\phi_0$ is an auxiliary scalar field. It is easy to check that the new Lagrangian density is mathematically equivalent to the initial one. This result follows from the use of the solution $\phi_0^2=\tilde R$ of the equation of motion of $\phi_0$ in the new $L_W$. Therefore, we obtain a new Weyl geometric Lagrangian containing a scalar degree of freedom, expressed as
\begin{equation}\label{alt3}
L_W= \left( \frac{\phi_0^2}{12 \xi^2}  \tilde R - \frac{\phi_0^4}{4!\,\xi^2} - \frac{1}{4} \tilde{W}_{\mu\nu} \tilde{W}^{\mu\nu}  \right) \sqrt{-\tilde{g}}.
\end{equation}
This Lagrangian represents the simplest gravitational Lagrangian density containing the Weyl gauge symmetry, as well as conformal invariance. As we have already mentioned, $L_W$ has a spontaneous breaking to an Einstein-Proca Lagrangian of the Weyl gauge field. Substituting into \cref{alt3} the expression of $\tilde{R}$ given in \cref{R}, after performing a gauge transformation and a redefinition of the physical and geometric variables, we obtain a Riemann geometry action, invariant under conformal transformation and given by \cite{Gh3,Gh4,Gh5} 
\bea\label{a3}
\mathcal{S} &=& \int \Bigg[ \frac{\phi^2}{12 \xi^2} \Big( R - 3\alpha\nabla_\mu \omega^\mu - \frac{3}{2} \alpha^2 \omega_\mu \omega^\mu \Big) \nonumber\\
&&- \frac{ \phi^4}{4! \,\xi^2} - \frac{1}{4} \tilde{W}_{\mu\nu} \tilde{W}^{\mu\nu} \Bigg]  \sqrt{-g} \, d^4x,
\eea

The field equations of this theory can be obtained by varying the action \eqref{a3} with respect to the metric tensor and are given by \cite{Yang}
\bea\label{b2a}
&&\frac{\phi ^{2}}{\xi ^{2}}\Big(R_{\mu \nu }-\frac{1}{2}Rg_{\mu \nu }\Big)+%
\frac{1}{\xi ^{2}}\Big(g_{\mu \nu }\Box -\nabla _{\mu }\nabla _{\nu }\Big)%
\phi ^{2}   \nonumber\\
&&-\frac{3\alpha }{2\xi ^{2}}\Big(\omega ^{\rho }\nabla _{\rho }\phi
^{2}g_{\mu \nu }-\omega _{\nu }\nabla _{\mu }\phi ^{2}-\omega _{\mu }\nabla
_{\nu }\phi ^{2}\Big)   \nonumber\\
&&+\frac{3\alpha ^{2}}{4\xi ^{2}}\phi ^{2}\Big(\omega _{\rho }\omega ^{\rho
}g_{\mu \nu }-2\omega _{\mu }\omega _{\nu }\Big) -6\tilde{W}_{\rho \mu }%
\tilde{W}_{\sigma \nu }g^{\rho \sigma }   \nonumber\\
&&+\frac{3}{2} \tilde{W}_{\rho\sigma} \tilde{W}^{\rho\sigma} g_{\mu \nu }+\frac{1}{4\xi ^{2}}%
\phi ^{4}g_{\mu \nu }=0.
\eea
Taking the trace of the above equation, we obtain
\begin{equation}  \label{b3n}
\Phi R+3\alpha \omega ^{\rho }\nabla _{\rho }\Phi -\Phi ^{2}-\frac{3}{2}%
\alpha ^{2}\Phi \omega _{\rho }\omega ^{\rho }-3\Box \Phi =0,
\end{equation}
where we have introduced the notation $\Phi \equiv \phi ^{2}$. By varying the action \eqref{a3} with respect to the scalar field $\phi$ we find
\begin{equation}  \label{b4}
R-3\alpha \nabla _{\rho }\omega ^{\rho }-\frac{3}{2}\alpha ^{2}\omega _{\rho
}\omega ^{\rho }-\Phi =0.
\end{equation}
The above relation represents the equation of motion of the scalar field $\phi$. From Eqs. \eqref{b3n} and \eqref{b4} we immediately obtain
\begin{equation}  \label{b5}
\Box \Phi-\alpha \nabla _{\rho }(\Phi\omega ^{\rho })=0.
\end{equation}
The equation of motion of the Weyl vector is obtained as
\begin{equation}  \label{Fmunu}
4\xi ^{2}\nabla _{\nu }\tilde{W}^{\mu \nu }+\alpha ^{2}\Phi\omega ^{\mu
}-\alpha \nabla ^{\mu }\Phi=0.
\end{equation}
Applying $\nabla_{\mu}$ to both sides of the above equation, we obtain \cref{b5}, a result that indicates the consistency of the field equations of the theory.

\subsection{Black hole solutions} \label{sec:bh}

We now introduce a static and spherically symmetric geometry, with coordinates $\left( t, r, \theta, \varphi \right)$. Thus, the line element can be written as
\begin{equation} \label{eq:metric0}
ds^{2}=e^{\nu (r)}dt^{2}-e^{\lambda (r)}dr^{2}-r^{2}d\Omega ^{2},
\end{equation}
where $d\Omega ^{2}=d\theta ^{2}+\sin ^{2}\theta d\varphi^{2}$. In the following, a prime denotes the derivative with respect to the radial coordinate $r$. Furthermore, we assume that the Weyl vector depends only on the radial coordinate $r$ and has only one nonvanishing component, so that $\omega _{\mu}$ is represented as $\omega _{\mu }=(0,\omega _{r}(r),0,0)$. Therefore, the one-form $\omega_{\mu}$ is closed and we have $\tilde{W}_{\mu \nu}\equiv 0$. We are in the special case of Weyl integrable geometry \cite{Romero_2012}. One can argue that this is the most physically relevant case of Weyl geometry, since the second clock effect does not arise \cite{Avalos2018}. This means that the ticking rates of clocks set by different timelike observers do not depend on their histories, and frequencies measured by different observers with the same $4$-velocities do not depend on the previous states of motion of the observers.

From \cref{Fmunu}, we obtain
\begin{equation}  \label{15}
\Phi^{\prime} =\alpha \Phi \omega _r.
\end{equation}
The gravitational field equations take the form \cite{Yang}
\begin{subequations}
    \bea\label{c15}
   && -1+e^{\lambda }-\frac{1}{4}e^{\lambda }r^{2}\Phi -\frac{2r\Phi ^{\prime }}{\Phi }+\frac{3r^{2}}{4}\frac{\Phi ^{\prime 2}}{\Phi ^{2}}+r\lambda ^{\prime } \nonumber\\
            && +\frac{r^{2}\lambda ^{\prime }}{2}\frac{\Phi ^{\prime }}{\Phi }-\frac{r^{2}\Phi ^{\prime \prime }}{\Phi }=0,
        \eea
        \bea\label{c16}
        &&1-e^{\lambda }+\frac{1}{4}e^{\lambda }r^{2}\Phi +\frac{2r\Phi ^{\prime }}{\Phi }+\frac{3r^{2}}{4}\frac{\Phi ^{\prime 2}}{\Phi ^{2}}\nonumber\\
            && +r\nu ^{\prime}\left( 1+\frac{r}{2}\frac{\Phi ^{\prime }}{\Phi }\right) =0,
        \eea
\end{subequations}
and
\begin{equation}\label{e22}
\begin{split}
&2(\nu ^{\prime }-\lambda ^{\prime })+(4-2r\lambda ^{\prime
}+2r\nu ^{\prime })\frac{\Phi ^{\prime }}{\Phi }    \\
&\quad + r\left( e^{\lambda }\Phi +4\frac{\Phi ^{\prime \prime }}{%
\Phi }-3\frac{\Phi ^{\prime 2}}{\Phi ^{2}}-\lambda ^{\prime }\nu ^{\prime
}+\nu ^{\prime 2}+2\nu ^{\prime \prime }\right) =0.
\end{split}
\end{equation}
The above system admits an exact solution, given by \cite{Yang}
\begin{equation}
e^{-\lambda }=e^{\nu }=1-\delta +\frac{\delta (2-\delta )}{3}\frac{r}{r_s}-%
\frac{r_s}{r}+C_{3}r^{2},  \label{metrW}
\end{equation}
where $\delta $ and $C_{3}$ are arbitrary integration constants, while $r_s=2M$ represents the gravitational mass of the compact object. For the scalar field, we obtain the expression
\begin{equation}
\Phi (r)=\frac{C_{1}}{\left( r+C_{2}r_s\right) ^{2}}=\frac{C_{1}}{r_s^{2}%
}\frac{1}{\left( \frac{r}{r_s}+C_{2}\right) ^{2}},  \label{40}
\end{equation}
where $C_{1}$ and $C_{2}$ are arbitrary integration constants. Finally, the radial component of the Weyl covector can be obtained as
\begin{equation} \label{eq:omega_r}
\omega _{r}=\frac{1}{\alpha }\frac{\Phi ^{\prime }}{\Phi }=-\frac{2}{\alpha }%
\frac{1}{r_s}\frac{1}{\frac{r}{r_s}+C_{2}},
\end{equation}
while its contravariant representation is given by
\bea
\omega ^{r} &=& g^{11}\omega _{r}=\frac{2}{\alpha }\frac{1}{r_s}\frac{%
e^{-\lambda }}{\frac{r}{r_s}+C_{2}}  \nonumber\\
&=& \frac{2}{\alpha r_s}\frac{ 1-\delta +\frac{\delta (2-\delta )}{3}%
\frac{r}{r_s}-\frac{r_s}{r}+C_{3}r_s^{2}\left( \frac{r}{r_s}\right)
^{2} }{\frac{r}{r_s}+C_{2} }.
\eea

We want to point out that similar black hole solutions can be found in conformal gravity \cite{M1}, and in de Rham-Gabadadze-Tolley (dRGT) massive gravity theory \cite{Pi}.

\section{WKB approximation for Maxwell's equations in Weyl geometry}\label{sect2}

In this section, we investigate the dynamics of the high-frequency electromagnetic waves in Weyl geometry. We perform a WKB analysis of Maxwell's equations, and we derive the equations of geometrical optics.

We consider Maxwell's equations for the vector potential
\begin{equation}
\left( \tilde{\nabla}^\nu \tilde{\nabla}_\mu - \delta^\nu_\mu \tilde{\nabla}%
^\sigma \tilde{\nabla}_\sigma \right) A_\nu = 0,
\end{equation}
and we fix the gauge by imposing the Lorenz gauge condition
\begin{equation}
\tilde{\nabla}_\mu A^\mu = 0.
\end{equation}

To describe the propagation of high-frequency electromagnetic waves, we
assume that the vector potential admits a WKB expansion
\begin{equation}  \label{eq:WKB_ansatz}
A_\mu = \left[ {A_0}_\mu + \epsilon {A_1}_\mu + \mathcal{O}(\epsilon^2) %
\right] e^{i S/\epsilon} ,
\end{equation}
where $\epsilon$ is a small expansion parameter related to the wavelength, $S$ is a real phase function, and ${A_i}_\mu$ are complex amplitudes. We define a wave vector as $k_\mu = \nabla_\mu S$, and the wave frequency $f$ measured by a timelike observer with $4$-velocity $t^\mu$ is $f = k_\mu t^\mu / \epsilon$. The definition of frequency does not suffer any modifications compared to the Riemannian case. Although vectors change their length under parallel transport in Weyl geometry, the vector $k_\mu$ that describes wave dynamics at the lowest order in the geometrical optics approximation is null and geodesic (as shown below). Therefore, $k_\mu$ will not be affected by length changes related to the nonmetricity of Weyl geometry. The observer $4$-velocity $t^\mu$ is not a dynamical quantity and represents an external choice relative to which we measure the frequency. Furthermore, note that frequency is defined with respect to a $4$-velocity (unit timelike vector), and not just any arbitrary timelike vector. Therefore, even if we decide to build a family of timelike observers by means of parallel transport, in Weyl geometry, the lengths of the vectors will generally change, and we would need to normalize the vectors after parallel transport to be able to use them in frequency measurements.

The validity of our high-frequency approximation is based on the same assumptions as in Riemannian geometry (see, for example, Ref. \cite[Sec. 22.5]{MTW}): the characteristic length scale over which the electromagnetic waves vary significantly (represented by the wavelength $\epsilon$) is considered to be much smaller than the characteristic length scale over which the properties of spacetime change significantly. For the black hole solutions introduced in \cref{sec:bh}, both the metric and the Weyl vector vary significantly over the same length scale given by the Schwarzschild radius $r_s$.

Our WKB analysis follows the same steps as in Ref. \cite[Sec. 3.2]{GSHE_rev}, and consists in inserting the WKB ansatz \eqref{eq:WKB_ansatz} into Maxwell's equations and the Lorenz gauge condition. Then the resulting equations will be analyzed at each order in $\epsilon$.

We start by inserting the WKB ansatz into the Lorenz gauge condition. At the
lowest two orders in $\epsilon$, we obtain
\begin{subequations}
\begin{align}
\mathcal{O}(\epsilon^{-1}):& \quad k^\mu {A_0}_\mu = 0,  \label{eq:Lorenz0}
\\
\mathcal{O}(\epsilon^0):& \quad \tilde{\nabla}^\mu {A_0}_\mu + i k^\mu {A_1}%
_\mu + \alpha {A_0}_\mu \omega^\mu = 0.  \label{eq:Lorenz1}
\end{align}
Note that the lowest-order equation is the same as in Riemann geometry, while the next-to-leading-order equation explicitly depends on the Weyl vector field $\omega^\mu$.

We continue our analysis by inserting the WKB ansatz into Maxwell's equations. Making use of Eq. \eqref{eq:Lorenz0}, at the lowest order in $\epsilon$, we obtain the geometrical optics dispersion relation
\end{subequations}
\begin{equation}
k_\mu k^\mu = 0.
\end{equation}
This is a Hamilton-Jacobi equation for the phase function $S$, which can be solved by using the method of characteristics \citep[Sec. 46]{Arnold_book}. For this purpose, we consider canonical coordinates $(x^\mu, p_\mu)$ on the cotangent bundle $T^*M$ and we define a Hamiltonian
\begin{equation}
H(x, p) = \frac{1}{2} g^{\alpha \beta} p_\alpha p_\beta = 0.
\end{equation}
The corresponding Hamilton's equations are
\begin{subequations}
\label{eq:geodesics}
\begin{align}
\dot{x}^\mu &= \frac{\partial H}{\partial p_\mu} = p^\mu, \\
\dot{p}_\mu &= -\frac{\partial H}{\partial x^\mu} = \tilde{\Gamma}%
^\alpha_{\beta \mu} p_\alpha p^\beta - \alpha \omega_\mu H. \label{eq:Eq_p0}
\end{align}
\end{subequations}
The derivation of \cref{eq:Eq_p0} above follows easily after expressing $\partial_\mu g^{\alpha \beta}$ in terms of $\tilde{\Gamma}^\lambda_{\mu \nu}$ and $\omega_\lambda$. This can be done using \cref{nmetr}, from which we immediately obtain $\partial_\mu g^{\alpha \beta} = \alpha \omega_\mu g^{\alpha \beta} - \tilde{\Gamma}^{\alpha}_{\mu \rho} g^{\rho \beta} - \tilde{\Gamma}^{\beta}_{\mu \rho} g^{\alpha \rho}$. The right-hand side of \cref{eq:Eq_p0} is obtained after performing the contractions with the momentum variable in $-\frac{1}{2}\partial_\mu g^{\alpha \beta} p_\alpha p_\beta$. However, the momentum is null ($H = 0$) and we can also use \cref{Con} to rewrite the above equations as
\begin{subequations}
\begin{align}
\dot{x}^\mu &= p^\mu, \\
\dot{p}_\mu &= \Gamma^\alpha_{\beta \mu} p_\alpha p^\beta.
\end{align}
\end{subequations}
Thus, we recovered the well-known result of geometrical optics that light rays follow the null geodesic equations of the background spacetime. Furthermore, the conformal invariance of null geodesics is reflected by the fact that the above equations do not depend on the Weyl geometry connection $\tilde{\Gamma}^\alpha_{\beta \mu}$, but only on the Levi-Civita connection $\Gamma^\alpha_{\beta \mu}$. The geodesic equations can also be derived in a covariant form by taking the covariant derivative of the geometrical optics dispersion relation:
\begin{equation}
0 = \tilde{\nabla}_\nu \left(\frac{1}{2} k_\mu k^\mu \right) = k^\mu \tilde{\nabla}_\nu k_\mu = k^\mu \tilde{\nabla}_\mu k_\nu = k^\mu \nabla_\mu k_\nu.
\end{equation}

At the next order in $\epsilon$, we obtain a transport equation for the amplitude ${A_0}_\mu$:
\begin{equation}  \label{eq:transp}
k^\nu \tilde{\nabla}_\nu {A_0}_\mu + \frac{1}{2} {A_0}_\mu \tilde{\nabla}%
^\nu k_\nu + \frac{\alpha}{2} k_\mu {A_0}_\nu \omega^\nu = 0.
\end{equation}
The last term in the above equation is not present in Riemann geometry. To analyze the above transport equation, we expand the complex amplitude ${A_0}_\mu$ as
\begin{equation}  \label{eq:field_expansion}
{A_0}_\mu = \sqrt{I} a_\mu,
\end{equation}
where $I = {\bar{A}_0}^{\mu} {A_0}_\mu$ is a real intensity, $a_\mu$ is a unit-complex polarization vector ($\bar{a}^\mu a_\mu = 1$). Then the transport equation \eqref{eq:transp} can be split into transport equations for the field intensity $I$ and the polarization vector $a_\mu$:
\begin{subequations}
\begin{align}
\tilde{\nabla}_\mu \left( I k^\mu \right) &= 2 \alpha I k_\mu \omega^\mu, \\
k^\nu \tilde{\nabla}_\nu a_\mu &= -\alpha k_{(\mu} a_{\nu)} \omega^\nu. \label{eq:transp1}
\end{align}
\end{subequations}

While in Riemann geometry the polarization vector is parallel transported along $k^\nu$, we see that this is no longer the case in Weyl geometry. Since we required the polarization vector to have unit norm, and keeping in mind that this is not generally conserved by parallel transport in Weyl geometry, it immediately follows that $a_\mu$ cannot satisfy a parallel transport equation. The additional term on the right-hand side of \cref{eq:transp1} ensures that the norm of $a_\mu$ is conserved in Weyl geometry.

It follows from \cref{eq:Lorenz0} that the polarization vector $a_\mu$ must be orthogonal to $k_\mu$. To further analyze the dynamics of the polarization vector, it is convenient to introduce a tetrad $\{ k_\mu, t_\mu, m_\mu, \bar{m}_\mu \}$ adapted to the wave vector $k_\mu$. Here, $t_\alpha$ is a real timelike vector, $m_\alpha$ and $\bar{m}_\alpha$ are complex null vectors, and the only nonzero contractions are $t_\alpha t^\alpha = 1$, $k_\alpha t^\alpha = \epsilon \omega$ and $m_\alpha \bar{m}^\alpha = -1$. Note that, due to the orthogonality relation $k_\alpha m^\alpha = 0$, the vector $m^\alpha = m^\alpha(x,k_{\mu}(x))$ is effectively a function of both $x$ and $k_\mu(x)$, similar to the Riemannian case \cite{O1}. This additional dependence can also be viewed as a general consequence of WKB approximations, where it is always the case that the amplitude in the WKB ansatz will be defined on the Lagrangian submanifold $(x, k_\mu(x) = \nabla_\mu S(x)) \subset T^*M$ \cite{Emmrich1996,MR1806388,Littlejohn}. Using this tetrad, the polarization vector can be expanded as
\begin{equation}
a_\mu = z_1 m_\mu + z_2 \bar{m}_\mu + z_3 k_\mu,
\end{equation}
where $z_i$ are complex scalar functions. The last term in the above equation represents a residual gauge degree of freedom not fixed by the Lorenz gauge, and we can ignore it in the following. The terms proportional to $z_1$ and $z_2$ describe the state of polarization of the electromagnetic wave, with circular polarization corresponding to $z_1 = 0$ or $z_2 = 0$.

Using \cref{eq:transp1}, we can derive a transport equation for the complex scalars $z_i$. It is convenient to introduce a Jones vector $z = (z_1 \,\, z_2)^T$, which will satisfy the transport equation
\begin{equation}  \label{eq:transp_z}
\dot{z} = k^\mu \tilde{\nabla}_\mu z = i k^\mu B_\mu \sigma_3 z,
\end{equation}
where $\sigma_3$ is the third Pauli matrix and $B_\mu = B_\mu(x, k(x))$ represents a Berry connection defined as

\begin{equation}
\begin{split}
B_\mu &= -i \left\{ \bar{m}^\nu(x, k(x)) \tilde{\nabla}_\mu \left[ m_\nu(x, k(x)) \right] - \frac{\alpha}{2}
\omega_\mu(x) \right\}
 .
\end{split}
\end{equation}
Note that compared to the Berry connection obtained in Riemann geometry \cite[Eq. 35]{GSHE_rev}, here there is an additional term proportional to $\omega_\mu$ in the above equation. This means that the Weyl vector field has a nontrivial contribution to the polarization dynamics. Furthermore, since $m_\nu$ and $\bar{m}_\nu$ are orthogonal to $k_\mu$, they will be functions of both $x^\mu$ and $k_\mu$. Then the action of the covariant derivative on $m_\nu$ and $\bar{m}_\nu$ should be understood as in Ref. \cite[Eq. 3.37]{O1}:
\begin{equation}
\begin{split}
    &k^\mu \tilde{\nabla}_\mu \left[ m_\nu (x, k(x)) \right] = k^\mu \bigg[ \frac{\partial m_\nu}{\partial x^\mu}(x,k(x)) \\
    &\qquad\quad - \tilde{\Gamma}_{\mu \nu}^\sigma m_\sigma(x,k(x))  + \tilde{\Gamma}^\sigma_{\alpha \mu} k_\sigma \frac{\partial m_\nu}{\partial k_\alpha} (x, k(x)) \bigg].
\end{split}
\end{equation}

The transport equation \eqref{eq:transp_z} can be integrated as
\begin{equation}
z(\tau) =
\begin{pmatrix}
e^{i \gamma(\tau)} & 0 \\
0 & e^{- i \gamma(\tau)}
\end{pmatrix}
z(0),
\end{equation}
where $\gamma$ is the Berry phase, defined as
\begin{equation}
\gamma(\tau) = \int_{0}^\tau d\tau^{\prime} k^\mu B_\mu.
\end{equation}
The Berry phase encodes the evolution of the polarization in a circular basis. Note that the state of circular polarization of an electromagnetic wave is conserved.

\section{Going beyond geometrical optics in Weyl geometry - the gravitational spin Hall effect of light}

In this section, we take into account the spin-orbit coupling for electromagnetic waves propagating in curved spacetime, and we derive the ray equations of the gravitational spin Hall effect of light in Weyl geometry. To provide a better context for this effect, we start with a brief review of the spin Hall effect of light in optics. Detailed reviews of the spin Hall effect of light in optics can be found in Refs. \cite{opt1,rev}, and a comparison between the optical and gravitational cases can be found in Ref. \cite{GSHE_rev}.

\subsection{Brief review of the spin Hall effect for light in optics}

We briefly review the spin Hall effect for light in optics, following the presentations given in Refs. \cite{B1,B2,B3,opt2} (different approaches can be found in Refs. \cite{Duval2006,Duval2007,Ruiz2015}). The geometrical optics approximation for the propagation of light is similar to the semiclassical limit of quantum mechanics \cite{B1}. To describe the propagation of light in some optical medium, the short-wavelength approximation is based on the assumption that the wavelength $\lambda$ of light is much smaller than the characteristic length scale $L$ of the variation of the medium, such that $\lambda / L  \ll 1$. In this approximation, the propagation of electromagnetic wave packets is effectively described as the motion of a point particle. The dynamical propagation of this wave packet can then be described using the canonical formalism on the phase space $({\bf r}, {\bf p})$. In the following, we introduce the dimensionless wave momentum ${\bf p}= \lambdabar {\bf k}$, where $\lambdabar = \lambda/2\pi$ and ${\bf k}$ is the average wave vector of the wave packet. The parameter $\lambdabar$ plays the same role as Planck's constant in the semiclassical approximation of quantum mechanics \cite{B1,B2}.

Electromagnetic waves do possess an intrinsic property - the polarization or spin, which determines the intrinsic angular momentum of light. The two spin eigenstates of light are given by the left-hand and right-hand circular polarizations of the photons, which are determined by the helicity $\sigma  = \pm 1$. For one photon, the spin angular momentum is $\sigma  {\bf p} / |{\bf p}|$.

In the $\lambdabar\rightarrow 0$ limit of the electromagnetic wave equations, the internal and external degrees of freedom of light are decoupled. Therefore, the propagation of lights is independent of the polarization, and the polarization vector is parallel transported along the light ray. This is similar to the geometrical optics description presented in \cref{sect2}. To take into account spin-orbit interactions between the internal and external degrees of freedom, one must go to the first-order approximation \cite{B1,B2,B3,opt2}. In this approximation, the orbital degrees of freedom (average position and momentum) and the polarization are coupled, and wave packets generally follow polarization-dependent trajectories.

The Lagrangian $L_{SOI}$ that describes the spin-orbit interaction is given by
\begin{equation} \label{SOI}
    L_{SOI}=-\lambdabar \sigma {\bf{A}}\left({\bf{p}}\right) \cdot \dot{\bf{p}},
\end{equation}
where $\bf{A}\left(\bf{p}\right)$ is the Berry connection, which has a purely geometric origin. The Lagrangian that describes the spin Hall effect of light in an inhomogeneous medium with refractive index $n\left(\bf{r}\right)$ is given by \cite{B1,B2}
\begin{equation}\label{Lagr}
    L = L_0 + L_{SOI} = \bf{p} \cdot \dot{\bf{r}}+n(\bf{r})-p - \lambdabar \sigma \bf{A}\left(\bf{p}\right) \cdot \dot{\bf{p}}
\end{equation}
where $L_0=\bf{p} \cdot \dot{\bf{r}}+n(\bf{r})-p$ is the Lagrangian of the geometrical optics limit $\lambdabar\rightarrow 0$. The role of the Berry connection in $L_{SOI}$ can be better understood by introducing a parametrization of the basis vectors of the ray coordinate frame of the form $\bf{t}=\bf{t}\left(\bf{p}\right) = \bf{p}/{|\bf{p}|}$, $\bf{v}=\bf{v}\left(\bf{p}\right)$, and $\bf{w}=\bf{w}\left(\bf{p}\right)$. Then the Berry connection can be written as
\begin{equation}
A_i={\bf{v}} \cdot \frac{\partial {\bf{w}}}{\partial p_i}.
\end{equation}
The wave polarization is generally described by a unit complex vector $\bf{e} = \bf{e}(\bf{p})$ orthogonal to the wave momentum $\bf{p}$. The space of all possible directions of $\bf{t} = \bf{p}/{|\bf{p}|}$ can be identified with the two-sphere $\bf{S}^2$. Thus, the polarization vector $\bf{e}(\bf{p})$ is tangent to $\bf{S}^2$, and its dynamics is described by the Berry connection as parallel transport over $\bf{S}^2$. One can also associate a curvature tensor with the Berry connection, given by \cite{B1,B2}
\begin{equation}
F_{ij}=\frac{\partial A_j}{\partial p_i}-\frac{\partial A_i}{\partial p_j}=\frac{\partial \bf{v}}{\partial p_i} \cdot \frac{\partial \bf{w}}{\partial p_j}-\frac{\partial \bf{v}}{\partial p_j} \cdot \frac{\partial \bf{w}}{\partial p_i}.
\end{equation}
The Berry curvature tensor is antisymmetric and one can associate a dual vector $\bf{F}= \frac{\partial }{\partial \bf{p}}\times \bf{A}$, so that $F_{ij}=\epsilon _{ijk}F_k$. For an electromagnetic wave, the Berry curvature is given by $\bf{F}=\bf{p}/|\bf{p}|^3$ \cite{B1}. Finally, the equations of motion of the polarized light ray, which describe the Hall spin effect of light, can be obtained from the Lagrangian (\ref{Lagr}) as \cite{B1}
\begin{subequations} \label{eq:SHEL}
    \begin{align}
        \dot{\bf{r}} &= \frac{\bf{p}}{|\bf{p}|}+\lambdabar_0\sigma\dot{\bf{p}}\times \bf{F} = \frac{\bf{p}}{|\bf{p}|} + \lambdabar \sigma \frac{\dot{\bf{p}}\times \bf{p}}{|\bf{p}|^3}, \\
        \dot{\bf{p}} &= \nabla n.
    \end{align}
\end{subequations}
Compared to the geometrical optics limit $\lambdabar\rightarrow 0$, there is an additional term in the equation for $\dot{\bf{r}}$. This additional term depends linearly on wavelength and its sign is determined by the state of circular polarization of the wave packet. Additionally, note that while the polarization dynamics is governed by the Berry connection, as was also the case in \cref{sect2}, the spin Hall equations are defined using the Berry connection. We will observe the same behavior when deriving the gravitational spin Hall equations in the next section.

The above spin Hall equations can already be used to infer results about the propagation of polarized light in gravitational fields. It is well known that the propagation of electromagnetic waves in curved spacetime can be analogously described as electromagnetic waves propagating in some dielectric medium. This analogy was first mentioned by Eddington \cite{Eddington} and was later developed by several authors \cite{Gordon,skrotskii, Balazs, Plebansky-Maxwell,deFelice}. Using this analog framework, the effect of curved spacetime on the propagation of light could be encoded by an inhomogeneous refractive index $n(\bf{r})$. In particular, it has been shown in Ref. \cite{GSHE_review} that the gravitational spin Hall equations in Schwarzschild spacetime, as first derived in Ref. \cite{G1}, can be obtained from \cref{eq:SHEL} by an appropriate choice of refractive index $n(\bf{r})$. The reverse statement of obtaining \cref{eq:SHEL} from the generally covariant form of the gravitational spin Hall equations has also been shown in Ref. \cite{O2} by using Gordon's optical metric.

\subsection{The gravitational spin Hall effect of light in Weyl geometry}

In this section, we go beyond the geometrical optics approximation of \cref{sect2} and we derive the gravitational spin Hall equations in Weyl geometry. We will follow the approach given in Ref. \cite{GSHE_rev}.

In the geometrical optics treatment presented in \cref{sect2}, the polarization dynamics is influenced by the geodesic rays followed by high-frequency electromagnetic waves, but there is no backreaction from the polarization onto the rays. In other words, spin-orbit interactions between the external (position and momentum) and internal (spin or polarization) degrees of freedom of the electromagnetic wave are not fully taken into account. This can be solved as in Ref. \cite{GSHE_rev}. First, we note that for circularly polarized electromagnetic waves, the WKB field will take the form
\begin{equation}
A_\alpha = \sqrt{I} m_\alpha e^{i (S+\epsilon \gamma)/\epsilon} \quad \text{%
or} \quad A_\alpha = \sqrt{I} \bar{m}_\alpha e^{i (S-\epsilon
\gamma)/\epsilon}.
\end{equation}
The above fields have a total phase factor $\tilde{S} = S + \epsilon s \gamma $, with $s = \pm 1$ depending on the state of circular polarization. While the geometrical optics geodesic ray equations \eqref{eq:geodesics} were derived by solving a Hamilton-Jacobi equation for the phase function $S$, higher-order corrections and spin-orbit interactions can be taken into account by defining an effective Hamilton-Jacobi equation for the total phase function $\tilde{S}$ \cite{GSHE_rev}. Using the results from the previous section, we can write this as
\begin{equation}
\frac{1}{2} (\tilde{\nabla}_\mu \tilde{S}) (\tilde{\nabla}^\mu \tilde{S}) -
\epsilon s (\tilde{\nabla}^\mu \tilde{S}) B_\mu = \mathcal{O}(\epsilon^2).
\end{equation}
As discussed in Refs. \cite[Box 25.3]{MTW} \cite[Sec. II]{Gerlach1969}, solving the Hamilton-Jacobi equation by means of point-particle ray equations (method of characteristics) is related to the principle of constructive interference. A localized wave packet can be constructed by taking a superposition of WKB plane waves with slightly different wave vectors. Then, the peak of intensity of this wave packet occurs where the waves interfere constructively and coincides with the ray trajectories given by the effective point-particle description used for solving the Hamilton-Jacobi equation. Thus, we obtain the polarization-dependent ray equations describing the gravitational spin Hall effect by applying the method of characteristics to the above effective Hamilton-Jacobi equations. In this way, we obtain the effective Hamiltonian
\begin{equation}
H (x, p) = \frac{1}{2} g^{\alpha \beta} p_\alpha p_\beta - \epsilon s p^\mu
B_\mu(x, p) = \mathcal{O}(\epsilon^2).
\end{equation}
The gravitational spin Hall equations can be derived by calculating Hamilton's equations. However, the above Hamiltonian is gauge dependent, in the sense that the Berry connection depends on the choice of complex vectors $m_\mu$ and $\bar{m}_\mu$. This gauge dependence can be removed by performing a coordinate transformation of the type introduced in Refs. \cite{Littlejohn,O1}:
\begin{subequations}
\begin{align}  \label{eq:coord}
x^\mu &\mapsto x^\mu + i \epsilon s \bar{m}^{\nu} \frac{\partial {m}_\nu}{\partial
p_\mu}, \\
p_\mu &\mapsto p_\mu - i \epsilon s \left( \bar{m}^{\nu} \frac{\partial m_\nu}{\partial x^\mu} - \bar{m}^{\nu} \tilde{\Gamma}_{\mu \nu}^\sigma m_\sigma - \frac{\alpha}{2} \omega_\mu \right).
\end{align}
After performing this coordinate transformation, the Hamiltonian reduces to
\end{subequations}
\begin{equation} \label{eq:H=0}
H (x, p) = \frac{1}{2} g^{\alpha \beta} p_\alpha p_\beta = \mathcal{O}%
(\epsilon^2),
\end{equation}
and the gauge-invariant equations of motion describing the gravitational spin Hall effect become
\begin{subequations}
\label{eq:gshe_eq}
\begin{align}
\dot{x}^\mu &= p^\mu + \frac{1}{p_\sigma t^\sigma} S^{\mu \nu} p^\rho \tilde{%
\nabla}_\rho t_\nu,  \label{eq:xdot} \\
\dot{x}^\nu \tilde{\nabla}_\nu p_\mu &= - \frac{1}{2} \tilde{R}_{\rho \sigma
\mu \nu} p^\nu S^{\rho \sigma}.  \label{eq:pdot}
\end{align}
In the above equations, $x^\mu(\tau)$ represents the worldline followed by the energy centroid of the electromagnetic wave packet, $p_\mu(\tau)$ represents the average momentum of the wave packet, and $t^\mu$ is a timelike vector field used to define the energy centroid of the wave packet. We can think of $t^\mu$ as representing the $4$-velocities of a family of timelike observers that describe the centroid and the dynamics of the wave packet (see Ref. \cite{O2} for a detailed discussion of the role of the observer in the gravitational spin Hall equations). The spin tensor $S^{\mu \nu}$ encodes the state of polarization and the angular momentum carried by the wave packet and is defined as
\end{subequations}
\begin{equation}  \label{eq:spin_tensor}
S^{\mu \nu} = 2 i \epsilon s \bar{m}^{[\mu} m^{\nu]} = \epsilon s \frac{
\varepsilon^{\mu \nu \rho \sigma} p_\rho t_\sigma }{p_\beta t^\beta}.
\end{equation}
The gravitational spin Hall equations of motion appear to have the same form as in Riemann geometry. However, the effect of Weyl geometry is hidden in the curvature term from \cref{eq:pdot}, since the symmetry properties of the curvature tensor are not the same in Riemann and Weyl geometry when $\tilde{W}_{\mu \nu} \neq 0$ [see \cref{eq:R_sym}]. Therefore, we must be careful when swapping the indices of the curvature tensor in \cref{eq:pdot}, as terms proportional to $\tilde{W}_{\mu \nu}$ can arise.

A better comparison between the gravitational spin Hall effect in Weyl and Riemann geometry can be achieved by expanding the Weyl covariant derivative and the Weyl curvature tensor in the spin Hall equations. We obtain
\begin{subequations} \label{eq:GSHE_weyl}
\begin{align}
\dot{x}^\mu &= p^\mu + \frac{1}{p \cdot t} S^{\mu \nu} p^\rho
\left( \nabla_\rho t_\nu - \frac{\alpha}{2} t_\rho \omega_\nu \right), \label{eq:GSHE_x}\\
\dot{x}^\nu \nabla_\nu p_\mu &= - \frac{1}{2} R_{\rho \sigma
\mu \nu} p^\nu S^{\rho \sigma} \nonumber\\
&\qquad- \frac{\alpha}{2 p \cdot t} p^\nu S^{\rho \sigma} \bigg( g_{\mu \sigma} t_\nu p^\gamma \nabla_\gamma \omega_\rho \nonumber\\
&\qquad- g_{\mu \sigma} \omega_\nu p^\gamma \nabla_\gamma t_\rho - g_{\mu \nu} \omega_\rho p^\gamma \nabla_\gamma t_\sigma \bigg) \label{eq:GSHE_p}.  
\end{align}
\end{subequations}
In the above form of the spin Hall equations, we can clearly see the effect of the Weyl geometry, given by the terms that contain the Weyl vector field $\omega_\mu$.

In the case of Riemann geometry, the spin Hall equations derived in Ref. \cite{O1} have been shown to be a special case of the Mathisson-Papapetrou equations \cite{O2}. This has the advantage that several known results for the Mathisson-Papapetrou equations, such as conservation laws, can also be applied in the context of the spin Hall equations. Furthermore, while the physical interpretation of the quantities $x^\mu(\tau)$ and $p_\mu(\tau)$ described by the spin Hall equations might not emerge in a transparent way from the WKB derivation, the connection with the Mathisson-Papapetrou equations clarifies this issue. In this context, these quantities are directly related to the stress-energy tensor of localized electromagnetic wave packets, and $x^\mu(\tau)$ represents the trajectory followed by the wave packet energy centroid, defined relative to a family of observers with $4$-velocity $t^\alpha$ \cite{O2}.

The same arguments also extend to the present case of Weyl geometry. A general derivation of the Mathisson-Papapetrou equations in Weyl geometry has been given in Ref. \cite{PhysRevD.90.084034}. Starting from \cite[Eqs. (106)-(107)]{PhysRevD.90.084034}, when we consider vanishing torsion ($T_{a b c} = 0$) and as matter fields the minimally coupled electromagnetic field (in the notation of \cite{PhysRevD.90.084034}, this implies $F = 1$, $h^{a b} = 0$, and $q^{a b c} = 0$; note also the sign difference due to different curvature conventions), we obtain
\begin{subequations} \label{eq:MPD_Weyl}
\begin{align}
    \dot{x}^\mu \nabla_\mu \mathcal{P}_\alpha  &= - \frac{1}{2}  R_{\alpha \beta \gamma \lambda } \dot{x}^\beta S^{\gamma \lambda}, \label{eq:MPD_P}\\
    \dot{x}^\mu \nabla_\mu S^{\alpha \beta} &=  \mathcal{P}^{\alpha} \dot{x}^{\beta} - \mathcal{P}^{\beta} \dot{x}^{\alpha}, \label{eq:MPD_S}
\end{align}
\end{subequations}
where $\mathcal{P}_\mu = p_\mu + \frac{\alpha}{2} \omega^\nu S_{\nu \mu}$. 

Following the same steps and using the same assumptions as in Ref. \cite[Sec. III.A]{O2}, it is straightforward to show that the spin Hall equations in the form given in \cref{eq:GSHE_weyl} are a special case of the above generalization of the Mathisson-Papapetrou equations (up to error terms of order $\epsilon^2$). The necessary assumptions are as follows \cite[Sec. III.A]{O2}:
\begin{enumerate}
	\item The worldline parameter $\tau$ is chosen such that $\dot{x}^\mu t_\mu = \mathcal{P}^\mu t_\mu$ for all time.
	\label{cond1}
	
	\item The momentum $p_\alpha$ is at least initially null.
        \label{cond2}
	
	\item The angular momentum satisfies $S^{\alpha \beta} p_\beta = 0$ at least initially, and $S^{\alpha \beta} t_\beta = 0$ for all time.
        \label{cond3}
        
	\item The magnitude of the angular momentum is at least initially given by $S^{\alpha \beta} S_{\alpha \beta} = 2 (s \epsilon)^2$.
        \label{cond4}
\end{enumerate}

In particular, \cref{eq:GSHE_x} can be obtained by imposing the spin supplementary condition $S^{\alpha \beta} t_\beta = 0$, differentiating it and using \cref{eq:MPD_S}, together with the choice of the worldline parameter $\tau$ mentioned above in condition \ref{cond1}. Equation \eqref{eq:MPD_P} can be rewritten in the form \eqref{eq:GSHE_p} by rearranging the terms. Finally, imposing the conditions \ref{cond3} and \ref{cond4} ensures that the spin tensor takes the form given in \cref{eq:spin_tensor}. This spin tensor will satisfy \cref{eq:MPD_S}. Thus, exactly as in the Riemannian case, the spin Hall equations in Weyl geometry are a particular case of the Mathisson-Papapetrou equations. Furthermore, this equivalence can also be viewed as an independent derivation of the spin Hall equations: since electromagnetic wave packets can be viewed as localized objects with conserved stress-energy tensor, their bulk motion can generally be described by the Mathisson-Papapetrou equations. The role of the observer vector field $t^\alpha$, as well as the interpretation of $x^\mu(\tau)$ as the worldline followed by the wave packet's energy centroid relative to $t^\alpha$, is also clear when viewed from the perspective of the Mathisson-Papapetrou equations, as was discussed in Ref. \cite{O2}.

When using the spin Hall equations, different families of timelike observers can be used to describe the dynamics of the same wave packet. In general, different observers will associate different energy centroids, average momenta, and spin tensors for the same wave packet. When changing observers, the relation between these quantities associated with the wave packet by different observers is best understood by examining the Mathisson-Papapetrou form of the equations. In this context, a change of observer is associated with a change of spin supplementary condition.

Given two timelike vector fields $t^\alpha$ and $T^\alpha$, we obtain two different sets of spin Hall equations by starting with the Mathisson-Papapetrou equations \eqref{eq:MPD_Weyl} and imposing two different spin supplementary conditions: $S^{\alpha \beta} t_\beta = 0$ or $\bar{S}^{\alpha \beta} T_\beta = 0$. Then, each observer will associate different energy centroids, average momenta, and spin tensors for the same wave packet. The observer $t^\alpha$ will describe the wave packet using the set of quantities $\{x^\mu, \mathcal{P}_\alpha, S^{\alpha \beta} \}$, while the observer $T^\alpha$ will describe the wave packet by a different set of quantities $\{\bar{x}^{\bar{\alpha}}, \bar{\mathcal{P}}_{\bar{\alpha}}, \bar{S}^{\bar{\alpha} \bar{\beta}} \}$ (bars on the indices indicate that the object is generally defined at a different spacetime point than the corresponding object without bars on the indices). Then, as shown in \cref{app:observer}, up to error terms of order $\epsilon^2$, these two sets of quantities are generally related by
\begin{subequations}
\begin{align}
    \bar{x}^{\bar{\mu}} &= \exp{}_{x^\mu}(\xi^\mu)  + \mathcal{O}(\epsilon^2), \\
    \bar{\mathcal{P}}_{\bar{\alpha}} &= g\indices{_{\bar{\alpha}}^{\alpha}} \mathcal{P}_\alpha + \mathcal{O}(\epsilon^2), \\
    \bar{S}^{\bar{\alpha} \bar{\beta}} &= g\indices{^{\bar{\alpha}}_{\alpha}} g\indices{^{\bar{\beta}}_{\beta}} ( S^{\alpha \beta} + \mathcal{P}^\alpha \xi^\beta - \mathcal{P}^\beta \xi^\alpha) + \mathcal{O}(\epsilon^2),
\end{align}
\end{subequations}
where $\exp$ is the exponential map on the tangent bundle and $g\indices{^{\bar{\alpha}}_{\alpha}}$ is the bitensor which parallel propagates vectors from $x^\alpha$ to $\bar{x}^{\bar{\alpha}}$ along the geodesic segment which connects those points (both $\exp$ and $g\indices{^{\bar{\alpha}}_{\alpha}}$ are defined with respect to the Levi-Civita connection $\Gamma^\lambda_{\mu \nu}$; see Ref. \cite{Poisson2011} for more details on these bitensors) and the shift vector $\xi^\mu$ is defined as
\begin{equation} 
    \xi^\mu = \frac{S^{\mu \nu} T_\nu}{\mathcal{P}_\sigma T^\sigma}.
\end{equation}
This transformation law takes the same form as in Riemannian geometry \cite{O2}, and all properties related to changes of observer in the spin Hall equations are described by the shift vector $\xi^\mu$. In particular, all the results obtained in \cite[Sec. IV]{O2} for the Riemannian case also apply now to the present case of spin Hall equations in Weyl geometry. 

The magnitude of the shift vector $\xi^\mu$ can be used to understand the displacement between energy centroids assigned by different observers to the same wave packet. Based on Ref. \cite[Sec. IV]{O2}, for most cases, the magnitude of $\xi^\mu$ is limited to one wavelength. However, there exist fine-tuned situations where $t^\alpha$ and $T^\alpha$ are related by particular boosts, for which the magnitude of $\xi^\mu$ is unbounded. This apparent paradox can be resolved by recalling that the momentum of the polarized electromagnetic wave packets that we describe by the spin Hall equations is not exactly null. This is only approximately true up to error terms of order $\epsilon^2$: $p_\alpha p^\alpha = \mathcal{P}_\alpha \mathcal{P}^\alpha = \mathcal{O}(\epsilon^2)$. As shown in Ref. \cite{O2}, the momentum will always be timelike when going one order higher in the WKB expansion, and different centroids associated to timelike objects are always going to be related by bounded displacements. 

Furthermore, note that $\xi^\mu$ is always zero in the spacetime regions where $t^\alpha \propto T^\alpha$. This means that energy centroids depend only locally on the family of timelike observers used to define it. In particular, if $t^\alpha$ and $T^\alpha$ coincide in the neighborhoods of a source and a receiver, any difference between $t^\alpha$ and $T^\alpha$ in the region between the source and receiver is irrelevant. The energy centroids associated at the source and the receiver will always coincide. Thus, physical meaning can be assigned to $t^\alpha$ or $T^\alpha$ only in the neighborhoods of the source and receiver, where it can be identified with the $4$-velocities of these objects. This is illustrated in \cite[Fig. 1]{O2}.

\begin{figure*}
    \centering
    \includegraphics[width=0.98\columnwidth]{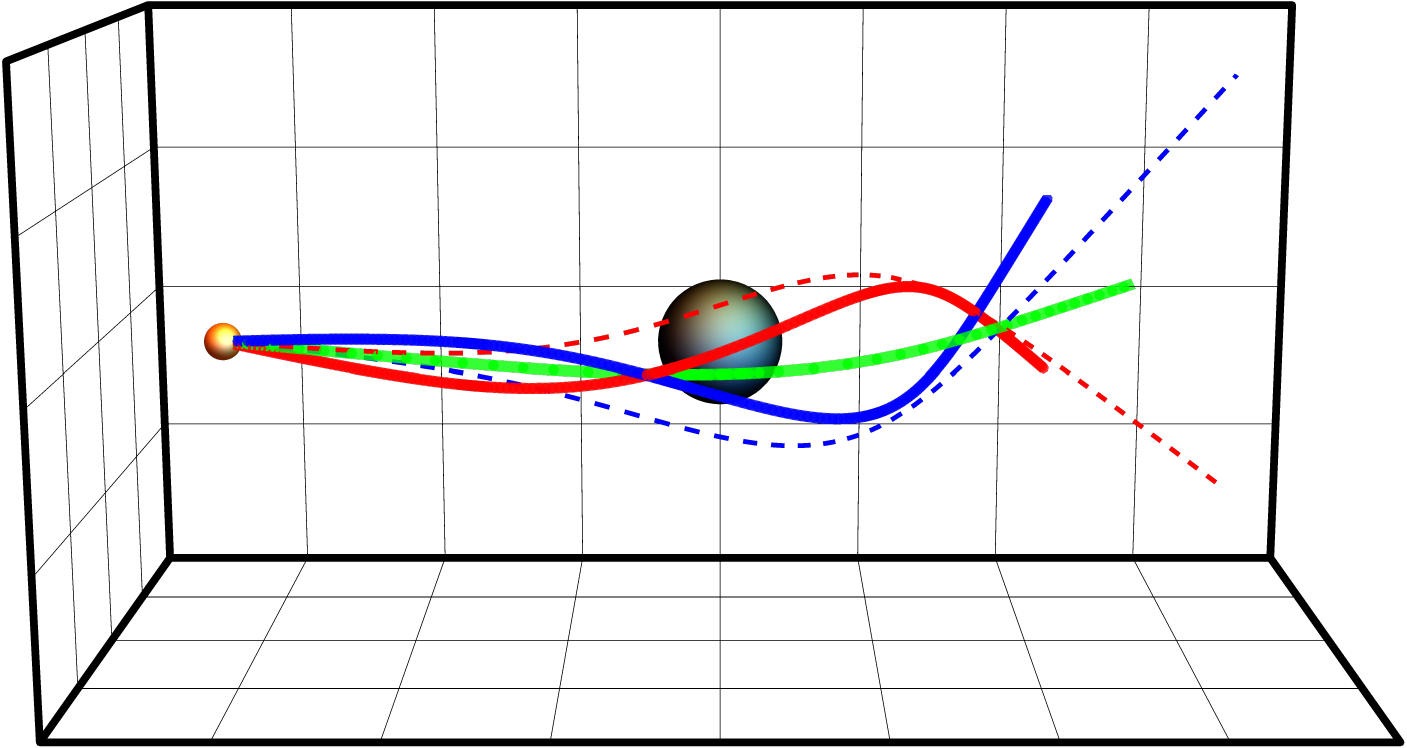}
    \hspace{0.01\columnwidth}
    \includegraphics[width=0.98\columnwidth]{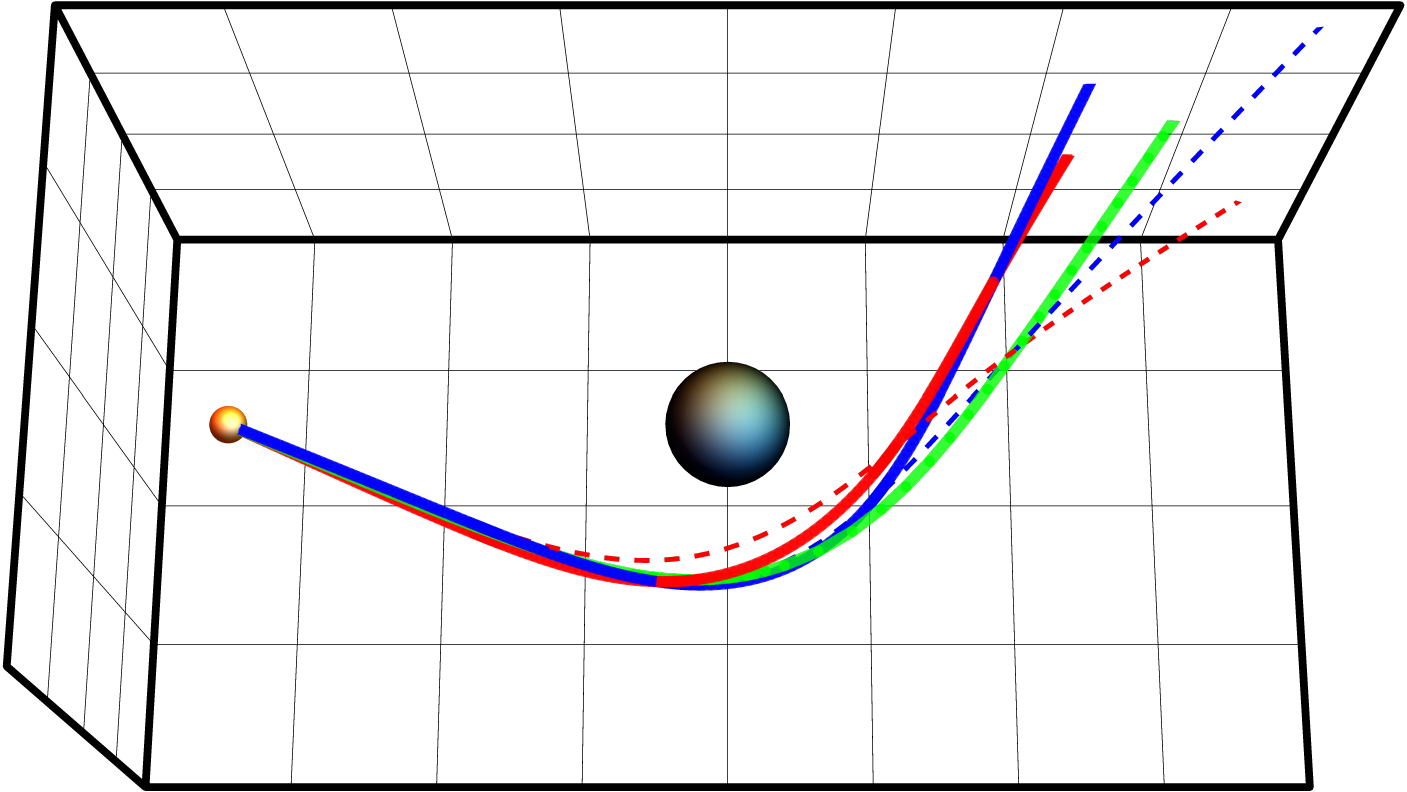}

    \vspace{0.2cm}

    \includegraphics[width=0.55\textwidth]{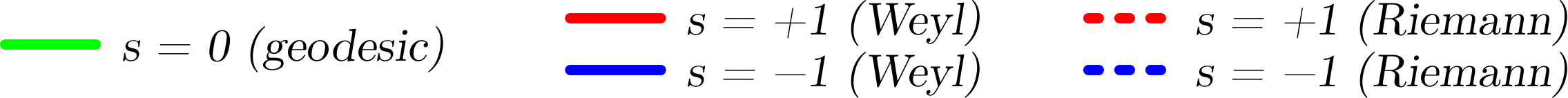}
    \caption{%
        Equatorial view (left) and top view (right) of light rays around a Schwarzschild black hole in Weyl geometry ($C_2 = C_3 = \delta = 0$). Gravitational spin Hall light rays of opposite circular polarization ($s = \pm 1$, red and blue lines) and a null geodesic ($s = 0$, green trajectory) are emitted with the same initial conditions from a source (orange sphere) at $r = 8 r_s$. For comparison, we also display the corresponding gravitational spin Hall rays in Riemann geometry (dashed lines). The individual components of the trajectory and their momenta are shown in \cref{fig:plot0_individual}.
    }
    \label{fig:plot0}
\end{figure*}

\vspace{1cm}

\section{Spin Hall effect for Weyl geometric black holes}\label{sect3}

In this section, we present some numerical examples of gravitational spin Hall trajectories near black holes in Weyl geometry. We consider the family of black hole solutions introduced in \cref{sec:bh}.

The gravitational spin Hall equations \eqref{eq:gshe_eq} can be rewritten in a more concrete form by considering the general metric given in \cref{eq:metric0} with $\nu(r) = -\lambda(r)$, a Weyl vector of the form $\omega_\mu = ( \omega_t(r), \omega_r(r), 0, 0 )$ and a choice of timelike vector field $t^\mu = (e^{-\nu(r)/2}, 0, 0, 0 )$. From a physical point of view, this particular choice of $t^\mu$ can be understood as having a point source of radiation at $x_s$ and with $4$-velocity $t^\mu(x_s)$ that emits electromagnetic wave packets. Then, an emitted wave packet is described by a family of timelike observers with $4$-velocity $t^\mu$, and the resulting trajectory from solving the spin Hall equations represents the energy centroid that this family of observers assigns for the given wave packet. A different family of observers would assign a different trajectory, since the energy centroid of a wave packet is observer dependent by definition. When changing $t^\mu$, the spin Hall equations will describe a different physical system, where the source has a different $4$-velocity and the emitted wave packet is assigned with a different energy centroid relative to another family of timelike observers. We provide an example of the observer dependence of spin Hall trajectories around Weyl-geometric black holes in \cref{app:observer}.

We denote the coordinate components of the worldline and the momentum by $x^\mu(\tau) = (t(\tau), r(\tau), \theta(\tau), \phi(\tau))$ and $p_\mu(\tau) = (p_t(\tau), p_r(\tau), p_\theta(\tau), p_\phi(\tau))$, where $\tau$ is an affine parameter. Furthermore, note that we can use the Hamiltonian constraint given in \cref{eq:H=0} to fix one of the components of $p_\mu$ and eliminate one of the eight spin Hall equations. Solving \cref{eq:H=0} for $p_t$ we obtain
\begin{equation}
    p_t = -\frac{e^{\frac{\nu}{2}} \sqrt{ p_{\phi }^2+p_{\theta }^2 \sin^2 \theta + e^{\nu} r^2 p_r^2 \sin^2 \theta } }{r \sin \theta},
\end{equation}
where the sign of $p_t$ is fixed so that $p^\mu$ is future directed. The gravitational spin Hall equations become
\begin{subequations}
\allowdisplaybreaks
\begin{align}
    \dot{t} &= -e^{-\nu} p_t, \\
    \dot{r} &= e^{\nu} p_r , \\
    \dot{\theta} &= \frac{p_{\theta }}{r^2} + \epsilon s \frac{ \alpha \omega_r+\nu'  }{2 r^2 p_t \sin \theta } p_{\phi }, \\
    \dot{\phi} &= \frac{p_{\phi }}{r^2 \sin^2 \theta} - \epsilon s \frac{ \alpha  \omega_r + \nu' }{2 r^2 p_t \sin \theta} p_{\theta } , 
\end{align}
\begin{widetext}
\begin{align}
   \dot{p}_r &= -\nu' e^{\nu} p_r^2 -\frac{\left(r \nu' - 2\right)\left( p_{\phi}^2 + p_{\theta}^2 \sin^2 \theta \right)}{2 r^3 \sin^2 \theta}, \\
   \dot{p}_\theta &= \frac{p_\phi^2 \cot \theta}{r^2 \sin^2 \theta} + \epsilon s \frac{ e^{\nu} p_r  \left[\alpha r^2 \omega _r'+ r \nu' \left(\alpha  r \omega_r+2\right)  - 2 \right] + 2 p_r  + p_{\theta} \cot \theta \left(\alpha  \omega_r+\nu'\right)}{2 r \sqrt{r^2 p_r^2 \sin^2 \theta +e^{-\nu (r)} \left( p_{\phi }^2 + p_{\theta }^2 \sin^2\theta \right)}} p_{\phi }, \\
   \dot{p}_\phi &= \epsilon s e^{\nu} \frac{ p_r p_{\theta } \left[\alpha r^2 e^{\nu} \omega_r'+r e^{\nu} \left(\alpha  r \omega_r+2\right) \nu'-2 e^{\nu}+2\right]+\cot \theta \left(p_{\phi}^2 \csc^2\theta + p_{\theta}^2 \right) \left(\alpha \omega_r+\nu'\right)}{2 r^2 p_t \csc \theta}.
\end{align}
\end{widetext}
\end{subequations}

\begin{figure*}[t!]
    \centering
    \includegraphics[width=0.32\textwidth]{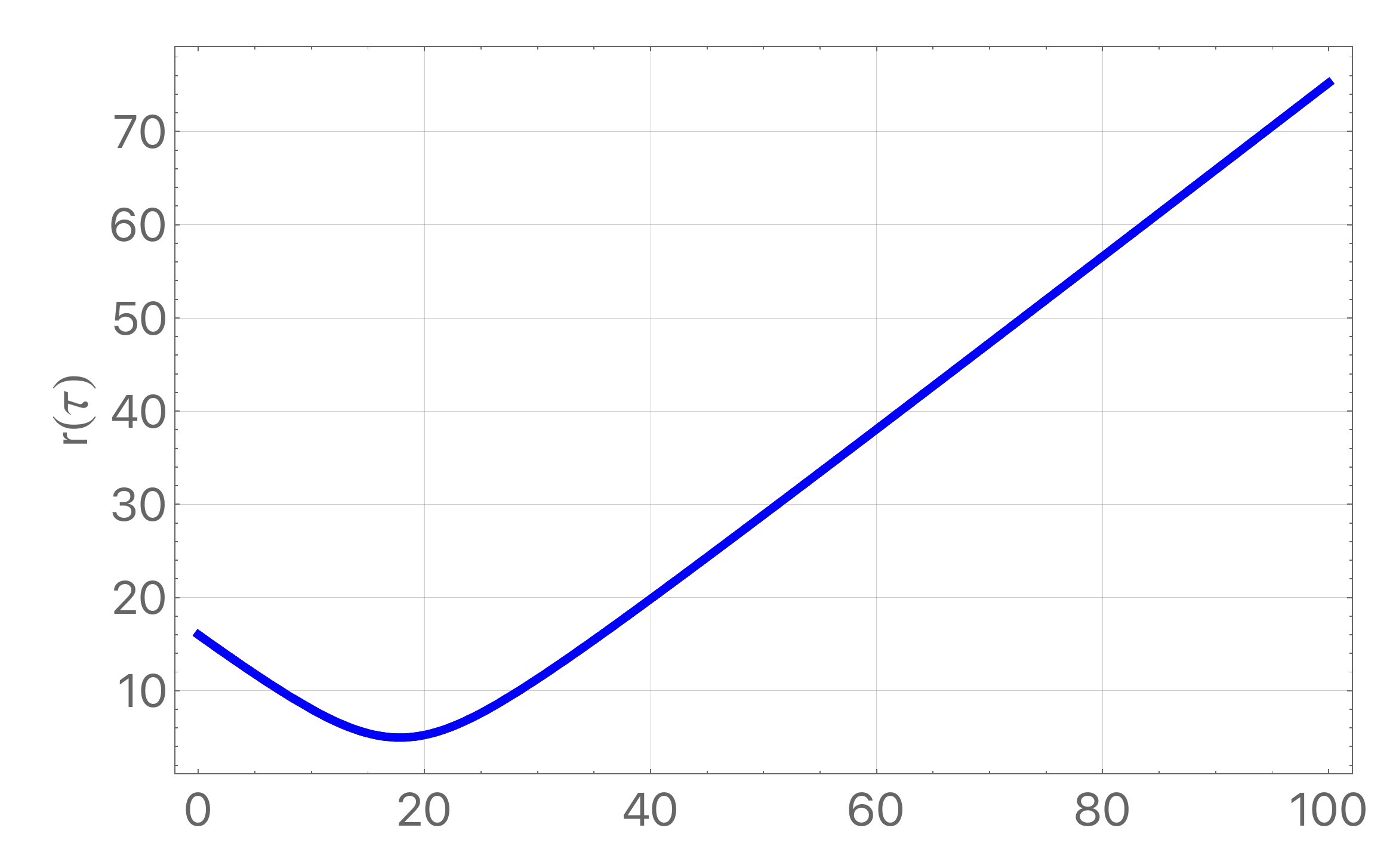}
    \hspace{0.01\columnwidth}
    \includegraphics[width=0.32\textwidth]{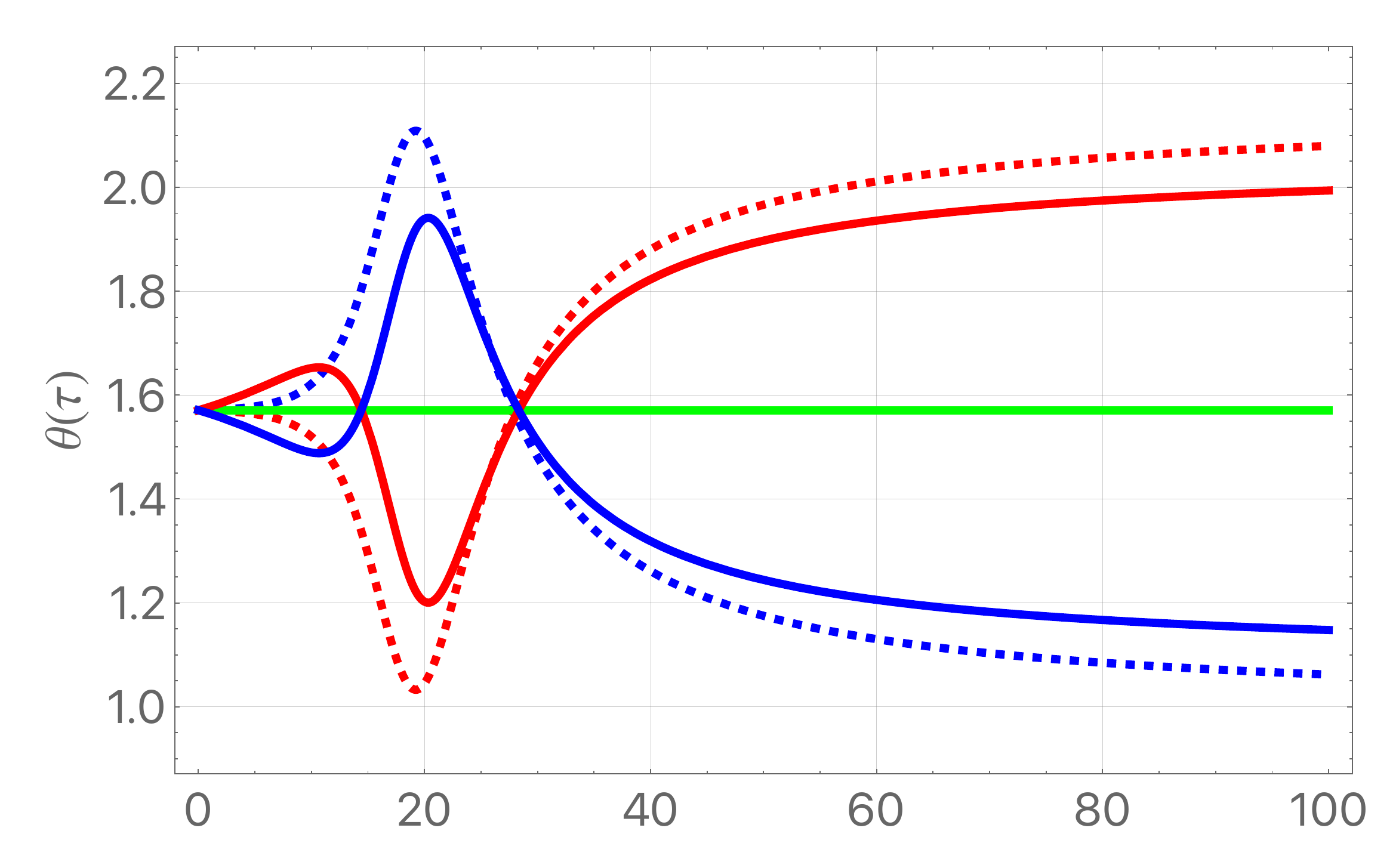}
    \hspace{0.01\columnwidth}
    \includegraphics[width=0.32\textwidth]{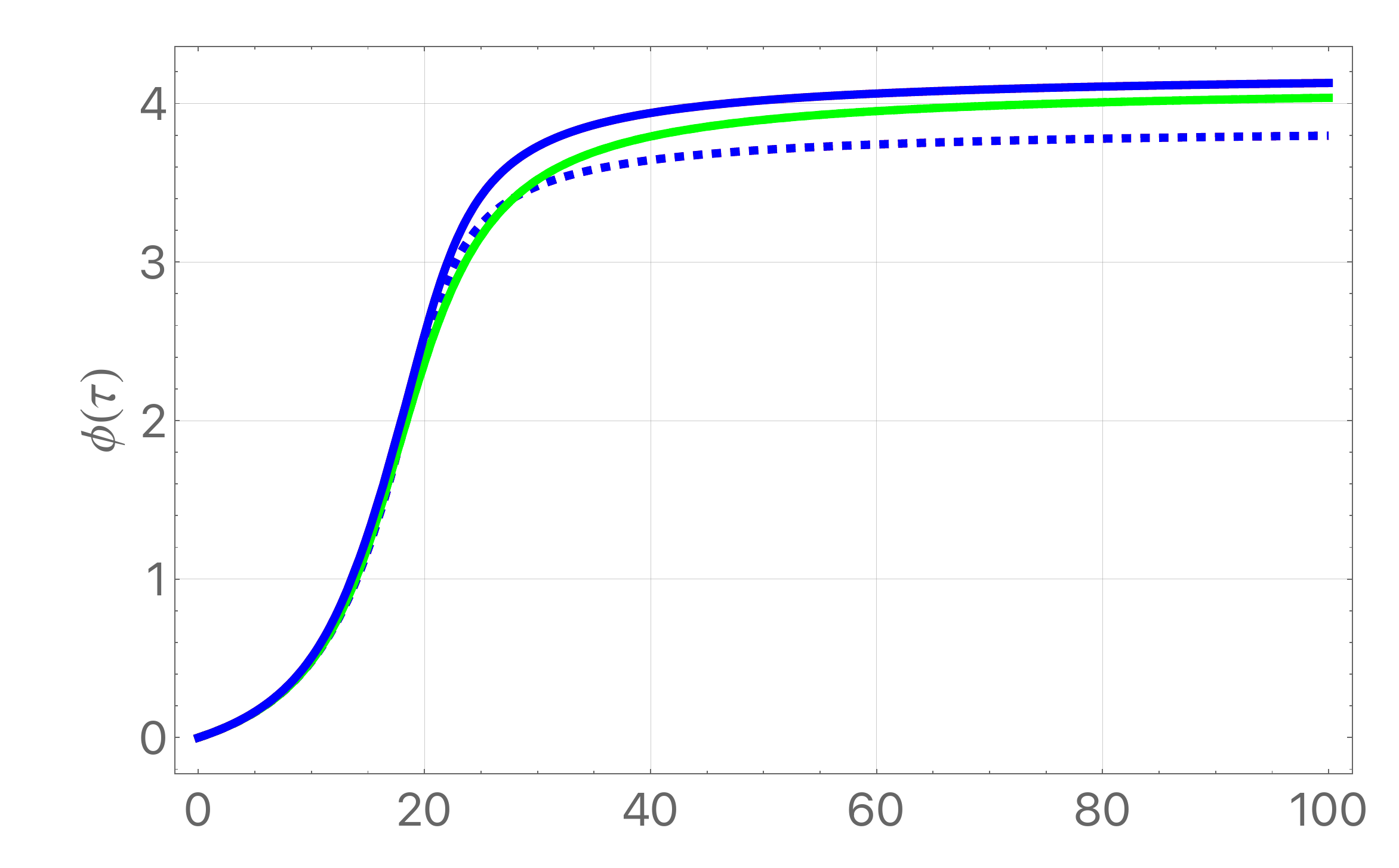}
    \includegraphics[width=0.32\textwidth]{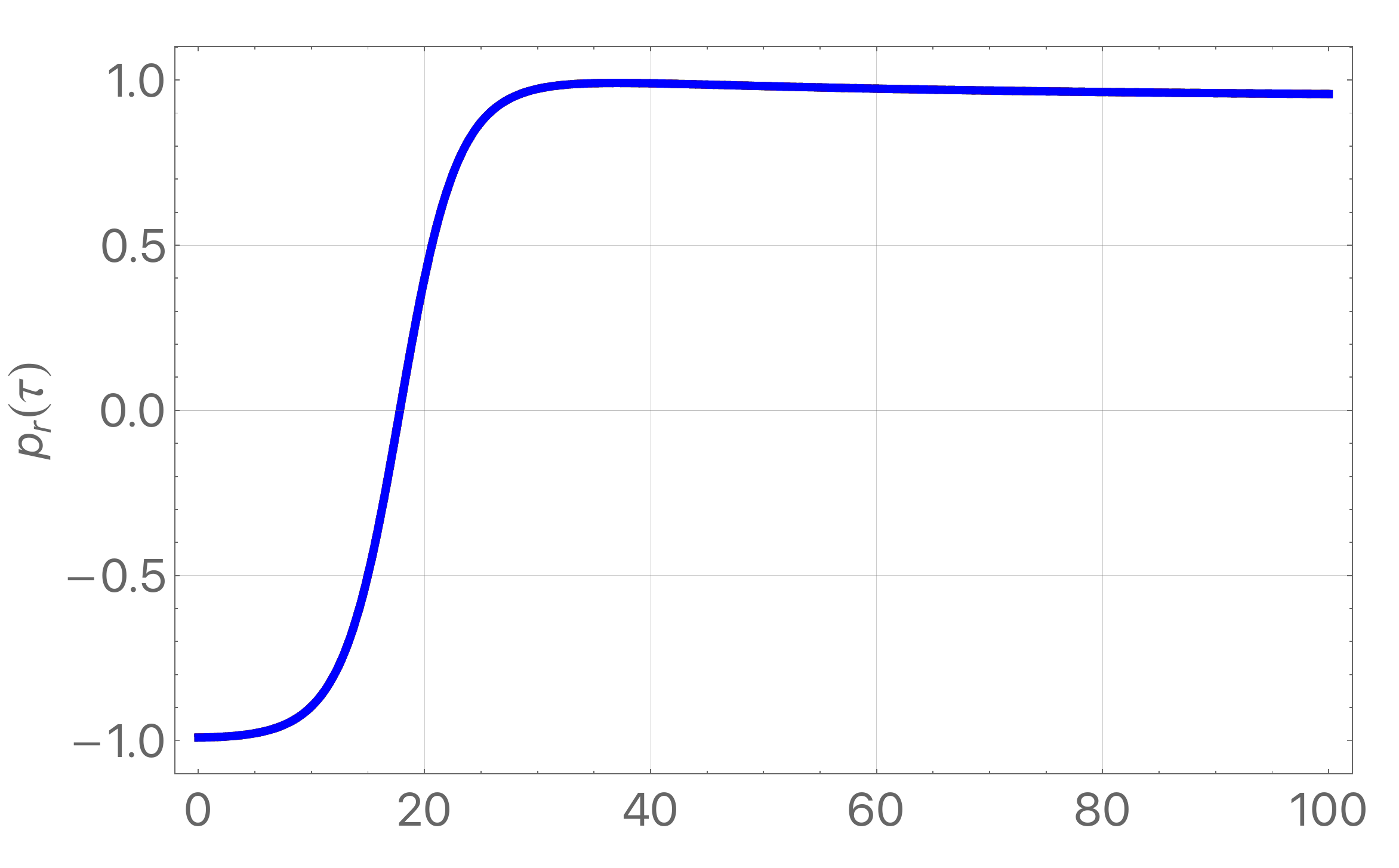}
    \hspace{0.01\columnwidth}
    \includegraphics[width=0.32\textwidth]{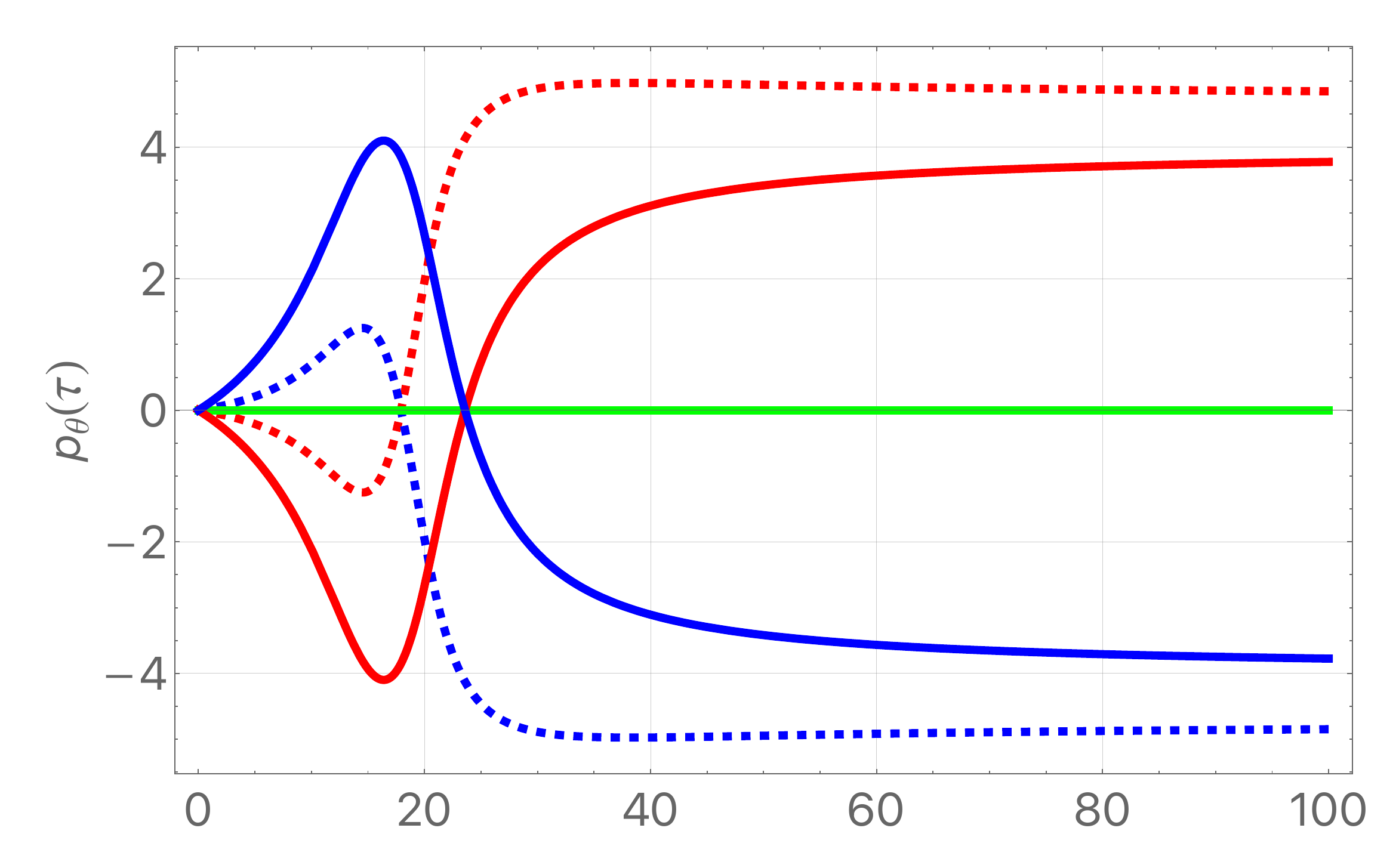}
    \hspace{0.01\columnwidth}
    \includegraphics[width=0.32\textwidth]{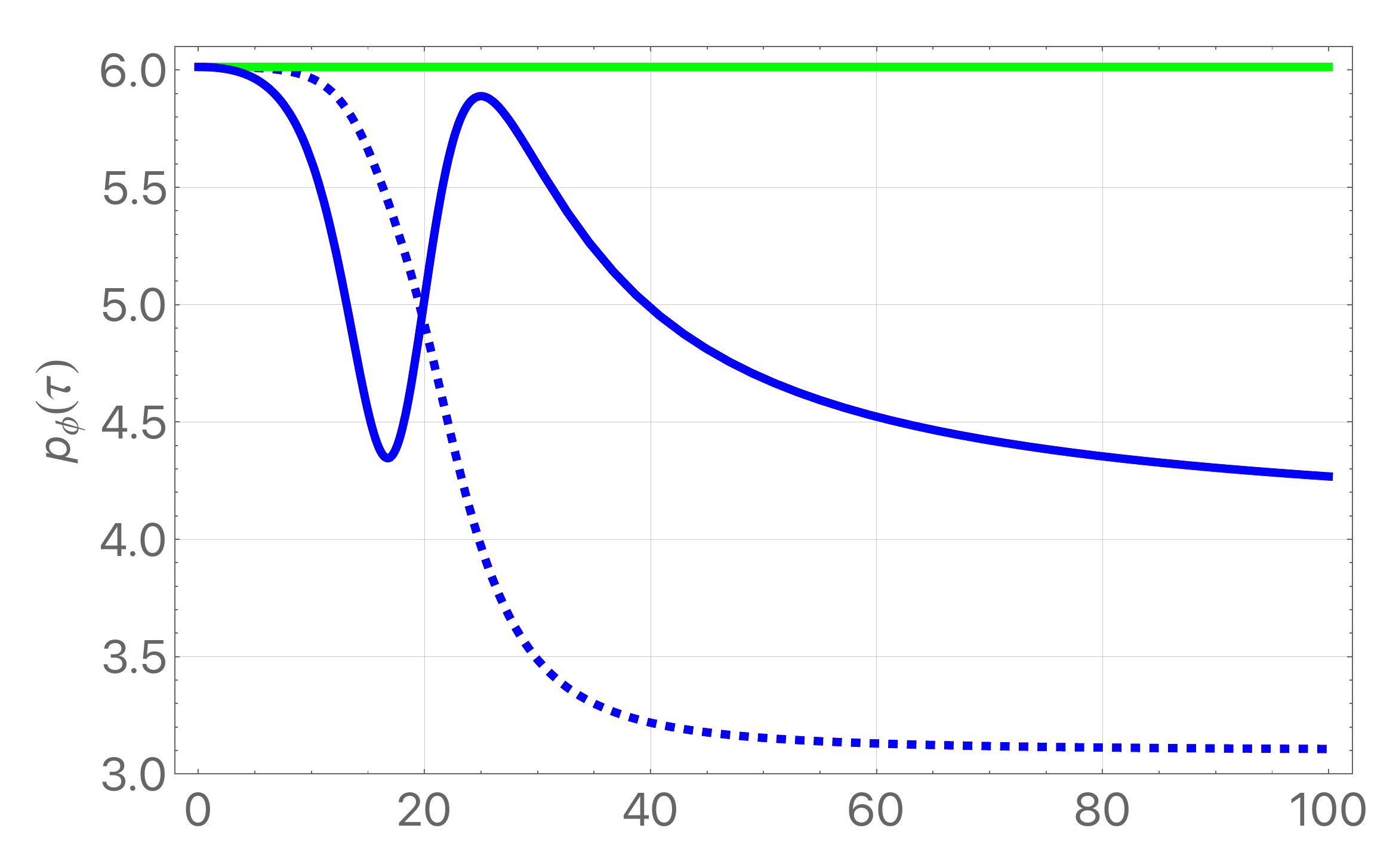}
    \includegraphics[width=0.55\textwidth]{plot0_legend.pdf}
    \caption{%
        Plots for the individual components of the trajectories $(r(\tau), \theta(\tau), \phi(\tau))$ and the momenta $(p_r(\tau), p_\theta(\tau), p_\phi(\tau))$ for the configuration given in \cref{fig:plot0}. Green lines represent geodesic motion, solid lines represent spin Hall trajectories of opposite circular polarization in Weyl geometry, and dashed lines represent spin Hall trajectories of opposite circular polarization in Riemann geometry. Red or green lines appear to be missing in certain plots due to perfect overlap.
    }
    \label{fig:plot0_individual}
\end{figure*}

Note that, in this particular case, the gravitational spin Hall equations do not depend on the time component $\omega_0(r)$ of the Weyl vector. We will numerically integrate the above equations using \texttt{Wolfram Mathematica} \cite{Mathematica}, with a straightforward extension of the code presented in Ref. \cite{Oanceathesis}. For some of the numerical examples presented below, we use unrealistically large values of $\epsilon$ (up to $\epsilon = 5$ in units of $M = 1$). This is done solely for the purpose of visualization, as otherwise the effect would be very small and would be hardly visible on some of the figures. It should be kept in mind that in physically relevant situations one should only consider $\epsilon < 1$, as the WKB expansion used to derive the spin Hall equations is not valid otherwise.

As a first example, we consider the black hole solution given in Eqs. \eqref{eq:metric0} and \eqref{metrW} with $C_2 = C_3 = \delta = 0$. In this case, the metric reduces to that of a Schwarzschild black hole. However, even in this case, the Weyl vector will be nonzero. Since this choice of metric represents a black hole solution in both Riemannian and Weyl geometry, it allows us to have a clear comparison between the two and to see how Weyl geometry will affect the propagation of light.

We have shown in the previous section that null geodesics do not depend on the Weyl vector, and take the same form as in Riemann geometry. However, this is no longer the case for polarized light rays. It can be clearly seen in Eq. \eqref{eq:GSHE_weyl} that the gravitational spin Hall equations have a nontrivial dependence on the Weyl vector [note that \cref{eq:GSHE_weyl} is independent of the Weyl gauge coupling constant $\alpha$]. Thus, in general, we expect that the gravitational spin Hall effect will be different in Weyl and Riemann geometry. 

To illustrate this difference, consider the numerical example in \cref{fig:plot0}. Here, we consider a source of light close to a Schwarzschild black hole, at $r = 8 r_s$, and we emit polarized light rays and geodesics with the same initial conditions. The green trajectory represents a null geodesic, which is the same in both Weyl and Riemann geometry. The red and blue trajectories represent finite-frequency light rays of opposite circular polarization, described by the spin Hall equations. The solid lines represent the gravitational spin Hall rays in Weyl geometry, while the dashed lines represent the gravitational spin Hall trajectories in Riemann geometry. The individual coordinate components of the worldlines, as well as the momenta, are shown in \cref{fig:plot0_individual}. 

Thus, we can clearly see that the gravitational spin Hall effect is different in Weyl and Riemann geometry. In this particular case ($C_2 = C_3 = \delta = 0$), the polarized light rays experience a stronger deflection toward the Weyl geometry black hole than the corresponding rays in Riemann geometry. This difference gradually fades away as we increase the value of $C_2$, and in the limit $C_2 \rightarrow \infty$ we obtain $\omega_\mu \rightarrow 0$ and the gravitational spin Hall rays of Weyl geometry converge to those of Riemann geometry.

\begin{figure*}[t!]
    \centering
    \includegraphics[width=0.98\columnwidth]{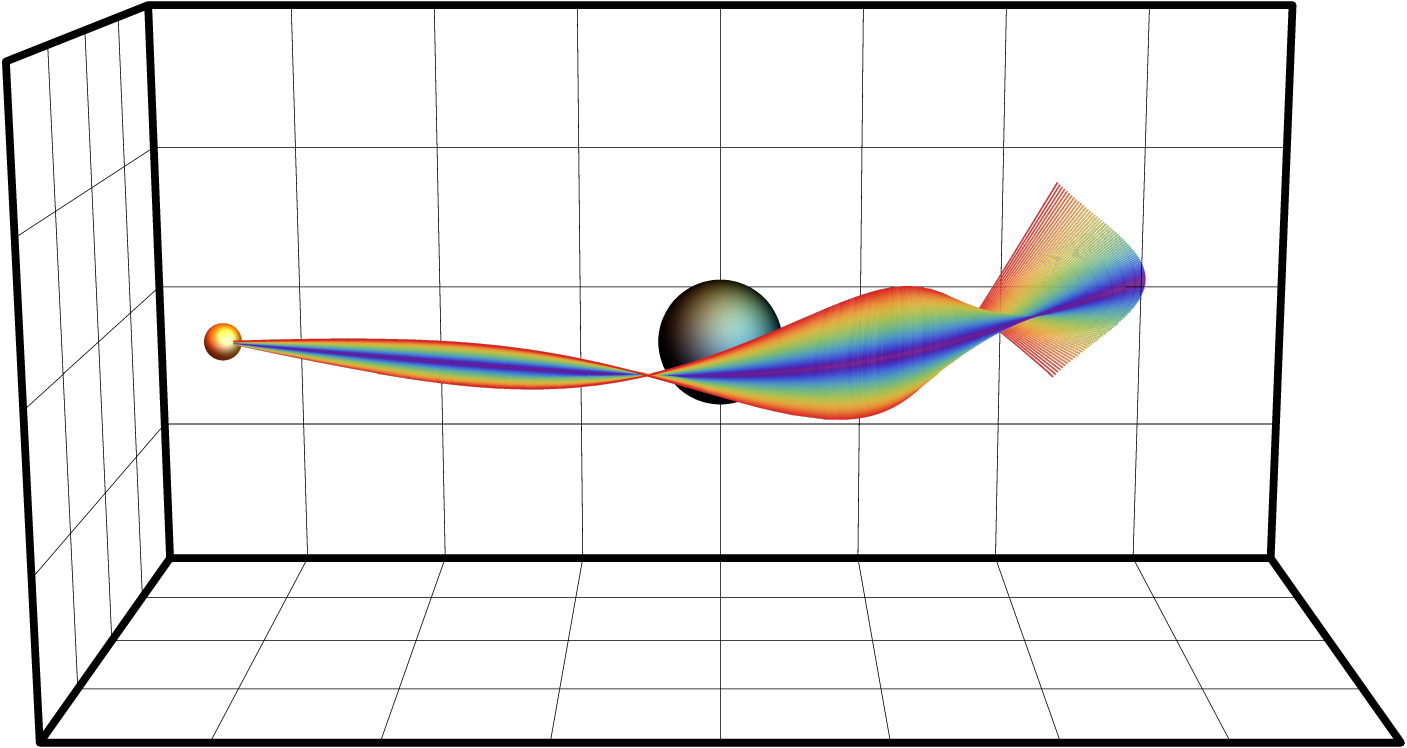}
    \hspace{0.01\columnwidth}
    \includegraphics[width=0.98\columnwidth]{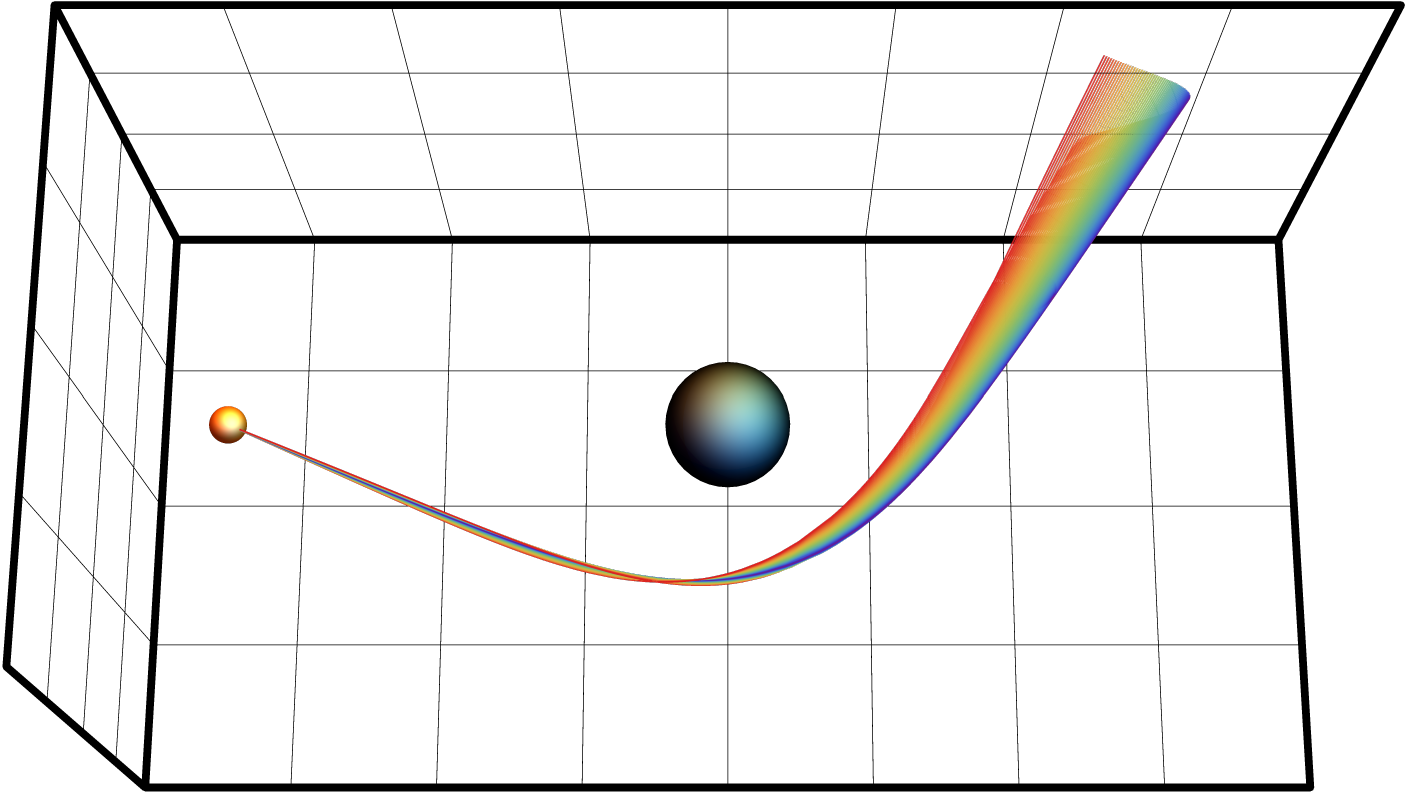}

    \vspace{0.2cm}

    \includegraphics[width=0.20\textwidth]{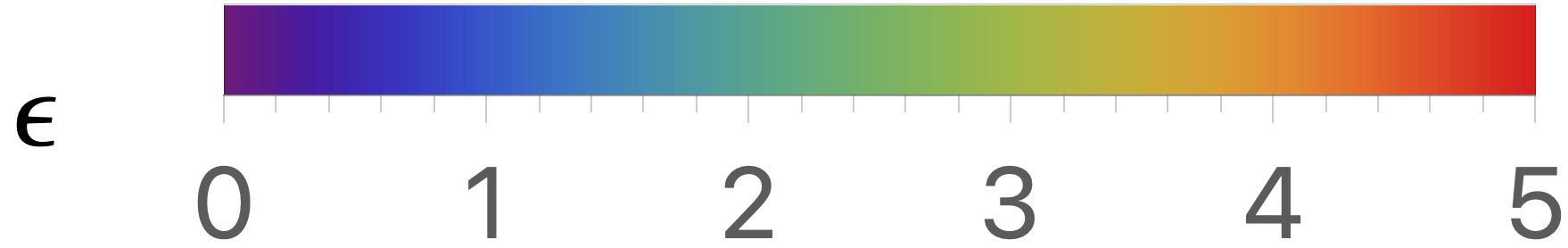}
    \caption{%
        Frequency dependence of the gravitational spin Hall effect of light in Weyl geometry. Equatorial view (left) and top view (right) of light rays of different frequencies (encoded by the colors of the rainbow) and opposite circular polarization, emitted by a source at $r = 8 r_s$ and traveling around a Schwarzschild black hole in Weyl geometry ($C_2 = C_3 = \delta = 0$).
    }
    \label{fig:plot1}
\end{figure*}

Using a similar setup, we also considered the frequency dependence of the gravitational spin Hall effect in Weyl geometry. This is illustrated in \cref{fig:plot1}, where rays of different frequencies, encoded by the colors of the rainbow, are emitted from a source at $r = 8 r_s$, close to a black hole with $C_2 = C_3 = \delta = 0$. There are two copies of the rainbow present in \cref{fig:plot1}, corresponding to the two states of opposite circular polarization ($s = \pm 1$) and separated by a null geodesic trajectory (violet color, corresponding to a wavelength zero). 

\begin{figure*}[t!]
    \centering
    \includegraphics[width=0.32\textwidth]{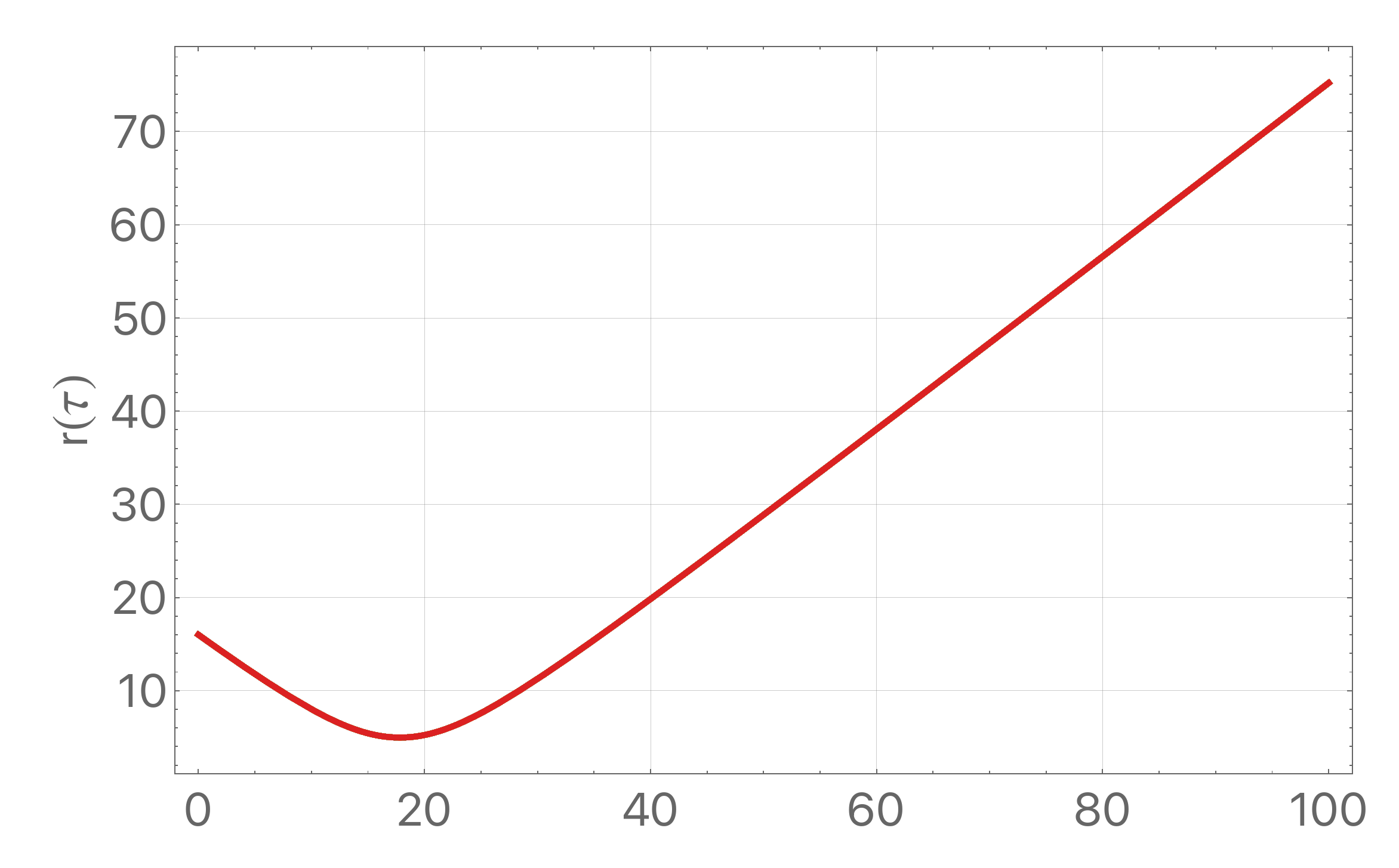}
    \hspace{0.01\columnwidth}
    \includegraphics[width=0.32\textwidth]{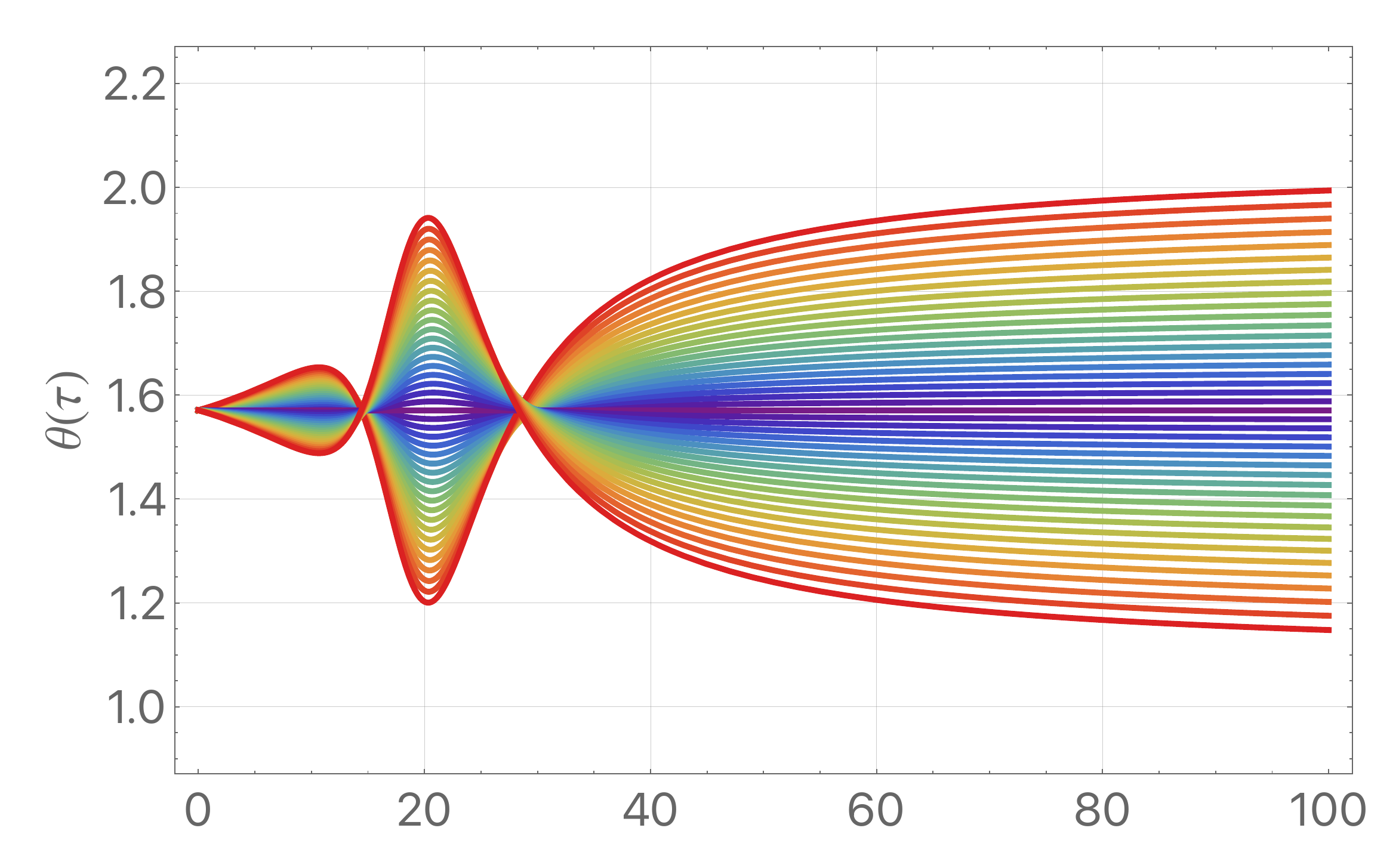}
    \hspace{0.01\columnwidth}
    \includegraphics[width=0.32\textwidth]{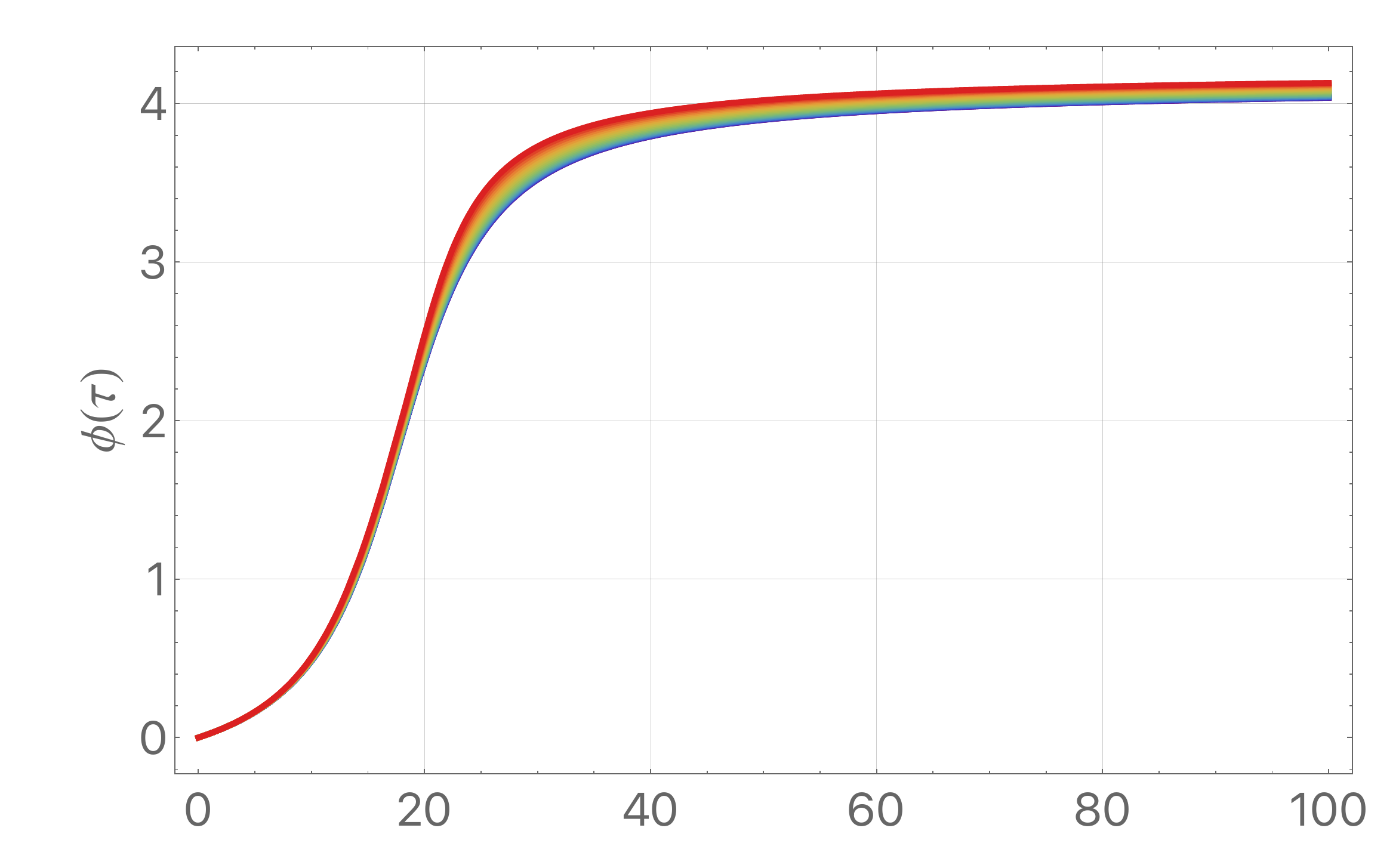}
    \includegraphics[width=0.32\textwidth]{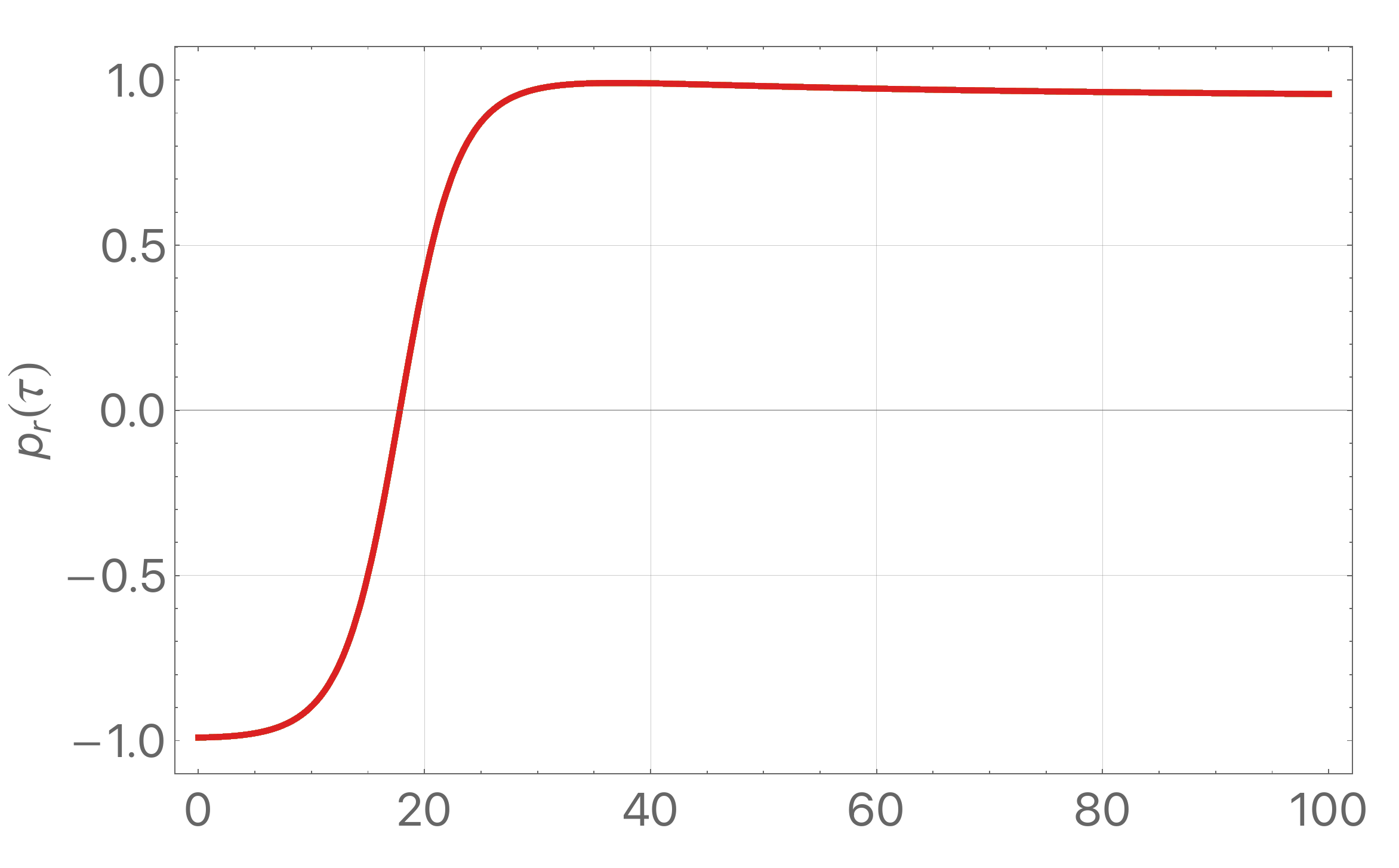}
    \hspace{0.01\columnwidth}
    \includegraphics[width=0.32\textwidth]{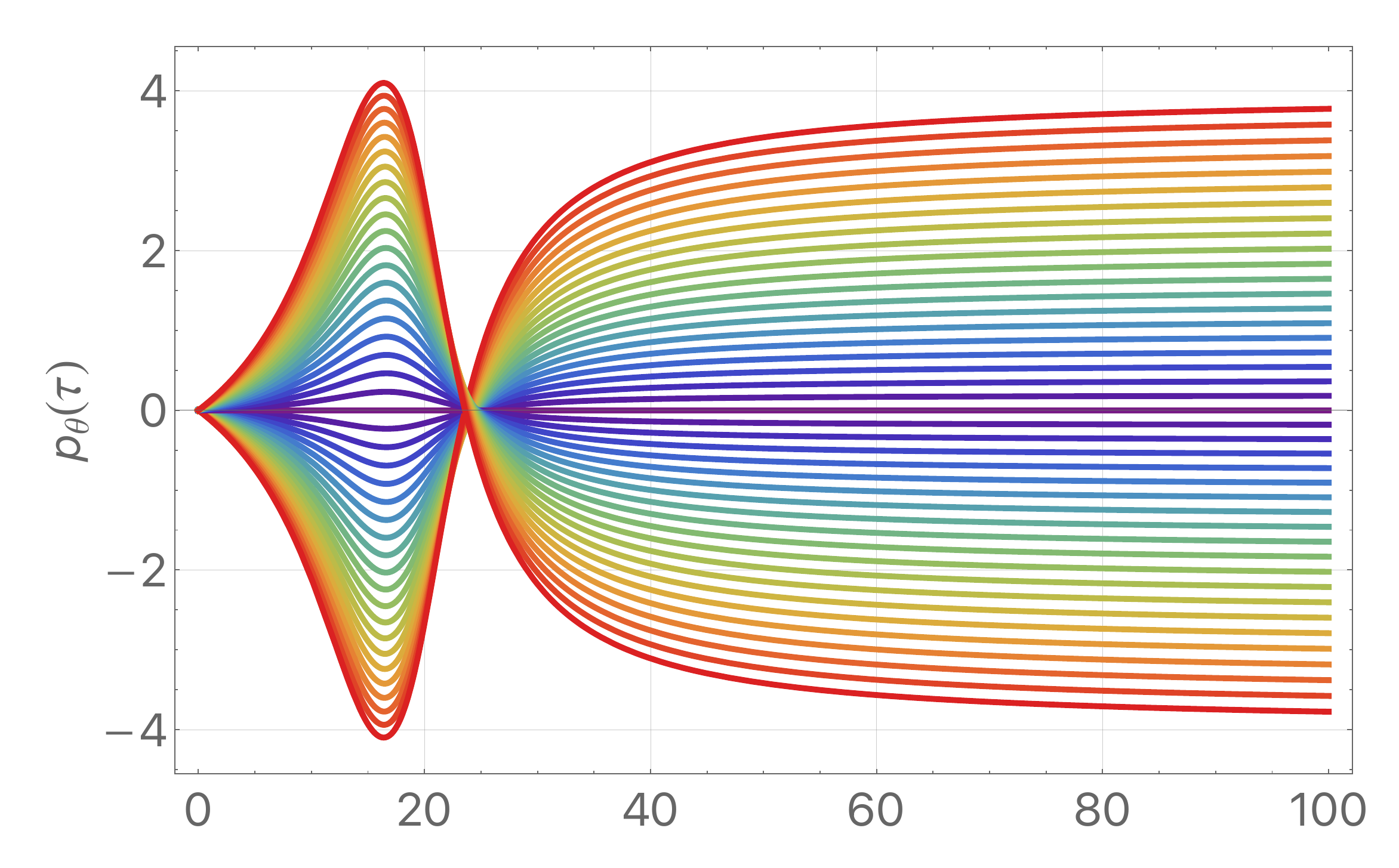}
    \hspace{0.01\columnwidth}
    \includegraphics[width=0.32\textwidth]{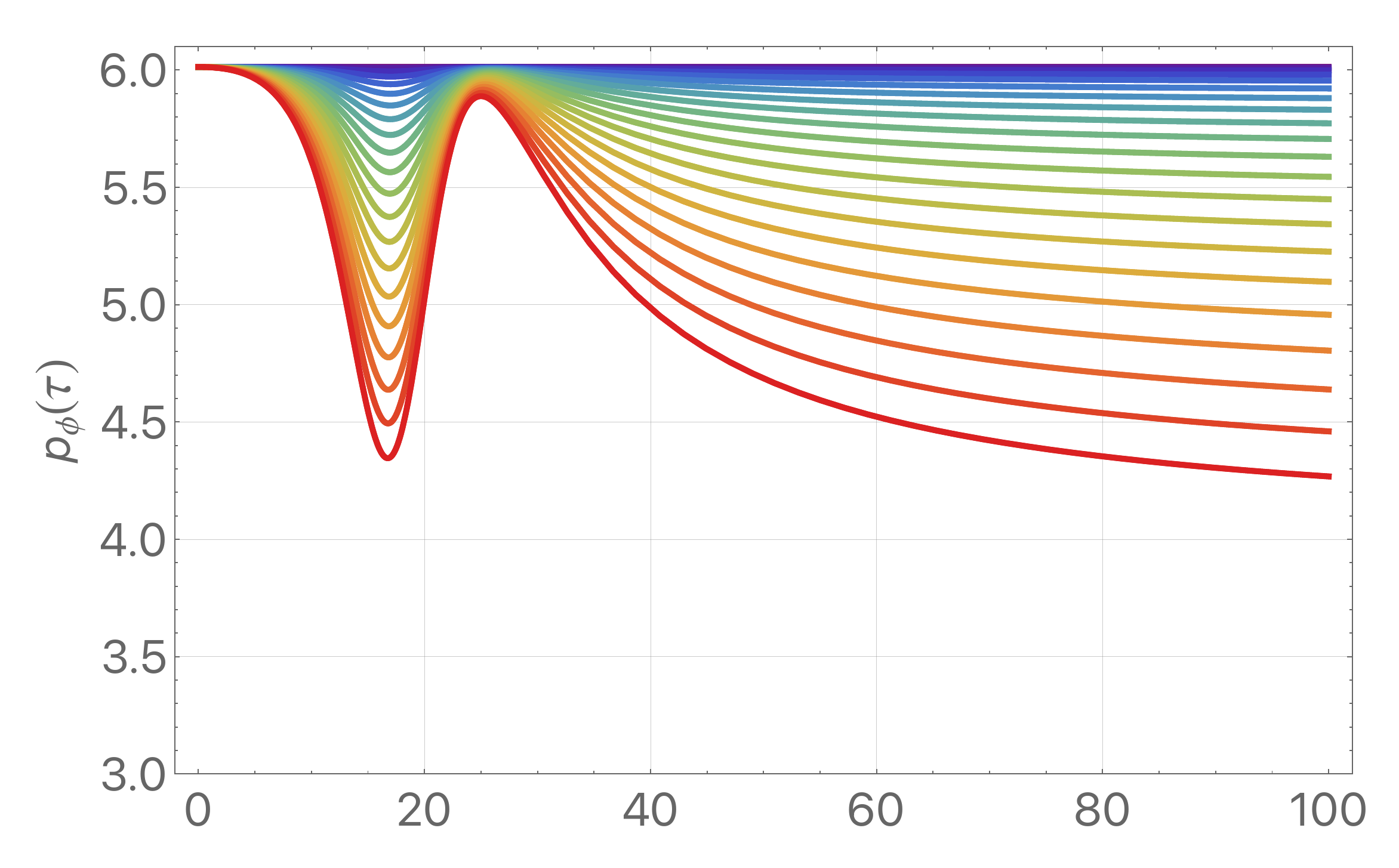}
    \includegraphics[width=0.20\textwidth]{plot4_legend.png}
    \caption{%
        Plots for the individual components of the trajectories $(r(\tau), \theta(\tau), \phi(\tau))$ and the momenta $(p_r(\tau), p_\theta(\tau), p_\phi(\tau))$ for the configuration given in \cref{fig:plot1}. Different wavelengths $\epsilon$ are encoded in the colors of the rainbow. Lines of certain colors appear to be missing on certain plots due to overlap.
    }
    \label{fig:plot1_individual}
\end{figure*}

The individual coordinate components of the worldlines and the momenta are shown in \cref{fig:plot1_individual}. As expected, light rays with small wavelengths, represented by blue colors, do not deviate too much from the null geodesic trajectory, whereas light rays with large wavelengths, represented by colors close to red, experience a strong deviation.

\begin{figure*}[t!]
    \centering
    \includegraphics[width=0.98\columnwidth]{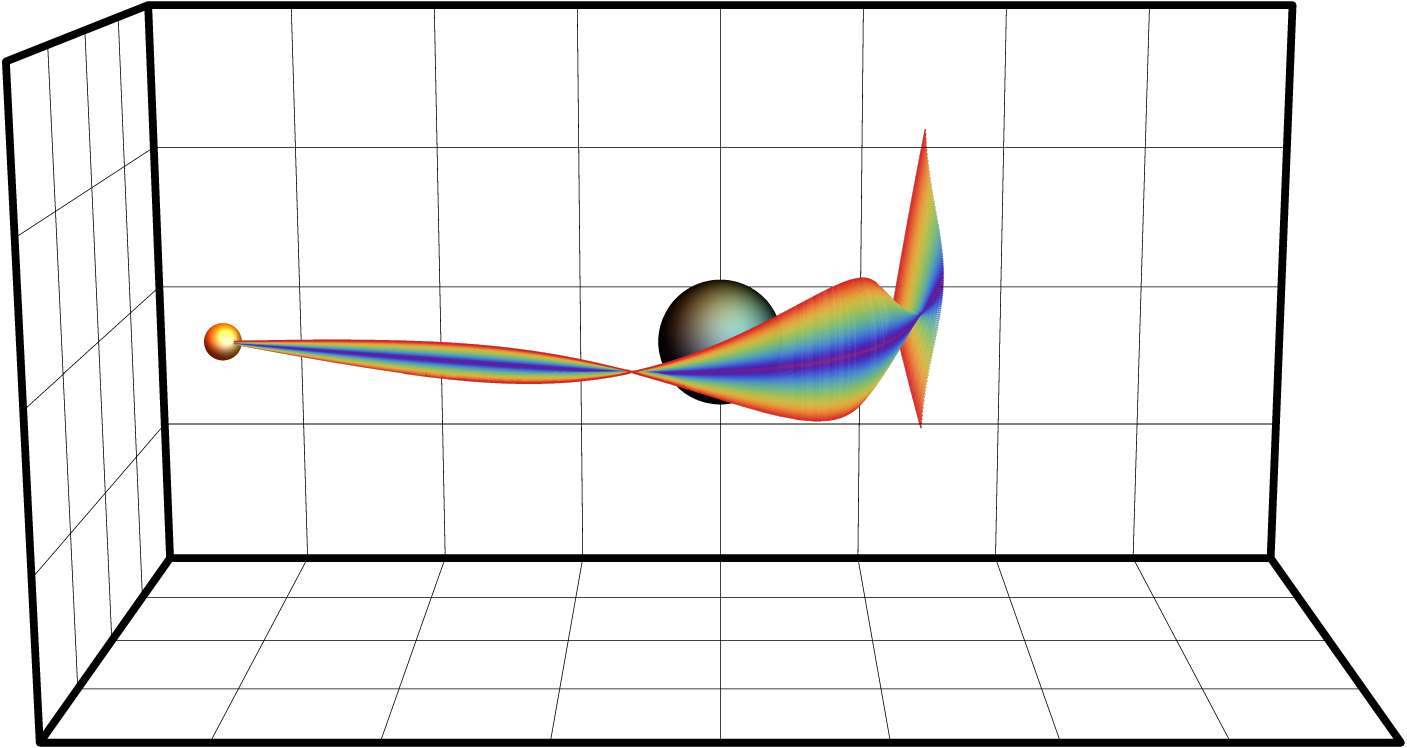}
    \hspace{0.01\columnwidth}
    \includegraphics[width=0.98\columnwidth]{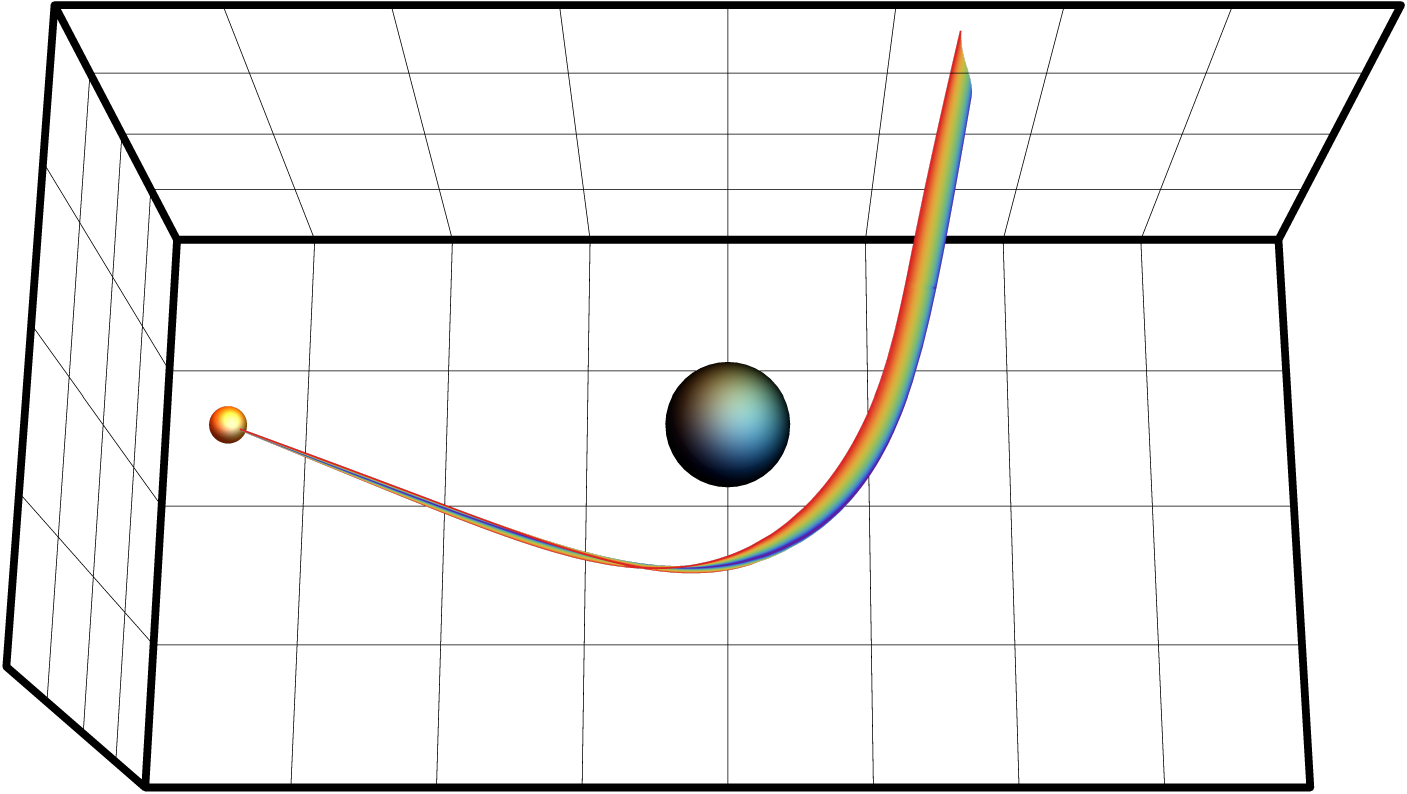}

    \vspace{0.2cm}

    \includegraphics[width=0.20\textwidth]{plot4_legend.png}
    \caption{%
        Frequency dependence of the gravitational spin Hall effect of light in Weyl geometry. Equatorial view (left) and top view (right) of light rays of different frequencies (encoded by the colors of the rainbow) and opposite circular polarization, emitted by a source at $r = 8 r_s$ and traveling around a black hole in Weyl geometry with parameters $C_2 = 0$, $C_3 = 5 \times 10^{-4}$, $\delta = 10^{-2}$.
    }
    \label{fig:plot2}
\end{figure*}

\begin{figure*}[t!]
    \centering
    \includegraphics[width=0.32\textwidth]{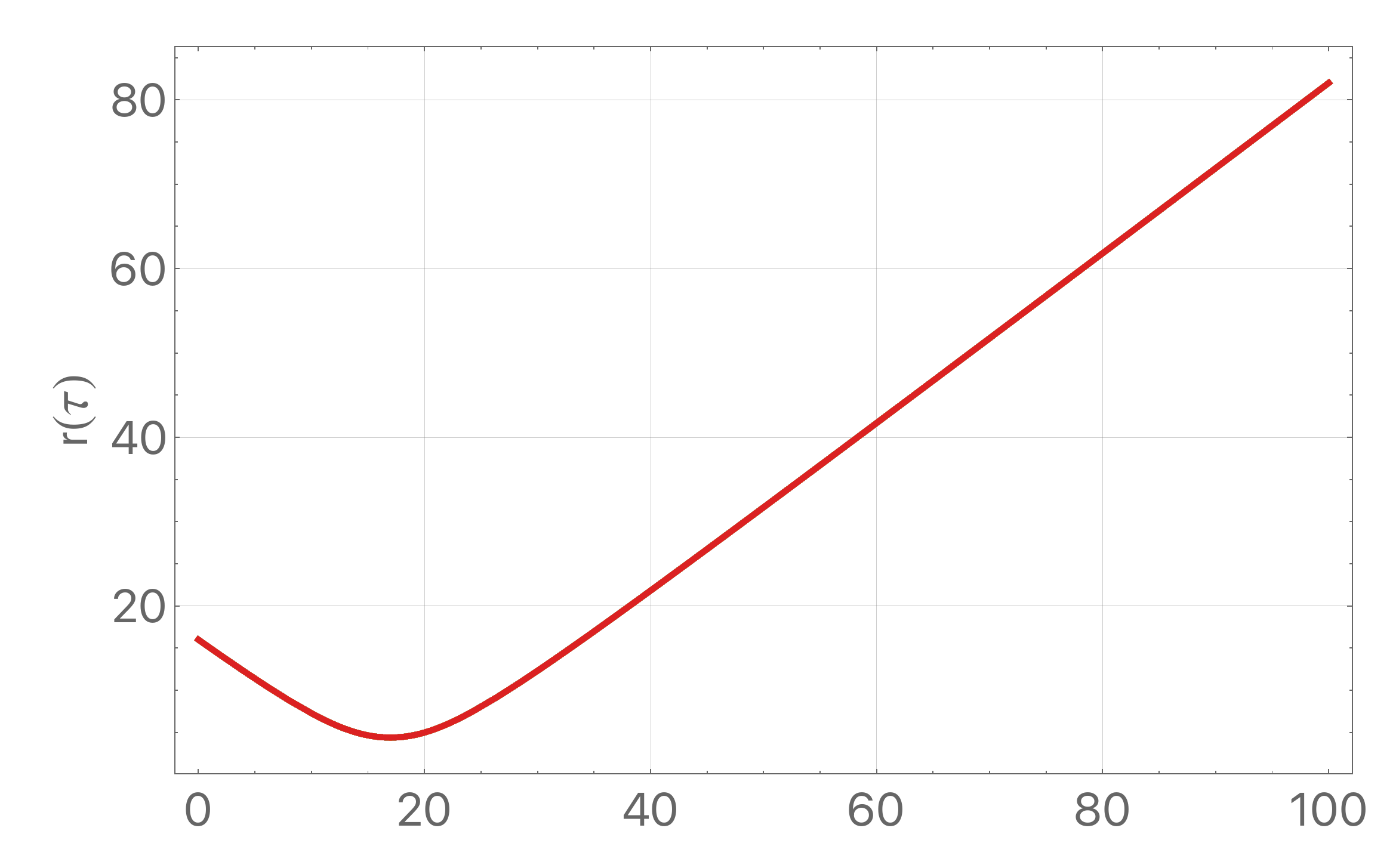}
    \hspace{0.01\columnwidth}
    \includegraphics[width=0.32\textwidth]{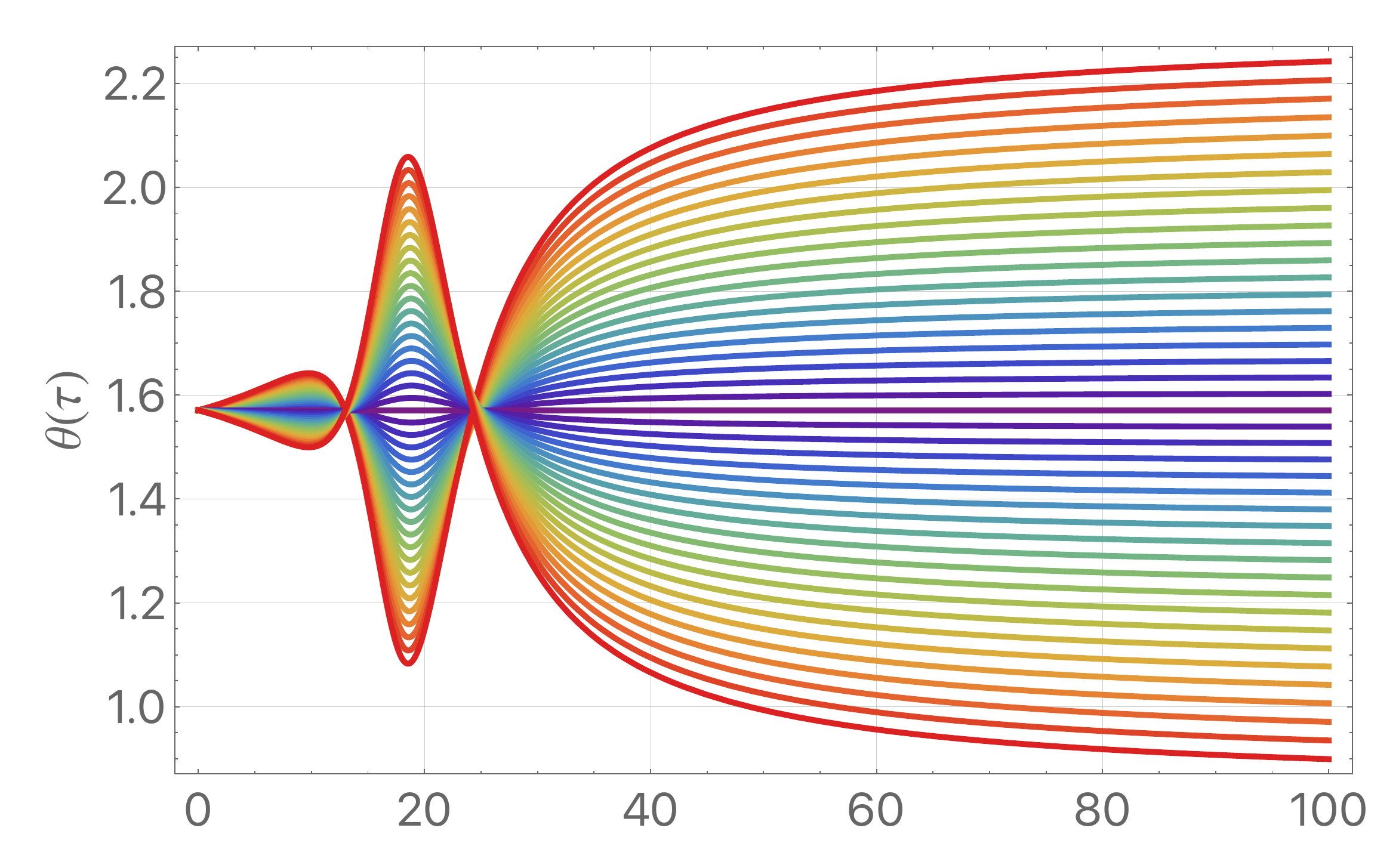}
    \hspace{0.01\columnwidth}
    \includegraphics[width=0.32\textwidth]{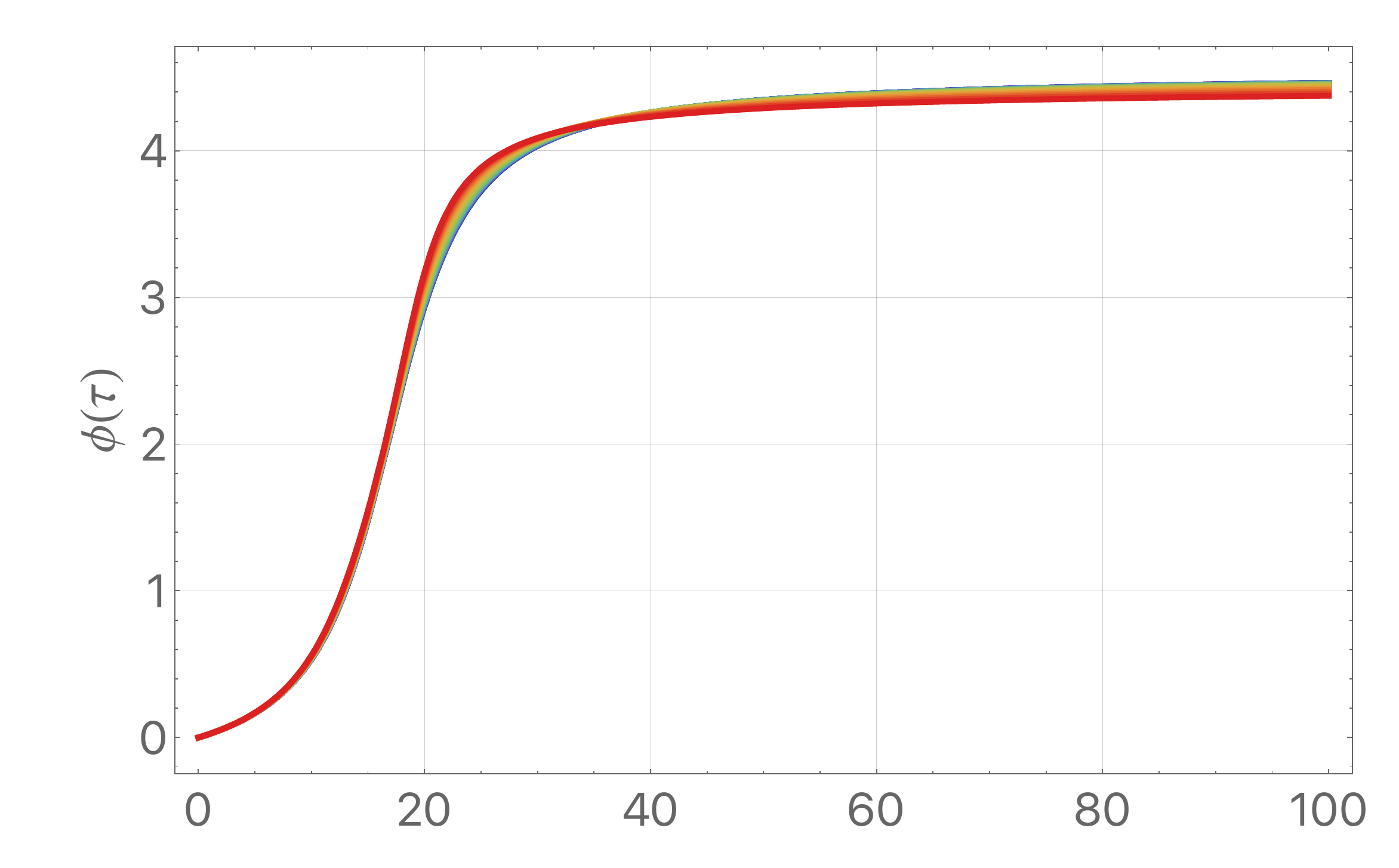}
    \includegraphics[width=0.32\textwidth]{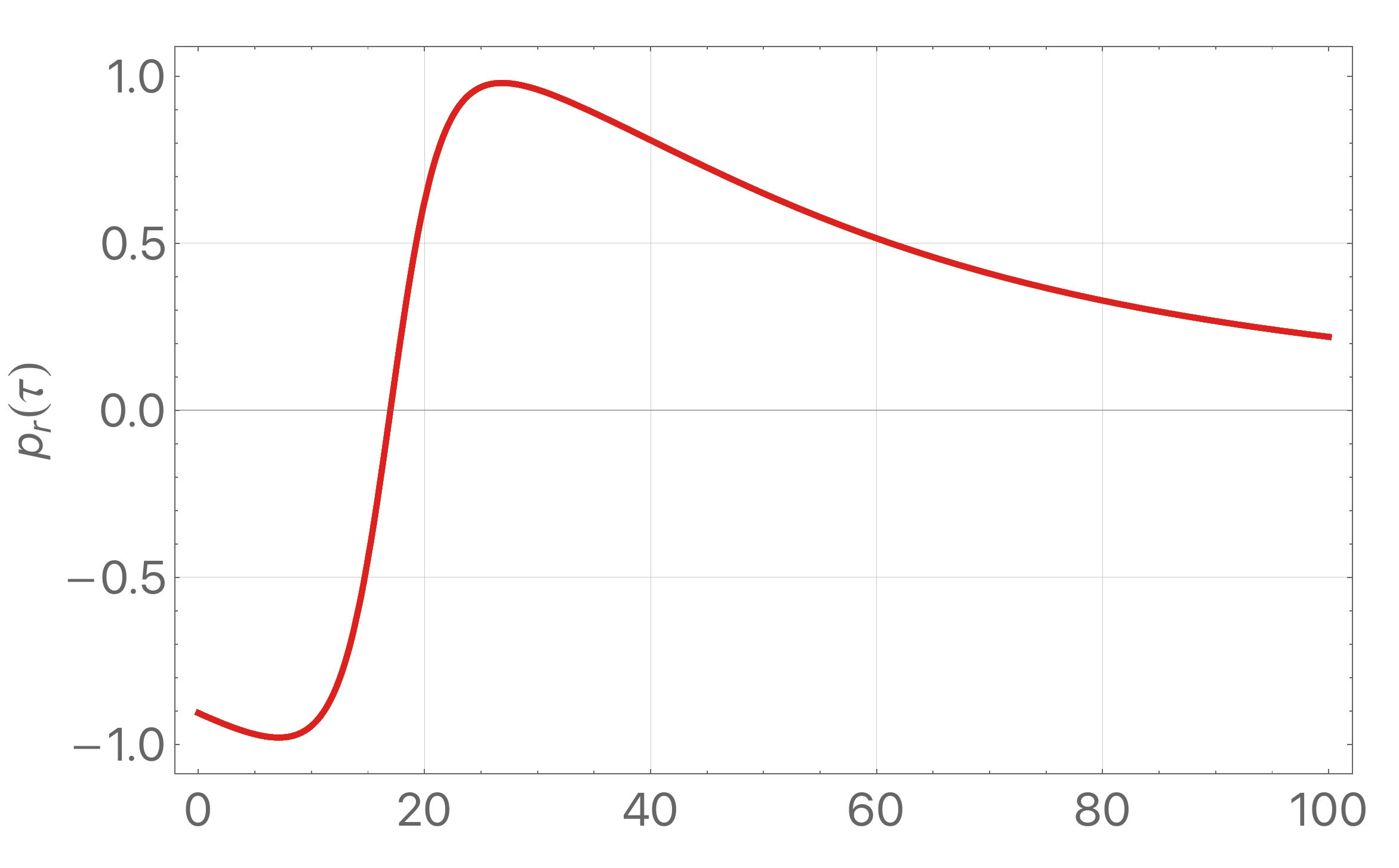}
    \hspace{0.01\columnwidth}
    \includegraphics[width=0.32\textwidth]{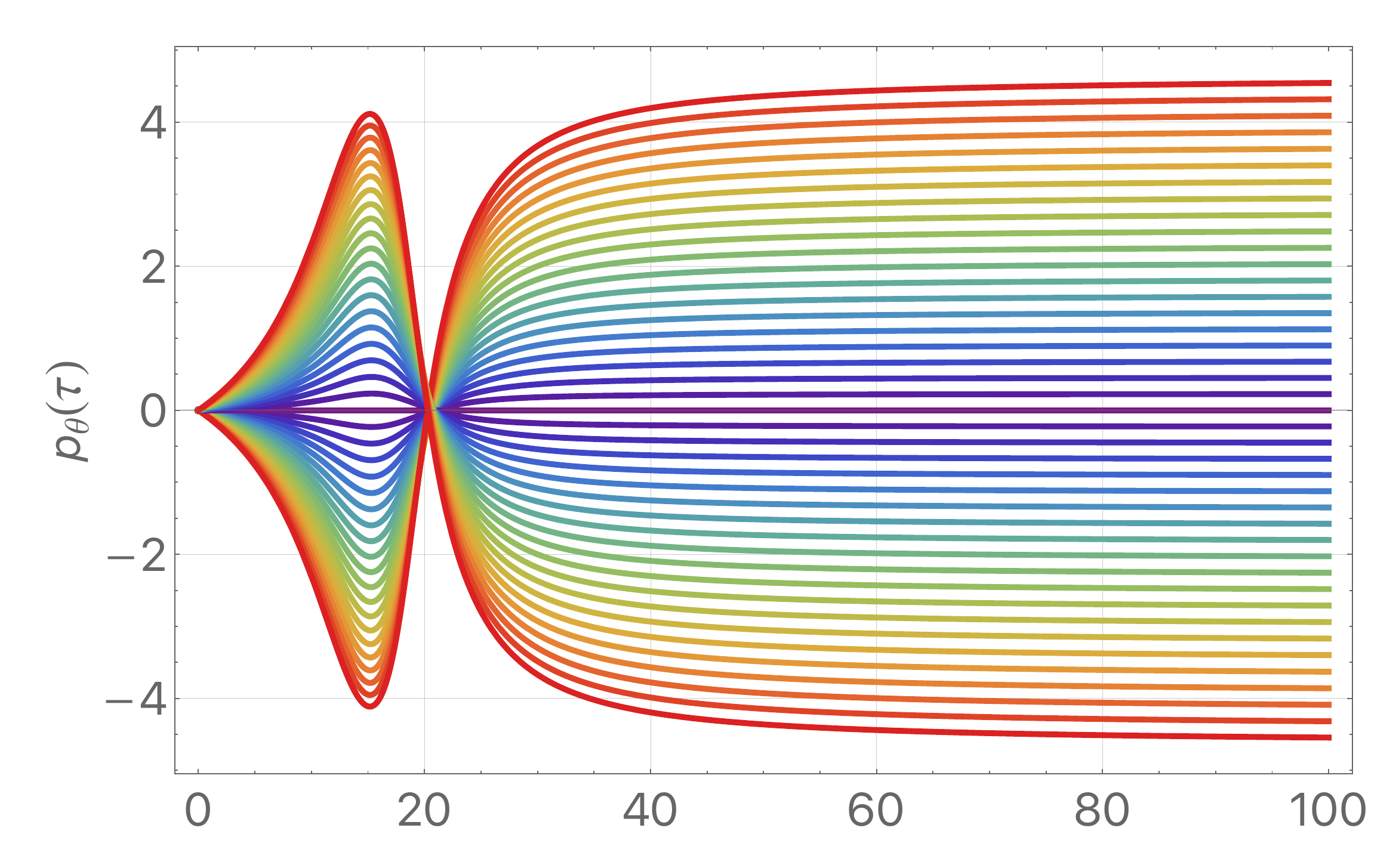}
    \hspace{0.01\columnwidth}
    \includegraphics[width=0.32\textwidth]{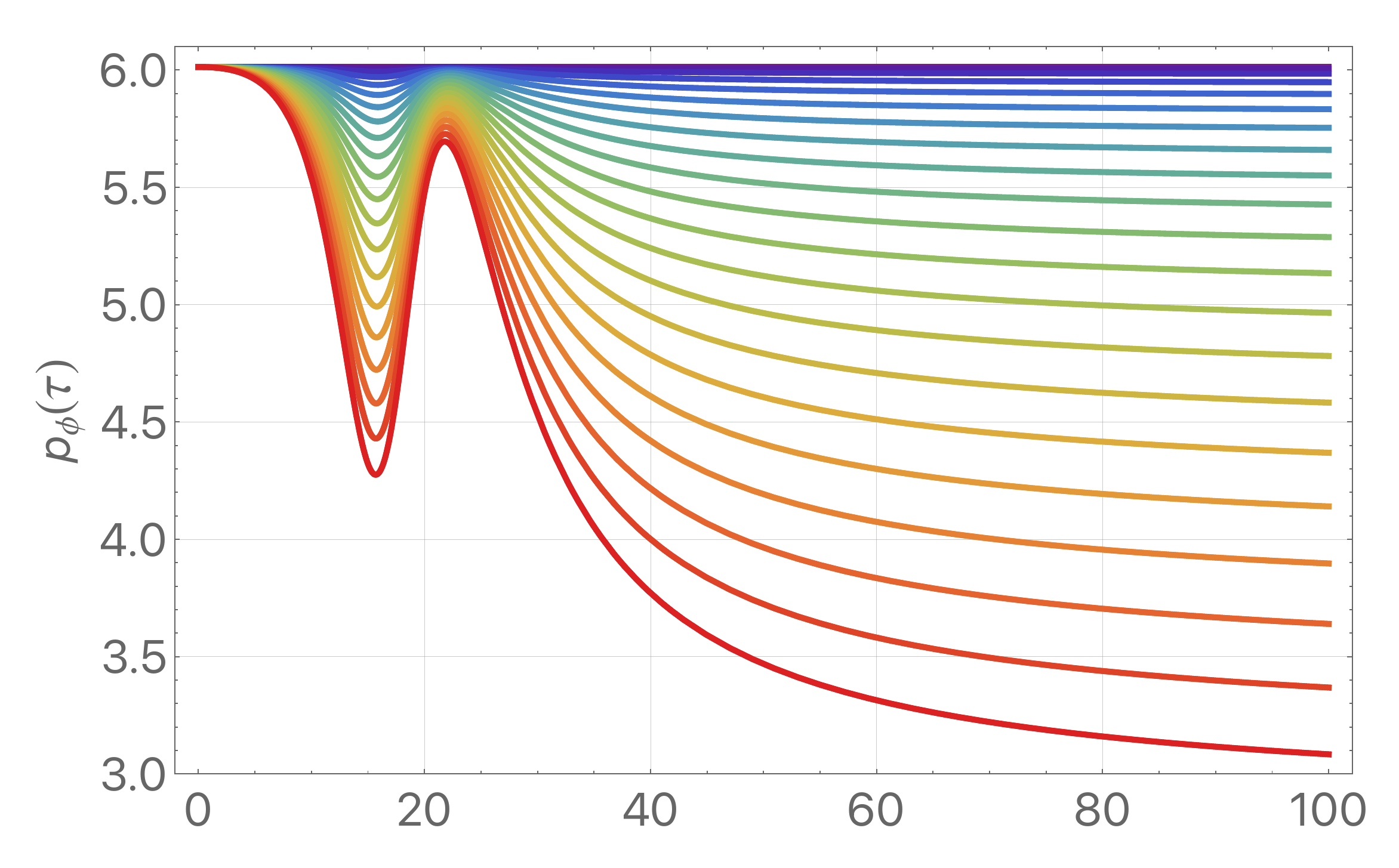}
    \includegraphics[width=0.20\textwidth]{plot4_legend.png}
    \caption{%
        Plots for the individual components of the trajectories $(r(\tau), \theta(\tau), \phi(\tau))$ and the momenta $(p_r(\tau), p_\theta(\tau), p_\phi(\tau))$ for the configuration given in \cref{fig:plot2}. Different wavelengths $\epsilon$ are encoded in the colors of the rainbow. Lines of certain colors appear to be missing on certain plots due to overlap.
    }
    \label{fig:plot2_individual}
\end{figure*}

Another example is shown in \cref{fig:plot2}, where we consider a more general black hole with $C_2 = 0$, $C_3 = 5 \times 10^{-4}$, and $\delta = 10^{-2}$. In this case, the general properties of the gravitational spin Hall effect of light remain mostly unchanged, with the exception of the overall magnitude of the deflection, which is now larger and is very sensitive to the values of the parameters $C_3$ and $\delta$. The individual coordinate components of the worldlines and the momenta are shown in \cref{fig:plot2_individual}.

\section{Discussions and final remarks}\label{sect4}

Weyl geometry is an interesting and natural extension of Riemannian geometry. Even though Weyl's original goal of formulating a successful unified theory of gravitation and electromagnetism was not fulfilled, the purely geometric framework introduced in this geometry has found many applications in physics and even engineering. For example, the mechanics of solids with distributed point defects can be formulated using Weyl geometry, the geometric object relevant to this distribution being the nonmetricity \cite{CM1}. The base manifold of a solid with distributed point defects, for a stress-free body, is a flat Weyl manifold, that is, a manifold with an affine connection that has a nonmetricity with a vanishing traceless part \cite{CM1}. Moreover, a large number of metric anomalies (intrinsic interstitials, vacancies, point stacking faults) arising from a distribution of point defects, as well as thermal deformations, biological growth, etc., are geometric in nature and can be analyzed using Weyl geometry \cite{CM2}.

In the present work, we have considered another aspect of Weyl geometry, namely, its effect on the propagation of electromagnetic waves in vacuum, in the presence of a gravitational field. More precisely, we have investigated the frequency-dependent propagation of light, which is the result of the spin-orbit coupling between the external and internal degrees of freedom of electromagnetic waves, and which is described physically with the help of the Berry phase \cite{opt1,opt2,opt3}. Spin-orbit coupling leads to the spin Hall effect for light, whose observational detection has opened new perspectives in the study of semiconductor spintronics/valleytronics,
in high-energy physics, and in condensed matter physics \cite{CM3}. It is also important to note that the spin Hall effect for light has a topological nature in the spin-orbit interaction similar to that of the standard spin Hall effect in electronic systems. The theory of the spin Hall effect of light was generalized to the case of Riemannian geometry in \cite{O1}, where the polarization-dependent ray equations describing the gravitational spin Hall effect of light were obtained. A numerical analysis of the polarization-dependent ray propagation in Schwarzschild geometry was also presented, and the magnitude of the effect was estimated. It is important to mention that the gravitational spin Hall effect for light is analogous to the spin Hall effect of light in inhomogeneous media, which has been observed experimentally. 

A particular and very interesting example of the gravitational spin Hall effect is represented by the deflection of light by a black hole that has the mass of the Sun and a gravitational (Schwarzschild) radius of the order of 3 km, $r_s\approx 3$ km. For a circularly polarized light ray coming from a distant source, passing
very near the surface of the Sun and then reaching Earth, it turns out that the distance of separation between the rays of opposite circular polarization would depend on the wavelength. For a wavelength of the order $\lambda \approx  10^{-9}$ m, the separation distance of the order $d \approx  10^{-15}$ m. For $\lambda \approx  1$ m, the separation distance is $d \approx 10^{-6}$ m \cite{O1}. This ray separation in standard Riemann geometry is very small, $10^{-6}$ times smaller than the wavelength of light. However, it is important to note that the rays are scattered by a finite angle, and after the reintersection point their separation increases linearly with the distance. On the other hand, much more massive compact objects, such as neutron stars or supermassive black holes, could generate a much stronger Riemannian gravitational spin Hall effect of light \cite{O1}.

It is interesting to point out that in Riemann geometry \cite{O2}, as well as in Weyl geometry, the gravitational spin Hall equations are a special case of the Mathisson-Papapetrou equations \cref{eq:MPD_Weyl} describing the motion of spinning bodies. To obtain an evolution equation for the worldline, one must impose the Corinaldesi-Papapetrou spin supplementary condition $S^{\alpha \beta} t_\beta = 0$, where $t_\beta$ is a timelike vector field and can be interpreted as a choice of family of observers relative to which energy centroids of wave packets are defined.

The study of the spin Hall effect for light in Weyl geometry is very much simplified by the conformal invariance of Maxwell's equations. This leads to the important result that in both Riemann and Weyl geometry the electromagnetic field equations take the same form, and thus one can efficiently use again the covariant WKB approximation to study the propagation of light in arbitrary Weyl geometries. However, the polarization-dependent ray equations \eqref{eq:GSHE_weyl} describing the gravitational spin Hall effect of light in Weyl geometry contain the curvature tensor of the Weyl manifold, which introduces a new degree of freedom for the description of motion, the Weyl vector $\omega _{\mu}$, and its covariant derivatives, respectively.

As an astrophysical application of the general formalism, we have considered the frequency-dependent motion of light in a black hole-type solution of the gravitational field equations in the simplest version of Weyl geometric gravity \cite{Yang}, in which the Weyl vector has only one nonzero component $\omega_1$. In this case the Weyl-type field equations do have an exact static spherically symmetric solution, which generalizes the Schwarzschild-de Sitter solutions of general relativity by introducing a new radial distance dependent linear term in the metric. This solution was used in Ref. \cite{Pi1} to propose an alternative geometric description of the galactic rotation curves and of the galactic properties that are usually attributed to the existence of dark matter. Within the framework of this model, an effective geometric mass term can be introduced, with an associated density profile. A comparison with a small selected sample of galactic rotation curves was also made by also considering an explicit breaking of the conformal invariance at galactic scales. The parameters of the black hole solution were fixed from the comparison with the observational data as $\gamma =2/C_2\approx 10^{-28}$ m$^{-1}$ and $C_1\approx 10^4$. For the integration constant $C_3$, it was assumed that it has values of the same order of magnitude as the cosmological constant. Hence, the preliminary investigations of \cite{Pi1} indicated that the Weyl geometric theory may represent a viable theoretical explanation for the galactic dynamics without invoking the existence of the mysterious dark matter. 

In the exact solution of Weyl geometric gravity, $C_1$, $C_2$, and $C_3$ represent integration constants, similar to the gravitational radius (or mass) in the Schwarzschild solution. Hence, their numerical values depend on the astrophysical system considered and may also depend on the mass of the compact object. We have studied numerically the spin Hall effect of light for this Weyl geometric type black hole solution, and investigated the motion of light in this metric, by also performing a detailed comparison with the similar effects in the Schwarzschild geometry. The numerical results indicate a strong effect of Weyl geometry on the polarized light dynamics and a significant increase in its magnitude compared to Riemann geometry. This effect is expected to increase with the distance from the source, and thus astrophysical observations of the spin Hall effect of light, as well as the possible detection of the deviations from the Schwarzschild/Kerr geometries may provide convincing evidence for the presence of the Weyl geometry in the Universe. Therefore, these results on the spin Hall effect of light lead to the possibility of directly constraining the Weyl geometric gravity theory by using astrophysical and astronomical observations of the motion of light emitted near compact objects. In the present work, we have introduced some basic tools necessary for a detailed comparison of the predictions of the spin Hall effect of light in the Weyl geometric gravity theory with the results of astrophysical observations.
  
\section*{Acknowledgments}

The work of T.H. is supported by a grant from the Romanian Ministry of Education and Research, CNCS-UEFISCDI, Project No. PN-III-P4-ID-PCE-2020-2255 (PNCDI III).

\appendix

\begin{widetext}
\section{The curvature tensor in Weyl geometry}\label{app}

In this appendix, we present the full details of the computation of the curvature tensor in Weyl geometry. Using the decomposition of the Weyl connection given in \cref{Con}, we can write
\bea
\tilde{R}\indices{^{\lambda }_{\mu \nu \sigma }} &=&\partial _{\nu
}\Gamma _{\mu \sigma }^{\lambda }-\partial _{\sigma }\Gamma _{\mu \nu
}^{\lambda }+\partial _{\nu }\Psi _{\mu \sigma }^{\lambda }-\partial
_{\sigma }\Psi _{\mu \nu }^{\lambda } + \left( \Gamma _{\rho \nu }^{\lambda }+\Psi _{\rho \nu
}^{\lambda }\right) \left( \Gamma _{\mu \sigma }^{\rho }+\Psi _{\mu \sigma
}^{\rho }\right)  - \left( \Gamma _{\rho \sigma }^{\lambda }+\Psi _{\rho
\sigma }^{\lambda }\right) \left( \Gamma _{\mu \nu }^{\rho }+\Psi _{\mu \nu
}^{\rho }\right)   \nonumber\\
&=&R\indices{^{\lambda }_{\mu \nu \sigma }}+\partial _{\nu }\Psi _{\mu
\sigma }^{\lambda }-\partial _{\sigma }\Psi _{\mu \nu }^{\lambda }+\Gamma
_{\rho \nu }^{\lambda }\Psi _{\mu \sigma }^{\rho }+\Gamma _{\mu \sigma
}^{\rho }\Psi _{\rho \nu }^{\lambda } -\Gamma _{\rho \sigma }^{\lambda }\Psi _{\mu \nu }^{\rho
}-\Gamma _{\mu \nu }^{\rho }\Psi _{\rho \sigma }^{\lambda }+\Psi _{\rho \nu
}^{\lambda }\Psi _{\mu \sigma }^{\rho }-\Psi _{\rho \sigma }^{\lambda }\Psi
_{\mu \nu }^{\rho }.
\eea
We have the following relations:
\begin{subequations}
\begin{align}
\partial _{\nu }\Psi _{\mu \sigma }^{\lambda } &= \frac{\alpha}{2} \Big(
\delta _{\mu }^{\lambda }\partial _{\nu }\omega _{\sigma }+\delta _{\sigma
}^{\lambda }\partial _{\nu }\omega _{\mu } -\omega ^{\lambda }\partial _{\nu
}g_{\mu \sigma }-g_{\mu \sigma }\partial _{\nu }\omega ^{\lambda }\Big) ,\\
\partial _{\sigma }\Psi _{\mu \nu }^{\lambda } &=\frac{\alpha}{2} \Big(
\delta _{\mu }^{\lambda }\partial _{\sigma }\omega _{\nu }+\delta _{\nu
}^{\lambda }\partial _{\sigma }\omega _{\mu }-\omega ^{\lambda }\partial
_{\sigma }g_{\mu \nu }-g_{\mu \nu }\partial _{\sigma }\omega ^{\lambda
}\Big) ,
\end{align}
\end{subequations}
Using the above equations, we can write
\begin{equation}
\partial _{\nu }\Psi _{\mu \sigma }^{\lambda }-\partial _{\sigma }\Psi
_{\mu \nu }^{\lambda } = \frac{\alpha}{2} \Big[ \tilde{W}_{\nu \sigma }\delta _{\mu
}^{\lambda }+\delta _{\sigma }^{\lambda }\partial _{\nu }\omega _{\mu
}-\delta _{\nu }^{\lambda }\partial _{\sigma }\omega _{\mu } -\omega ^{\lambda }\left( \partial _{\nu }g_{\mu \sigma }-\partial
_{\sigma }g_{\mu \nu }\right) -g_{\mu \sigma }\partial _{\nu }\omega
^{\lambda }+g_{\mu \nu }\partial _{\sigma }\omega ^{\lambda }\Big].
\end{equation}
Next, we obtain
\begin{subequations}
    \begin{align}
        \Gamma _{\rho \nu }^{\lambda }\Psi _{\mu \sigma }^{\rho }&=\frac{\alpha}{2}
        \left( \Gamma _{\mu \nu }^{\lambda }\omega _{\sigma }+\Gamma _{\sigma \nu
        }^{\lambda }\omega _{\mu }-\Gamma _{\rho \nu }^{\lambda }g_{\mu \sigma }\omega ^{\rho }\right) , \\
        \Gamma _{\mu \sigma }^{\rho }\Psi _{\rho \nu }^{\lambda }&=\frac{\alpha}{2}
\left( \Gamma _{\mu \sigma }^{\lambda }\omega _{\nu }+\Gamma _{\mu \sigma
}^{\rho }\delta _{\nu }^{\lambda }\omega _{\rho }-\Gamma _{\nu \mu \sigma
} \omega ^{\lambda }\right) , \\
        \Gamma _{\rho \sigma }^{\lambda }\Psi _{\mu \nu }^{\rho }&=\frac{\alpha}{2}
\left( \Gamma _{\mu \sigma }^{\lambda }\omega _{\nu }+\Gamma _{\nu \sigma
}^{\lambda }\omega _{\mu }-\Gamma _{\rho \sigma }^{\lambda }g_{\mu \nu
}\omega ^{\rho }\right) , \\
        \Gamma _{\mu \nu }^{\rho }\Psi _{\rho \sigma }^{\lambda }&=\frac{\alpha}{2}
\left( \Gamma _{\mu \nu }^{\lambda }\omega _{\sigma }+\Gamma _{\mu \nu
}^{\rho }\delta _{\sigma }^{\lambda }\omega _{\rho }-\Gamma _{\sigma \mu
\nu }\omega ^{\lambda }\right) .
    \end{align}
\end{subequations}
Therefore, we find
\begin{equation}
\begin{split}
&\Gamma _{\rho \nu }^{\lambda }\Psi _{\mu \sigma }^{\rho }+\Gamma _{\mu
\sigma }^{\rho }\Psi _{\rho \nu }^{\lambda }-\Gamma _{\rho \sigma }^{\lambda
}\Psi _{\mu \nu }^{\rho }-\Gamma _{\mu \nu }^{\rho }\Psi _{\rho \sigma
}^{\lambda } =\frac{\alpha}{2} \Big( \Gamma _{\mu \sigma }^{\rho }\delta _{\nu
}^{\lambda }\omega _{\rho }-\Gamma _{\mu \nu }^{\rho }\delta _{\sigma
}^{\lambda }\omega _{\rho } +\Gamma _{\rho \sigma }^{\lambda }g_{\mu \nu
}\omega ^{\rho }-\Gamma _{\rho \nu }^{\lambda }g_{\mu \sigma }\omega ^{\rho
} +\Gamma _{\sigma \mu \nu }\omega ^{\lambda }-\Gamma _{\nu \mu \sigma
}\omega ^{\lambda }\Big).
\end{split}
\end{equation}
Then
\bea
&&\partial _{\nu }\Psi _{\mu \sigma }^{\lambda }-\partial _{\sigma }\Gamma
_{\mu \nu }^{\lambda }+\Gamma _{\rho
\nu }^{\lambda }\Psi _{\mu \sigma }^{\rho }+\Psi _{\rho \nu }^{\lambda
}\Gamma _{\mu \sigma }^{\rho }-\Gamma _{\rho \sigma }^{\lambda }\Psi _{\mu
\nu }^{\rho }-\Gamma _{\mu \nu }^{\rho }\Psi _{\rho \sigma }^{\lambda }\nonumber\\
&=&\frac{\alpha}{2} \Bigg(
\tilde{W}_{\nu \sigma }\delta _{\mu }^{\lambda }+\delta _{\sigma }^{\lambda
}\partial _{\nu }\omega _{\mu }-\delta _{\nu }^{\lambda }\partial _{\sigma
}\omega _{\mu }-\omega ^{\lambda }\partial _{\nu }g_{\mu \sigma }+\omega
^{\lambda }\partial _{\sigma }g_{\mu \nu }-g_{\mu \sigma }\partial _{\nu
}\omega ^{\lambda }+g_{\mu \nu }\partial _{\sigma }\omega ^{\lambda }\nonumber\\
&&-\Gamma _{\rho \nu }^{\lambda }g_{\mu \sigma }\omega ^{\rho }+\Gamma
_{\rho \sigma }^{\lambda }g_{\mu \nu }\omega ^{\rho }+\Gamma _{\mu \sigma
}^{\rho }\delta _{\nu }^{\lambda }\omega _{\rho }-\Gamma _{\mu \nu }^{\rho
}\delta _{\sigma }^{\lambda }\omega _{\rho }+\Gamma _{\sigma ,\mu \nu
}\omega ^{\lambda }-\Gamma _{\nu ,\mu \sigma }\omega ^{\lambda }\Bigg).
\eea
With the use of the relation
\begin{equation}
\Gamma _{\sigma \mu \nu }-\Gamma _{\nu \mu \sigma }=\partial _{\nu
}g_{\sigma \mu }-\partial _{\sigma }g_{\mu \nu },
\end{equation}
we simplify the above equation, thus obtaining
\bea
&&\partial _{\nu }\Psi _{\mu \sigma }^{\lambda }-\partial _{\sigma }\Gamma
_{\mu \nu }^{\lambda }+\Gamma _{\rho
\nu }^{\lambda }\Psi _{\mu \sigma }^{\rho }+\Psi _{\rho \nu }^{\lambda
}\Gamma _{\mu \sigma }^{\rho }-\Gamma _{\rho \sigma }^{\lambda }\Psi _{\mu
\nu }^{\rho }-\Gamma _{\mu \nu }^{\rho }\Psi _{\rho \sigma }^{\lambda }  \nonumber\\
&&=\frac{\alpha}{2} \Big( \tilde{W}_{\nu \sigma }\delta _{\mu }^{\lambda }+\delta
_{\sigma }^{\lambda }\nabla _{\nu }\omega _{\mu }-\delta _{\nu }^{\lambda
}\nabla _{\sigma }\omega _{\mu }+g_{\mu \nu }\nabla _{\sigma }\omega
^{\lambda }-g_{\mu \sigma }\nabla _{\nu }\omega ^{\lambda }\Big) .
\eea
The term $\Psi _{\rho \nu }^{\lambda }\Psi _{\mu \sigma }^{\rho }-\Psi
_{\rho \sigma }^{\lambda }\Psi _{\mu \nu }^{\rho }$ can be represented in
the form
\begin{equation}
\begin{split}
\Psi _{\rho \nu }^{\lambda }\Psi _{\mu \sigma }^{\rho }-\Psi _{\rho \sigma
}^{\lambda }\Psi _{\mu \nu }^{\rho }=\frac{\alpha^{2}}{4}\Big[ \left(
g_{\mu \nu }\omega _{\rho }\omega ^{\rho }-\omega _{\nu }\omega _{\mu
}\right) \delta _{\sigma }^{\lambda }-\left( g_{\mu \sigma }\omega _{\rho
}\omega ^{\rho }-\omega _{\mu }\omega _{\sigma }\right) \delta _{\nu
}^{\lambda }+\left( g_{\mu \sigma }\omega _{\nu }-g_{\mu \nu }\omega
_{\sigma }\right) \omega ^{\lambda }\Big].
\end{split}
\end{equation}
Finally, we obtain the curvature tensor in the Weyl conformal geometry as
\bea
\tilde{R}\indices{^{\lambda }_{\mu \nu \sigma }} =R\indices{^{\lambda }_{\mu \nu \sigma }} &+&\frac{\alpha}{2} \Big[ \tilde{W}_{\nu \sigma }\delta _{\mu }^{\lambda
}+\left( \delta _{\sigma }^{\lambda }\nabla _{\nu }-\delta _{\nu }^{\lambda
}\nabla _{\sigma }\right) \omega _{\mu }+\left( g_{\mu \nu }\nabla _{\sigma
}-g_{\mu \sigma }\nabla _{\nu }\right) \omega ^{\lambda }\Big] \nonumber\\
&&+\frac{\alpha ^{2}}{4}\Big[ \left( \omega ^{2}g_{\mu \nu }-\omega _{\mu
}\omega _{\nu }\right) \delta _{\sigma }^{\lambda }-\left( \omega ^{2}g_{\mu
\sigma }-\omega _{\mu }\omega _{\sigma }\right) \delta _{\nu }^{\lambda
}+\left( g_{\mu \sigma }\omega _{\nu }-g_{\mu \nu }\omega _{\sigma }\right)
\omega ^{\lambda }\Big],
\eea
where we used the notation $\omega ^2=\omega _\rho \omega ^\rho$.

\end{widetext}

\section{The observer dependence of the spin Hall trajectories in Weyl geometry} \label{app:observer}

In this appendix, we show how the spin Hall trajectories depend on the choice of observer. The approach here closely follows the discussion in Ref. \cite{O2}, where the observer dependence of the spin Hall equations in Riemannian geometry was presented. We start by introducing the general transformation law between energy centroids, average momenta, and spin tensors associated with the same wave packet by different observers. Then, we present an explicit example by analyzing the spin Hall equations for different families of observers.

\subsection{Change of observer for the spin Hall equations}

Similarly to Riemannian geometry, changes of observer for the spin Hall equations are best understood by examining the change of spin supplementary condition for the Mathisson-Papapetrou form of the equations \cite{O2}. Given two timelike vector fields $t^\alpha$ and $T^\alpha$, representing two different families of observers that can describe the dynamics of electromagnetic wave packets, the corresponding spin Hall equations can be obtained by starting with the Mathisson-Papapetrou equations \eqref{eq:MPD_Weyl} and imposing different spin supplementary conditions: $S^{\alpha \beta} t_\beta = 0$ or $\bar{S}^{\alpha \beta} T_\beta = 0$. Then, each family of observers will generally provide a different description for the same electromagnetic wave packet by assigning different energy centroids, average momenta, and spin tensors:
\begin{subequations}
\begin{align}
    t^\alpha &: \qquad \{x^\mu, \mathcal{P}_\alpha, S^{\alpha \beta} \}, \\
    T^\alpha &: \qquad \{\bar{x}^{\bar{\alpha}}, \bar{\mathcal{P}}_{\bar{\alpha}}, \bar{S}^{\bar{\alpha} \bar{\beta}} \}.
\end{align}    
\end{subequations}
We use bars on the indices to emphasize that these objects might be defined at different spacetime points than the objects without bars on the indices (similar to the notation used in Refs. \cite{vines2016canonical,O2}). In the Riemannian case it has been shown that, under arbitrary changes of timelike observers, the two sets of quantities $\{x^\mu, \mathcal{P}_\alpha, S^{\alpha \beta} \}$ and $\{\bar{x}^{\bar{\mu}}, \bar{\mathcal{P}}_{\bar{\alpha}}, \bar{S}^{\bar{\alpha} \bar{\beta}} \}$ are related by the transformation \cite{O2}
\begin{subequations} \label{eq:obs_transf}
\begin{align}
    \bar{x}^{\bar{\mu}} &= \exp{}_{x^\mu}(\xi^\mu)  + \mathcal{O}(\epsilon^2), \label{eq:delta_x1}\\
    \bar{\mathcal{P}}_{\bar{\alpha}} &= g\indices{_{\bar{\alpha}}^{\alpha}} \mathcal{P}_\alpha + \mathcal{O}(\epsilon^2), \label{eq:delta_p1}\\
    \bar{S}^{\bar{\alpha} \bar{\beta}} &= g\indices{^{\bar{\alpha}}_{\alpha}} g\indices{^{\bar{\beta}}_{\beta}} ( S^{\alpha \beta} + \mathcal{P}^\alpha \xi^\beta - \mathcal{P}^\beta \xi^\alpha) + \mathcal{O}(\epsilon^2), \label{eq:delta_S1}
\end{align}
\end{subequations}
where $\exp$ is the exponential map on the tangent bundle and $g\indices{^{\bar{\alpha}}_{\alpha}}$ is the bitensor which parallel propagates vectors from $x^\alpha$ to $\bar{x}^{\bar{\alpha}}$ along the geodesic segment which connects those points (both $\exp$ and $g\indices{^{\bar{\alpha}}_{\alpha}}$ are defined with respect to the Levi-Civita connection $\Gamma^\lambda_{\mu \nu}$; see Ref. \cite{Poisson2011} for more details on bitensors), and the shift vector $\xi^\mu$ is defined as
\begin{equation} \label{eq:xi1}
    \xi^\mu = \frac{S^{\mu \nu} T_\nu}{\mathcal{P}_\sigma T^\sigma}.
\end{equation}
Note that $\xi^\mu T_\mu = \xi^\mu t_\mu = 0$. Also, $\xi^\mu = 0$ for $T^\mu = f(x) t^\mu$, where $f(x)$ is any real, smooth, and nonzero scalar function. 

We show now by direct calculations that, using the Mathisson-Papapetrou form of the spin Hall equations in Weyl geometry, changes of observers are also governed by the same transformation laws defined above. Consider first a timelike vector field $t^\alpha$. The corresponding spin Hall equations can be derived by imposing the spin supplementary condition $S^{\alpha \beta} t_\beta = 0$ for the Mathisson-Papapetrou equations \eqref{eq:MPD_Weyl}, as well as choosing a worldline parameter such that $\dot{x}^\mu t_\mu = \mathcal{P}^\mu t_\mu$. We obtain
\begin{subequations} \label{eq:obs1}
\begin{align}
    \dot{x}^\mu &= \mathcal{P}^\mu + \frac{1}{\mathcal{P} \cdot t} S^{\mu \nu} \dot{x}^\sigma \nabla_\sigma t_\nu, \label{eq:dot_x}\\
    \dot{x}^\mu \nabla_\mu \mathcal{P}_\alpha  &= - \frac{1}{2}  R_{\alpha \beta \gamma \lambda } \dot{x}^\beta S^{\gamma \lambda}, \label{eq:P}\\
    \dot{x}^\mu \nabla_\mu S^{\alpha \beta} &=  \mathcal{P}^{\alpha} \dot{x}^{\beta} - \mathcal{P}^{\beta} \dot{x}^{\alpha}. \label{eq:spin_eq1}
\end{align}
\end{subequations}
Also, in this case the spin tensor is defined as
\begin{equation}
    S^{\alpha \beta} = \epsilon s \frac{
\varepsilon^{\alpha \beta \rho \sigma} p_\rho t_\sigma }{p \cdot t} = \epsilon s \frac{
\varepsilon^{\alpha \beta \rho \sigma} \mathcal{P} \cdot t }{\mathcal{P} \cdot t} + \mathcal{O}(\epsilon^2),
\end{equation}
and satisfies \cref{eq:spin_eq1}. 

On the other hand, if we start with a different timelike vector field $T^\alpha$ by imposing the spin supplementary condition $\bar{S}^{\bar{\alpha} \bar{\beta}} T_{\bar{\beta}} = 0$ and a choice of the worldline parameter such that $\dot{\bar{x}}^{\bar{\mu}} T_{\bar{\mu}} = \bar{\mathcal{P}}^{\bar{\mu}} T_{\bar{\mu}}$, then we obtain
\begin{subequations} \label{eq:obs2}
\begin{align}
    \dot{\bar{x}}^{\bar{\mu}} &= \bar{\mathcal{P}}^{\bar{\mu}} + \frac{1}{\bar{\mathcal{P}} \cdot T} \bar{S}^{\bar{\mu} \bar{\nu}} \dot{\bar{x}}^{\bar{\sigma}} \nabla_{\bar{\sigma}} T_{\bar{\nu}}, \label{eq:bar_dot_x}\\
    \dot{\bar{x}}^{\bar{\mu}} \nabla_{\bar{\mu}} \bar{\mathcal{P}}_{\bar{\alpha}}  &= - \frac{1}{2}  R_{\bar{\alpha} \bar{\beta} \bar{\gamma} \bar{\lambda} } \dot{\bar{x}}^{\bar{\beta}} \bar{S}^{\bar{\gamma} \bar{\lambda}}, \label{eq:bar_P} \\
    \dot{\bar{x}}^{\bar{\mu}} \nabla_{\bar{\mu}} \bar{S}^{\bar{\alpha} \bar{\beta}} &=  \bar{\mathcal{P}}^{\bar{\alpha}} \dot{\bar{x}}^{\bar{\beta}} - \bar{\mathcal{P}}^{\bar{\beta}} \dot{\bar{x}}^{\bar{\alpha}}. \label{eq:spin_eq2}
\end{align}
\end{subequations}
In this case, the spin tensor satisfying \cref{eq:spin_eq2} is
\begin{equation} \label{eq:S2}
    \bar{S}^{\bar{\alpha} \bar{\beta}} = \epsilon s \frac{
\varepsilon^{\bar{\alpha} \bar{\beta} \bar{\rho} \bar{\sigma}} \bar{p}_{\bar{\rho}} T_{\bar{\sigma}} }{ \bar{p} \cdot T} = \epsilon s \frac{
\varepsilon^{\bar{\alpha} \bar{\beta} \bar{\rho} \bar{\sigma}} \bar{\mathcal{P}}_{\bar{\rho}} T_{\bar{\sigma}} }{ \bar{\mathcal{P}} \cdot T} + \mathcal{O}(\epsilon^2).
\end{equation}

In the following, we show that, up to error terms of order $\epsilon^2$, Eqs. \eqref{eq:obs1} and \eqref{eq:obs2} are consistent with the transformations given in Eqs. \eqref{eq:obs_transf} and \eqref{eq:xi1}. We start with the transformation law \eqref{eq:delta_x1} for energy centroids. Based on \cite[Eq. 3.5]{vines2016canonical}, the derivative of \cref{eq:delta_x1} can be expanded as
\begin{equation} \label{eq:dotx_bar_dotx}
    \dot{\bar{x}}^{\bar{\mu}} = g\indices{^{\bar{\mu}}_{\mu}} \left[ \dot{x}^\mu + \dot{x}^\nu \nabla_\nu \xi^\mu + \mathcal{O}(\xi^2) \right].
\end{equation}
Note that $\mathcal{O}(\xi) = \mathcal{O}(S) = \mathcal{O}(\epsilon)$, so in the above equation we are ignoring terms of order $\epsilon^2$. The covariant derivative of the shift vector $\xi^\mu$ along the worldline $x^\nu$ is
\begin{align} \label{eq:dxi}
    \dot{x}^\nu \nabla_\nu \xi^\mu &= \frac{\dot{x} \cdot T}{\mathcal{P} \cdot T} \mathcal{P}^\mu + \frac{1}{\mathcal{P} \cdot T} S^{\mu \nu} \dot{x}^\sigma \nabla_\sigma T_\nu \nonumber \\
    &\qquad- \dot{x}^\mu - \frac{\mathcal{P}^\nu \dot{x}^{\sigma} \nabla_\sigma T_\nu }{\mathcal{P} \cdot T} \xi^\mu + \mathcal{O}(\epsilon^2).
\end{align}
Using the expression above, we obtain
\begin{align} \label{eq:dotx_3}
     \dot{\bar{x}}^{\bar{\mu}} &= g\indices{^{\bar{\mu}}_{\mu}} \bigg[ \frac{\dot{x} \cdot T}{\mathcal{P} \cdot T} \mathcal{P}^\mu + \frac{1}{\mathcal{P} \cdot T} S^{\mu \nu} \dot{x}^\sigma \nabla_\sigma T_\nu \nonumber \\
    &\qquad\qquad + \frac{\mathcal{P}^\nu \dot{x}^{\sigma} \nabla_\sigma T_\nu }{\mathcal{P} \cdot T} \xi^\mu + \mathcal{O}(\epsilon^2) \bigg]. 
\end{align}
The first term on the right-hand side of the above equation can be rewritten as
\begin{align}
    \frac{\dot{x} \cdot T}{\mathcal{P} \cdot T} \mathcal{P}^\mu &= \frac{\left(\dot{\bar{x}}^\nu - \dot{x}^\sigma \nabla_\sigma \xi^\nu \right) T_\nu}{\mathcal{P} \cdot T} \mathcal{P}^\mu \nonumber \\
    &= \left( 1 - \frac{T_\nu \dot{x}^\sigma \nabla_\sigma \xi^\nu }{\mathcal{P} \cdot T} \right) \mathcal{P}^\mu.
\end{align}
Furthermore, using \cref{eq:delta_S1} we can also express the spin tensor as $S^{\mu \nu} = \bar{S}^{\mu \nu} - \mathcal{P}^\mu \xi^\nu + \mathcal{P}^\nu \xi^\mu + \mathcal{O}(\epsilon^2)$. Using these expressions, \cref{eq:dotx_3} becomes
\begin{align}
     \dot{\bar{x}}^{\bar{\mu}} &= g\indices{^{\bar{\mu}}_{\mu}} \bigg[ \mathcal{P}^\mu + \frac{1}{\mathcal{P} \cdot T} \bar{S}^{\mu \nu} \dot{x}^\sigma \nabla_\sigma T_\nu \nonumber \\
    &\qquad\qquad - 2 \frac{ \mathcal{P}^\mu }{\mathcal{P} \cdot T} \dot{x}^\sigma \nabla_\sigma \left( \xi^\nu T_\nu \right) + \mathcal{O}(\epsilon^2) \bigg]   \nonumber  \\
    &= g\indices{^{\bar{\mu}}_{\mu}} \bigg[ \mathcal{P}^\mu + \frac{1}{\mathcal{P} \cdot T} \bar{S}^{\mu \nu} \dot{x}^\sigma \nabla_\sigma T_\nu + \mathcal{O}(\epsilon^2) \bigg] \nonumber \\
    &=  \bar{\mathcal{P}}^{\mu} + \frac{1}{\mathcal{P} \cdot T} \bar{S}^{\bar{\mu} \nu} \dot{x}^\sigma \nabla_\sigma T_\nu + \mathcal{O}(\epsilon^2) .
\end{align}
Up to error terms of order $\epsilon^2$, the above equation is the same as \cref{eq:bar_dot_x} if we note the following. First, $\mathcal{P} \cdot T = \mathcal{P}_\mu T^\mu = \bar{\mathcal{P}}_{\bar{\mu}} T^{\bar{\mu}}$ since the scalar product is invariant if both vectors are parallel transported with respect to the Levi-Civita connection. Second, since $\dot{\bar{x}}^\sigma = \dot{x}^\sigma + \mathcal{O}(\epsilon)$ by \cref{eq:dotx_bar_dotx}, we can replace $\dot{x}^\sigma$ with $\dot{\bar{x}}^\sigma$ in the second term on the right-hand side. Finally, using \cite[Eq. (6.11)]{Poisson2011} or \cite[Eq. (3.8)]{vines2016canonical} we have $\nabla_\sigma g\indices{_\nu^{\bar{\nu}}} = \mathcal{O}(\xi) = \mathcal{O}(\epsilon)$ and we can write
\begin{align}
    \bar{S}^{\bar{\mu} \nu} \dot{\bar{x}}^\sigma \nabla_\sigma T_\nu &= \bar{S}^{\bar{\mu} \nu} \dot{\bar{x}}^\sigma g\indices{_\sigma^{\bar{\sigma}}} g\indices{_{\bar{\sigma}}^\sigma} \nabla_\sigma \left( g\indices{_\nu^{\bar{\nu}}} T_{\bar{\nu}} \right) \nonumber \\
    &= \bar{S}^{\bar{\mu} \nu} g\indices{_\nu^{\bar{\nu}}} \dot{\bar{x}}^{\bar{\sigma}}  \nabla_{\bar{\sigma}} T_{\bar{\nu}} + \mathcal{O}(\epsilon^2) \nonumber \\
    &= \bar{S}^{\bar{\mu} \bar{\nu}} \dot{\bar{x}}^{\bar{\sigma}}  \nabla_{\bar{\sigma}} T_{\bar{\nu}} + \mathcal{O}(\epsilon^2).
\end{align}
Thus, up to error terms of order $\epsilon^2$, Eqs. \eqref{eq:dot_x} and \eqref{eq:bar_dot_x} are consistent under the transformation given in \cref{eq:delta_x1}. 

Next, we examine the transformation law for the momenta. Expanding the covariant derivative of \cref{eq:delta_p1} as in \cite[Eq. (3.7)]{vines2016canonical}, we obtain
\begin{align}
    \dot{\bar{x}}^{\bar{\mu}} \nabla_{\bar{\mu}} \bar{\mathcal{P}}_{\bar{\alpha}} &= g\indices{_{\bar \alpha}^\alpha} \left[ \dot{x}^\mu \nabla_\mu \mathcal{P}_\alpha - R\indices{_\alpha^\beta_{\gamma \lambda} } \mathcal{P}_\beta \dot{x}^\gamma \xi^\lambda + \mathcal{O}(\epsilon^2) \right].
\end{align}
Since $\dot{x}^\mu = \mathcal{P}^\mu + \mathcal{O}(\epsilon)$ and $\dot{\bar{x}}^\sigma = \dot{x}^\sigma + \mathcal{O}(\epsilon)$, we can rewrite the above expression as
\begin{align}
    \dot{\bar{x}}^{\bar{\mu}} \nabla_{\bar{\mu}} \bar{\mathcal{P}}_{\bar{\alpha}} &= g\indices{_{\bar \alpha}^\alpha} \left[ \dot{x}^\mu \nabla_\mu \mathcal{P}_\alpha - R\indices{_\alpha_\beta_{\gamma \lambda} }\dot{x}^\beta \mathcal{P}^{[\gamma}  \xi^{\lambda]} + \mathcal{O}(\epsilon^2) \right] \nonumber \\
    &= g\indices{_{\bar \alpha}^\alpha} \left[ - \frac{1}{2}  R_{\alpha \beta \gamma \lambda } \dot{x}^\beta  \bar{S}^{\gamma \lambda} + \mathcal{O}(\epsilon^2) \right] \nonumber \\
    &= - \frac{1}{2}  R_{\bar{\alpha} \bar{\beta} \bar{\gamma} \bar{\lambda} } \dot{\bar{x}}^{\bar{\beta}} \bar{S}^{\bar{\gamma} \bar{\lambda}} + \mathcal{O}(\epsilon^2).
\end{align}
Thus, up to error terms of order $\epsilon^2$, Eqs. \eqref{eq:P} and \eqref{eq:bar_P} are consistent with the transformation law in \cref{eq:delta_p1}.

Finally, we examine the transformation law for the spin tensors. Expanding the covariant derivative of \cref{eq:delta_S1}, we obtain
\begin{equation}
    \dot{\bar{x}}^{\bar{\mu}} \nabla_{\bar{\mu}} \bar{S}^{\bar{\alpha} \bar{\beta}} = g\indices{^{\bar{\alpha}}_{\alpha}} g\indices{^{\bar{\beta}}_{\beta}} \dot{x}^\sigma \nabla_\sigma ( S^{\alpha \beta} + \mathcal{P}^\alpha \xi^\beta - \mathcal{P}^\beta \xi^\alpha) + \mathcal{O}(\epsilon^2).
\end{equation}
Note that the terms that involve covariant derivatives of bitensors do not appear in the above expression because they are of order $\epsilon^2$. Using Eqs. \eqref{eq:spin_eq1}, \eqref{eq:dxi} and \eqref{eq:dotx_bar_dotx}, we obtain
\begin{align}
    \dot{\bar{x}}^{\bar{\mu}} \nabla_{\bar{\mu}} \bar{S}^{\bar{\alpha} \bar{\beta}} &=  g\indices{^{\bar{\alpha}}_{\alpha}} g\indices{^{\bar{\beta}}_{\beta}} \big[\mathcal{P}^{\alpha} (\dot{x}^{\beta} + \dot{x}^\sigma \nabla_\sigma \xi^\beta) \nonumber \\
    &\qquad- \mathcal{P}^{\beta} (\dot{x}^{\alpha} + \dot{x}^\sigma \nabla_\sigma \xi^\alpha)] + \mathcal{O}(\epsilon^2) \nonumber \\
    &= \bar{\mathcal{P}}^{\bar{\alpha}} \dot{\bar{x}}^{\bar{\beta}} - \bar{\mathcal{P}}^{\bar{\beta}} \dot{\bar{x}}^{\bar{\alpha}} + \mathcal{O}(\epsilon^2).
\end{align}
Thus, up to error terms of order $\epsilon^2$, Eqs. \eqref{eq:spin_eq1} and \eqref{eq:spin_eq2} are consistent with the transformation law in \cref{eq:delta_S1}.

\subsection{Examples of different observers}

For the black hole spacetimes introduced in \cref{sec:bh} [setting $\nu(r) = - \lambda(r)$], we can define an orthonormal tetrad
\begin{subequations}
    \begin{align}
        e_0 &= e^{-\nu(r)/2} \partial_t, \\
        e_1 &= e^{\nu(r)/2} \partial_r, \\
        e_2 &= \frac{1}{r} \partial_\theta, \\
        e_3 &= \frac{1}{r \sin \theta} \partial_\phi,
    \end{align}
\end{subequations}
where $e_0$ is a unit timelike vector and $e_i$ are unit spacelike vectors.

In \cref{sect3}, we consider the spin Hall effect of light in the black hole spacetimes introduced in \cref{sec:bh}. As discussed in the main text, the spin Hall trajectories represent the dynamics of energy centroids of electromagnetic wave packets, relative to a family of timelike observers with $4$-velocity $t^\mu$. For the examples presented in \cref{sect3}, our choice is $t^\mu = (e_0)^\mu = e^{-\nu(r)/2} \partial_t$. This represents a family of static observers in the considered black hole spacetimes, and the physical system described in Figs. \ref{fig:plot0}-\ref{fig:plot2} is that of a static source of light that emits circularly polarized wave packets. Then, the resulting spin Hall trajectories describe the dynamics of the energy centroids of these wave packets, as seen by a family of static timelike observers with $4$-velocities $t^\mu$.

\begin{figure*}
    \centering
    \includegraphics[width=0.98\columnwidth]{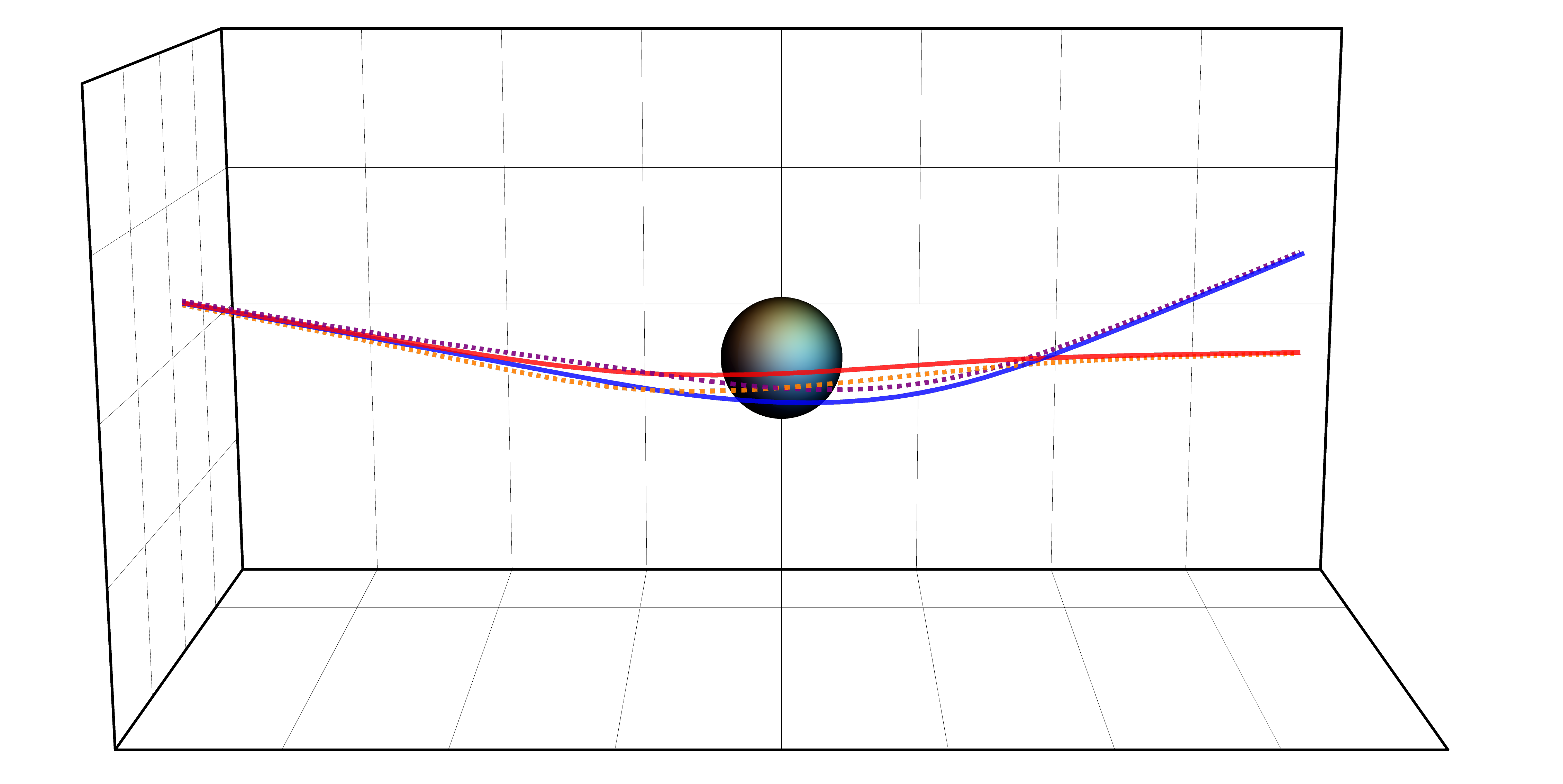}
    \hspace{0.01\columnwidth}
    \includegraphics[width=0.98\columnwidth]{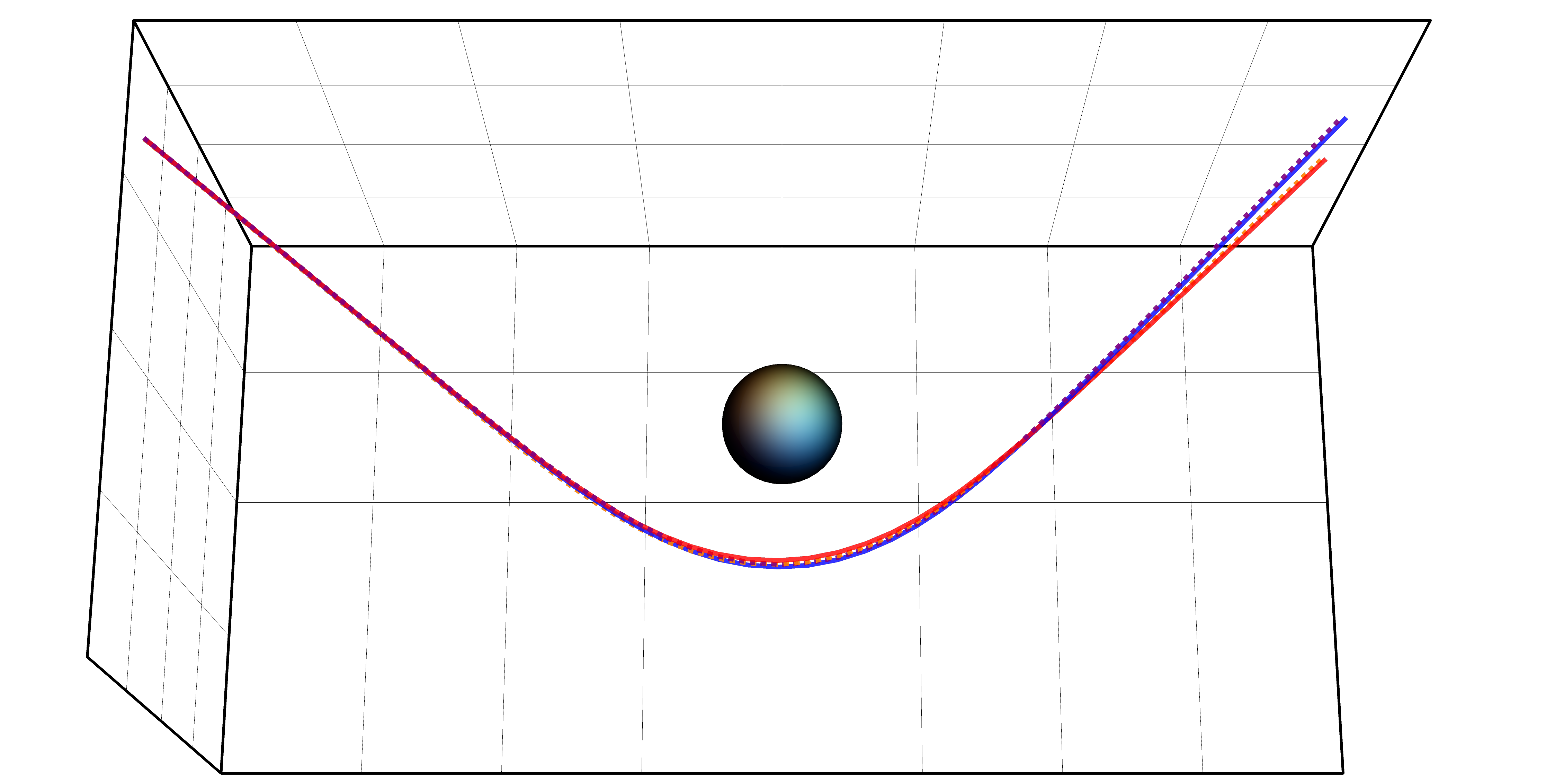}

    \vspace{0.2cm}

    \includegraphics[width=0.40\textwidth]{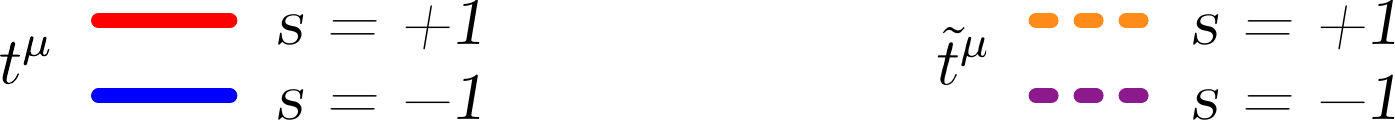}
    \caption{%
        Equatorial view (left) and top view (right) of spin Hall trajectories around a Schwarzschild black hole in Weyl geometry ($C_2 = C_3 = \delta = 0$), as seen by two families of observers with $4$-velocities $t^\mu$ (solid lines) and $\tilde{t}^\mu$ (dashed lines). These vector fields are defined in \cref{eq:observers}, with $v(r) = - \sqrt{r_s/r}$. All rays are emitted with the same initial conditions from a source very far away from the black hole, at $r = 10^4 r_s$, so that at the point of emission we have $t^\mu \propto \tilde{t}^\mu$.
    } \label{fig:obs}
\end{figure*} 

However, one can consider a different choice of a timelike vector field in the spin Hall equations, corresponding to the description of a different physical scenario. One simple example could be to consider a timelike vector field $\tilde{t}^\mu$ related to $t^\mu = (e_0)^\mu$ by a radial boost:
\begin{equation}
    \tilde{t} = \frac{e_0 + v e_1}{\sqrt{1 - v^2}},
\end{equation}
where the boost velocity $v = v(r)$ satisfies $v^2 < 1$. Another possibility could be to consider a family of free-falling observers moving on radial geodesics. These will be represented by a timelike vector field satisfying the geodesic equation
\begin{equation} \label{eq:free_fall_geo}
    T^\mu \tilde{\nabla}_\mu T^\alpha = 0.
\end{equation}
To solve this equation, we can parametrize $T^\alpha$ as
\begin{equation} \label{eq:T_param}
    T = f(r) \left[e_0 + g(r) e_1 \right],
\end{equation}
where $f(r)$ and $g(r)$ are two scalar functions to be determined by solving the above geodesic equation. Note that for $T$ to be a timelike vector, we must impose $f(r) \neq 0$ and $g^2(r) < 1$. Inserting \cref{eq:T_param} into \cref{eq:free_fall_geo}, we obtain
\begin{subequations}
\begin{align}
    f'(r) &= - \left[ \frac{\nu'(r)}{2} + \alpha \omega_r (r) \right] f(r) ,\\    
    g'(r) &= - \left[ \frac{\nu'(r)}{2} + \frac{\alpha}{2} \omega_r (r) \right] \frac{1 - g^2(r)}{g(r)}. 
\end{align}
\end{subequations}
Note that using \cref{eq:omega_r} we have $\omega_r = \frac{1}{\alpha }\frac{\Phi ^{\prime }}{\Phi } = \frac{1}{\alpha } \left( \ln{ \Phi} \right)'$. Then, the solutions to the above differential equations are
\begin{subequations}
\begin{align}
    f(r) &= \frac{ n_1 }{\Phi(r) e^{\nu(r)/2}} ,\\    
    g(r) &= \pm \sqrt{1 - e^{n_2} \Phi(r) e^{\nu(r)}}, 
\end{align}
\end{subequations}
where $\nu(r)$ is defined in \cref{metrW}, $\Phi(r)$ is defined in \cref{40} (here with $C_1 = 1$), and $n_{1,2}$ are integration constants.

To illustrate how the spin Hall trajectories change under a change of observer, we consider the explicit example of a Schwarzschild black hole in Weyl geometry ($C_2 = C_3 = \delta = 0$), and the timelike vector fields $t^\mu$, $\tilde{t}^\mu$, and $T^\mu$ introduced above. For this spacetime, the vectors take the form
\begin{subequations} \label{eq:observers}
\begin{align}
    t &= \frac{1}{\sqrt{1- \frac{r_s}{r}}} \partial_t, \\
    \tilde{t} &= \frac{1}{\sqrt{\left(1 - v^2\right) \left( 1- \frac{r_s}{r} \right)}} \partial_t + v \sqrt{\frac{1- \frac{r_s}{r} }{1 - v^2} } \partial_r, \\
    T &= \frac{n_1 r^3}{r - r_s} \partial_t + n_1 \sqrt{r^4 - e^{n_2} r (r - r_s)} \partial_r.
\end{align}   
\end{subequations}

As a first example, we start by comparing spin Hall trajectories corresponding to the observers $t^\mu$ and $\tilde{t}^\mu$ with $v(r) = - \sqrt{r_s/r}$. These two different choices of observers will be used in the spin Hall equations \cref{eq:GSHE_weyl}. Then, we consider a source of light very far away from the black hole. We take $r_{source} = 10^4 r_s$, such that $t^\mu (r = r_{source}) \approx \tilde{t}^\mu (r = r_{source})$ (this ensures that the relativistic Hall effect \cite{Relativistic_Hall,Stone2015} is negligible on the initial wave packet prescribed at the source for the two observers). The resulting spin Hall trajectories, which describe the dynamics of the energy centroids relative to $t^\mu$ and $\tilde{t}^\mu$, are shown in \cref{fig:obs}. We can see that the spin Hall trajectories relative to $\tilde{t}^\mu$ are slightly different from the trajectories relative to $t^\mu$. This difference is more pronounced near the black hole, where the difference between $t^\mu$ and $\tilde{t}^\mu$ is more significant. As the light rays move away from the black hole, there is no significant difference between the two sets of trajectories. Thus, we see that the observer-dependent effects on the trajectories are small and fade out as the light rays move away from the black hole, whereas gravity produces a finite scattering angle, and the separation between rays of opposite circular polarization continues to grow away from the black hole.

\begin{figure*}
    \centering
    \includegraphics[width=0.98\columnwidth]{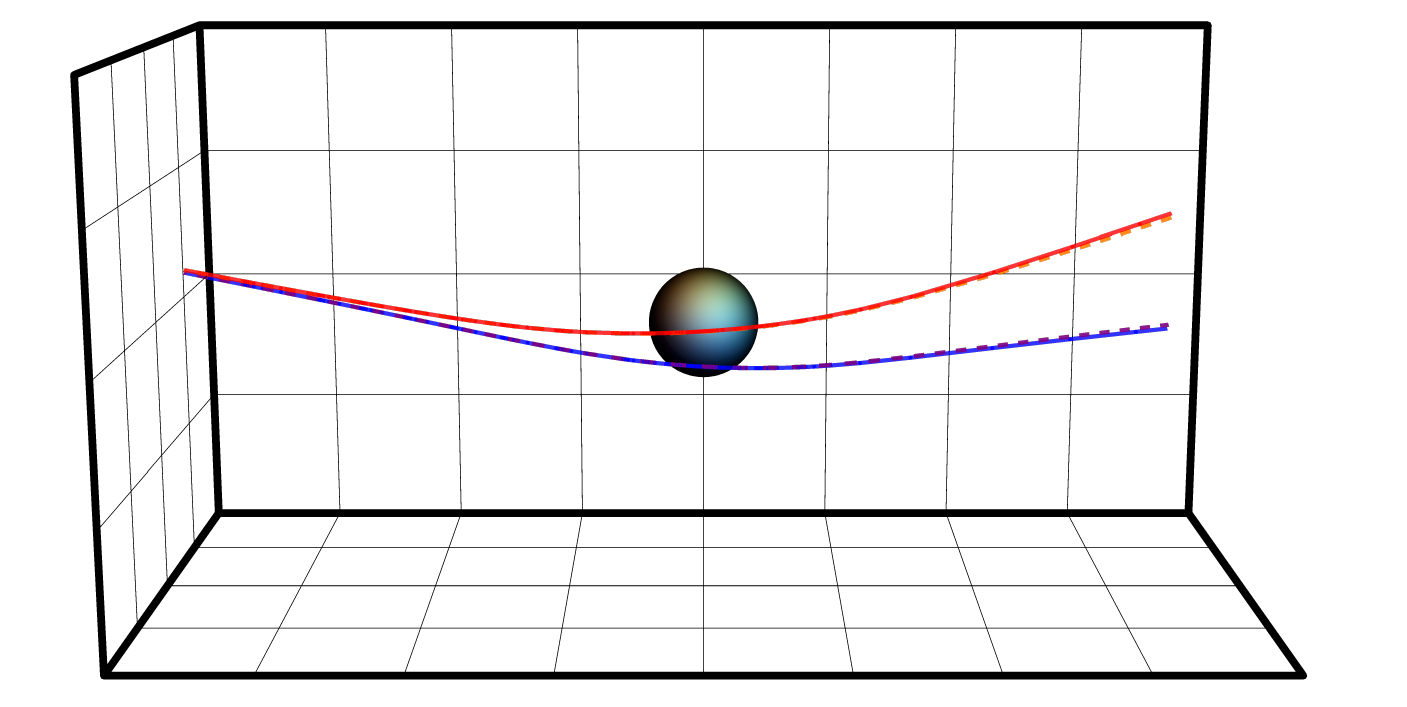}
    \hspace{0.01\columnwidth}
    \includegraphics[width=0.98\columnwidth]{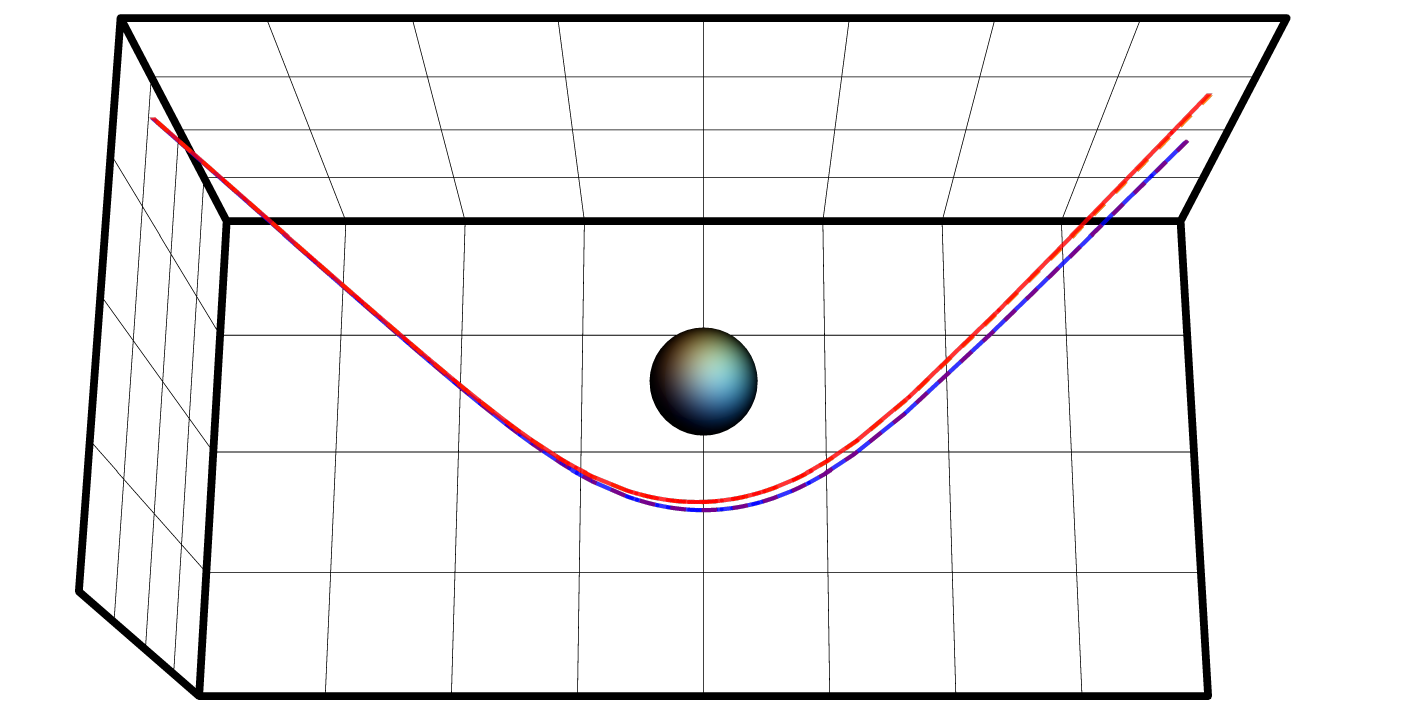}

    \vspace{0.2cm}

    \includegraphics[width=0.40\textwidth]{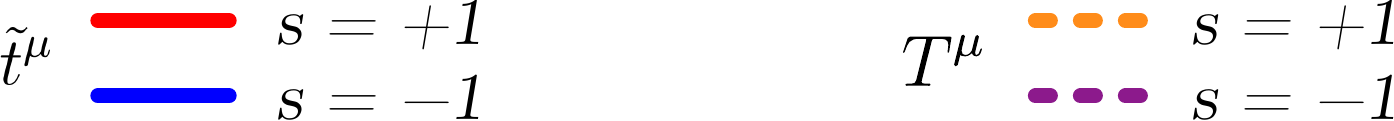}
    \caption{%
        Equatorial view (left) and top view (right) of spin Hall trajectories around a Schwarzschild black hole in Weyl geometry ($C_2 = C_3 = \delta = 0$), as seen by two families of observers with $4$-velocities proportional to $\tilde{t}^\mu$ (solid lines) and $T^\mu$ (dashed lines). These vector fields are defined in \cref{eq:observers}, with $e^{n_2} = 2$ and $v$ given in \cref{eq:v_boost}. All rays are emitted with the same initial conditions from a source far away from the black hole, at $r = 10^2 r_s$, and the constants $n_2$ and $v$ are chosen such that at the point of emission we have $\tilde{t}^\mu \propto T^\mu$.
    } \label{fig:obs2}
\end{figure*} 

As a second example, we will consider spin Hall trajectories relative to the radially free-falling observer $T^\mu$. However, in this case, it is not easy to compare the different trajectories corresponding to $t^\mu$ and $T^\mu$. The reason behind this difficulty arises because when making such a comparison, we would like to have $t^\mu (x_{source}) \propto T^\mu (x_{source})$. This condition is required to ensure that, at the location of the source of radiation, using the same initial conditions for both sets of spin Hall equations (those relative to $t^\mu$ and those relative to $T^\mu$) corresponds to initially prescribing the same electromagnetic wave packet. Otherwise, prescribing the same initial centroid at $x_{source}$ relative to both $t^\mu$ and $T^\mu$, we could end up describing the dynamics of different wave packets. In the previous example, this problem was solved by placing the source very far from the black hole, where $t^\mu (x_{source}) \approx \tilde{t}^\mu (x_{source})$. If we now insist on comparing spin Hall trajectories with respect to $t^\mu$ and $T^\mu$, then we would need to fix the integration constants $n_{1,2}$ such that $t^\mu (x_{source}) \propto T^\mu (x_{source})$. This can be achieved by setting 
\begin{equation}
    e^{n_2} = \frac{r_{source}^3}{r_{source} - r_s}.
\end{equation}
In this case, $T (x_{source}) \propto t (x_{source}) \propto \partial_t$. However, depending on the value of the integration constant $n_2$, the vector field $T^\mu$ will be undefined when $r^4 - e^{n_2} r (r - r_s) < 0$. Thus, $T^\mu$ will be undefined when $r_s < r_{source} < \frac{3 r_s}{2}$ and
\begin{equation}
   r_{source} < r < \frac{r_{source}}{2} \left( \sqrt{\frac{r_{source} + 3 r_s}{r_{source} - r_s}} -1 \right),
\end{equation}
or when $r_{source} > \frac{3 r_s}{2}$ and 
\begin{equation}
    \frac{r_{source}}{2} \left( \sqrt{\frac{r_{source} + 3 r_s}{r_{source} - r_s}} -1 \right) < r < r_{source}.
\end{equation}
Thus, if we insist on setting the integration constant $n_2$ so that at some spacetime point $T (x_{source}) \propto t (x_{source}) \propto \partial_t$, then the vector field $T^\mu$ will be undefined in certain regions of spacetime. As a consequence, such an ill-defined vector field cannot be used in the spin Hall equations. On the other hand, if the integration constant $n_2$ is small enough such that $r^4 - e^{n_2} r (r - r_s) > 0$ for all $r>r_s$, then the vector field $T^\mu$ is well defined and there is no problem to use it in the spin Hall equations. 

As an alternative, we can compare the spin Hall trajectories corresponding to the observers $\tilde{t}^\mu$ and $T^\mu$. These two different choices of observers will be used in the spin Hall equations \cref{eq:GSHE_weyl}. In this case, we can fix $n_2$ small enough so that $T^\mu$ is well defined for all $r>r_s$, and then we can also pick $v(r)$ so that $\tilde{t}^\mu (x_{source}) \propto T^\mu (x_{source})$. To illustrate a concrete example, we choose $e^{n_2} = 2$. In this case, $T^\mu$ is well defined for all $r>r_s$. If we place a source of light at $r_{source} = 10^2 r_s$, then we can have $\tilde{t}^\mu (x_{source}) \propto T^\mu (x_{source})$ if we choose the boost velocity
\begin{equation} \label{eq:v_boost}
    v = \sqrt{1 - e^{n_2} \frac{r_{source} - r_s}{r^3_{source}}}.
\end{equation}
An example of the spin Hall trajectories corresponding to these two families of observers is presented in \cref{fig:obs2}. As can be seen in this figure, there is very little difference between the two sets of trajectories.

\bibliography{references}

\begin{thebibliography}{109}%
\makeatletter
\providecommand \@ifxundefined [1]{%
 \@ifx{#1\undefined}
}%
\providecommand \@ifnum [1]{%
 \ifnum #1\expandafter \@firstoftwo
 \else \expandafter \@secondoftwo
 \fi
}%
\providecommand \@ifx [1]{%
 \ifx #1\expandafter \@firstoftwo
 \else \expandafter \@secondoftwo
 \fi
}%
\providecommand \natexlab [1]{#1}%
\providecommand \enquote  [1]{``#1''}%
\providecommand \bibnamefont  [1]{#1}%
\providecommand \bibfnamefont [1]{#1}%
\providecommand \citenamefont [1]{#1}%
\providecommand \href@noop [0]{\@secondoftwo}%
\providecommand \href [0]{\begingroup \@sanitize@url \@href}%
\providecommand \@href[1]{\@@startlink{#1}\@@href}%
\providecommand \@@href[1]{\endgroup#1\@@endlink}%
\providecommand \@sanitize@url [0]{\catcode `\\12\catcode `\$12\catcode
  `\&12\catcode `\#12\catcode `\^12\catcode `\_12\catcode `\%12\relax}%
\providecommand \@@startlink[1]{}%
\providecommand \@@endlink[0]{}%
\providecommand \url  [0]{\begingroup\@sanitize@url \@url }%
\providecommand \@url [1]{\endgroup\@href {#1}{\urlprefix }}%
\providecommand \urlprefix  [0]{URL }%
\providecommand \Eprint [0]{\href }%
\providecommand \doibase [0]{https://doi.org/}%
\providecommand \selectlanguage [0]{\@gobble}%
\providecommand \bibinfo  [0]{\@secondoftwo}%
\providecommand \bibfield  [0]{\@secondoftwo}%
\providecommand \translation [1]{[#1]}%
\providecommand \BibitemOpen [0]{}%
\providecommand \bibitemStop [0]{}%
\providecommand \bibitemNoStop [0]{.\EOS\space}%
\providecommand \EOS [0]{\spacefactor3000\relax}%
\providecommand \BibitemShut  [1]{\csname bibitem#1\endcsname}%
\let\auto@bib@innerbib\@empty
\bibitem [{\citenamefont {Wald}(1984)}]{Wald}%
  \BibitemOpen
  \bibfield  {author} {\bibinfo {author} {\bibfnamefont {R.~M.}\ \bibnamefont
  {Wald}},\ }\href@noop {} {\emph {\bibinfo {title} {General relativity}}}\
  (\bibinfo  {publisher} {University of Chicago Press},\ \bibinfo {address}
  {Chicago, USA},\ \bibinfo {year} {1984})\BibitemShut {NoStop}%
\bibitem [{\citenamefont {C{\^o}t{\'e}}\ \emph {et~al.}(2019)\citenamefont
  {C{\^o}t{\'e}}, \citenamefont {Faraoni},\ and\ \citenamefont {Giusti}}]{Far}%
  \BibitemOpen
  \bibfield  {author} {\bibinfo {author} {\bibfnamefont {J.}~\bibnamefont
  {C{\^o}t{\'e}}}, \bibinfo {author} {\bibfnamefont {V.}~\bibnamefont
  {Faraoni}},\ and\ \bibinfo {author} {\bibfnamefont {A.}~\bibnamefont
  {Giusti}},\ }\bibfield  {title} {\bibinfo {title} {{Revisiting the conformal
  invariance of Maxwell's equations in curved spacetime}},\ }\href
  {https://doi.org/10.1007/s10714-019-2599-x} {\bibfield  {journal} {\bibinfo
  {journal} {General Relativity and Gravitation}\ }\textbf {\bibinfo {volume}
  {51}},\ \bibinfo {pages} {117} (\bibinfo {year} {2019})}\BibitemShut
  {NoStop}%
\bibitem [{\citenamefont {Weyl}(1918)}]{Weyl}%
  \BibitemOpen
  \bibfield  {author} {\bibinfo {author} {\bibfnamefont {H.}~\bibnamefont
  {Weyl}},\ }\bibfield  {title} {\bibinfo {title} {{Gravitation und
  Elektrizit\"at}},\ }\href
  {https://doi.org/https://doi.org/10.1007/978-3-663-19510-8_11} {\bibfield
  {journal} {\bibinfo  {journal} {Sitzungsberichte der Königlich Preussischen
  Akademie der Wissenschaften zu Berlin}\ ,\ \bibinfo {pages} {465}} (\bibinfo
  {year} {1918})}\BibitemShut {NoStop}%
\bibitem [{\citenamefont {Scholz}(2018)}]{Scholz}%
  \BibitemOpen
  \bibfield  {author} {\bibinfo {author} {\bibfnamefont {E.}~\bibnamefont
  {Scholz}},\ }\bibinfo {title} {{The Unexpected Resurgence of Weyl Geometry in
  late 20th-Century Physics}},\ in\ \href
  {https://doi.org/10.1007/978-1-4939-7708-6_11} {\emph {\bibinfo {booktitle}
  {{Beyond Einstein: Perspectives on Geometry, Gravitation, and Cosmology in
  the Twentieth Century}}}},\ \bibinfo {editor} {edited by\ \bibinfo {editor}
  {\bibfnamefont {D.~E.}\ \bibnamefont {Rowe}}, \bibinfo {editor}
  {\bibfnamefont {T.}~\bibnamefont {Sauer}},\ and\ \bibinfo {editor}
  {\bibfnamefont {S.~A.}\ \bibnamefont {Walter}}}\ (\bibinfo  {publisher}
  {Springer New York},\ \bibinfo {address} {New York, NY},\ \bibinfo {year}
  {2018})\ pp.\ \bibinfo {pages} {261--360}\BibitemShut {NoStop}%
\bibitem [{\citenamefont {Penrose}(2010)}]{C1}%
  \BibitemOpen
  \bibfield  {author} {\bibinfo {author} {\bibfnamefont {R.}~\bibnamefont
  {Penrose}},\ }\href@noop {} {\emph {\bibinfo {title} {{Cycles of time: an
  extraordinary new view of the universe}}}}\ (\bibinfo  {publisher} {Random
  House},\ \bibinfo {address} {London, UK},\ \bibinfo {year}
  {2010})\BibitemShut {NoStop}%
\bibitem [{\citenamefont {Gurzadyan}\ and\ \citenamefont {Penrose}(2013)}]{C2}%
  \BibitemOpen
  \bibfield  {author} {\bibinfo {author} {\bibfnamefont {V.~G.}\ \bibnamefont
  {Gurzadyan}}\ and\ \bibinfo {author} {\bibfnamefont {R.}~\bibnamefont
  {Penrose}},\ }\bibfield  {title} {\bibinfo {title} {{On CCC-predicted
  concentric low-variance circles in the CMB sky}},\ }\href
  {https://doi.org/10.1140/epjp/i2013-13022-4} {\bibfield  {journal} {\bibinfo
  {journal} {The European Physical Journal Plus}\ }\textbf {\bibinfo {volume}
  {128}},\ \bibinfo {pages} {22} (\bibinfo {year} {2013})}\BibitemShut
  {NoStop}%
\bibitem [{\citenamefont {Bars}\ \emph {et~al.}(2013)\citenamefont {Bars},
  \citenamefont {Steinhardt},\ and\ \citenamefont {Turok}}]{C3}%
  \BibitemOpen
  \bibfield  {author} {\bibinfo {author} {\bibfnamefont {I.}~\bibnamefont
  {Bars}}, \bibinfo {author} {\bibfnamefont {P.~J.}\ \bibnamefont
  {Steinhardt}},\ and\ \bibinfo {author} {\bibfnamefont {N.}~\bibnamefont
  {Turok}},\ }\bibfield  {title} {\bibinfo {title} {{Cyclic cosmology,
  conformal symmetry and the metastability of the Higgs}},\ }\href
  {https://doi.org/https://doi.org/10.1016/j.physletb.2013.08.071} {\bibfield
  {journal} {\bibinfo  {journal} {Physics Letters B}\ }\textbf {\bibinfo
  {volume} {726}},\ \bibinfo {pages} {50} (\bibinfo {year} {2013})}\BibitemShut
  {NoStop}%
\bibitem [{\citenamefont {Penrose}(2014)}]{C4}%
  \BibitemOpen
  \bibfield  {author} {\bibinfo {author} {\bibfnamefont {R.}~\bibnamefont
  {Penrose}},\ }\bibfield  {title} {\bibinfo {title} {{On the Gravitization of
  Quantum Mechanics 2: Conformal Cyclic Cosmology}},\ }\href
  {https://doi.org/10.1007/s10701-013-9763-z} {\bibfield  {journal} {\bibinfo
  {journal} {Foundations of Physics}\ }\textbf {\bibinfo {volume} {44}},\
  \bibinfo {pages} {873} (\bibinfo {year} {2014})}\BibitemShut {NoStop}%
\bibitem [{\citenamefont {Tod}(2015)}]{C5}%
  \BibitemOpen
  \bibfield  {author} {\bibinfo {author} {\bibfnamefont {P.}~\bibnamefont
  {Tod}},\ }\bibfield  {title} {\bibinfo {title} {{The equations of Conformal
  Cyclic Cosmology}},\ }\href {https://doi.org/10.1007/s10714-015-1859-7}
  {\bibfield  {journal} {\bibinfo  {journal} {General Relativity and
  Gravitation}\ }\textbf {\bibinfo {volume} {47}},\ \bibinfo {pages} {17}
  (\bibinfo {year} {2015})}\BibitemShut {NoStop}%
\bibitem [{\citenamefont {’t Hooft}(2015)}]{C6}%
  \BibitemOpen
  \bibfield  {author} {\bibinfo {author} {\bibfnamefont {G.}~\bibnamefont {’t
  Hooft}},\ }\bibfield  {title} {\bibinfo {title} {{Local conformal symmetry:
  The missing symmetry component for space and time}},\ }\href
  {https://doi.org/10.1142/S0218271815430014} {\bibfield  {journal} {\bibinfo
  {journal} {International Journal of Modern Physics D}\ }\textbf {\bibinfo
  {volume} {24}},\ \bibinfo {pages} {1543001} (\bibinfo {year}
  {2015})}\BibitemShut {NoStop}%
\bibitem [{\citenamefont {'t~Hooft}(2018)}]{C7}%
  \BibitemOpen
  \bibfield  {author} {\bibinfo {author} {\bibfnamefont {G.}~\bibnamefont
  {'t~Hooft}},\ }\bibfield  {title} {\bibinfo {title} {Singularities, horizons,
  firewalls, and local conformal symmetry},\ }in\ \href
  {https://doi.org/10.1007/978-3-319-94256-8_1} {\emph {\bibinfo {booktitle}
  {2nd Karl Schwarzschild Meeting on Gravitational Physics}}},\ \bibinfo
  {editor} {edited by\ \bibinfo {editor} {\bibfnamefont {P.}~\bibnamefont
  {Nicolini}}, \bibinfo {editor} {\bibfnamefont {M.}~\bibnamefont {Kaminski}},
  \bibinfo {editor} {\bibfnamefont {J.}~\bibnamefont {Mureika}},\ and\ \bibinfo
  {editor} {\bibfnamefont {M.}~\bibnamefont {Bleicher}}}\ (\bibinfo
  {publisher} {Springer International Publishing},\ \bibinfo {address} {Cham},\
  \bibinfo {year} {2018})\ pp.\ \bibinfo {pages} {1--12}\BibitemShut {NoStop}%
\bibitem [{\citenamefont {Karananas}\ and\ \citenamefont {Monin}(2016)}]{Kaz}%
  \BibitemOpen
  \bibfield  {author} {\bibinfo {author} {\bibfnamefont {G.~K.}\ \bibnamefont
  {Karananas}}\ and\ \bibinfo {author} {\bibfnamefont {A.}~\bibnamefont
  {Monin}},\ }\bibfield  {title} {\bibinfo {title} {Weyl vs. conformal},\
  }\href {https://doi.org/https://doi.org/10.1016/j.physletb.2016.04.001}
  {\bibfield  {journal} {\bibinfo  {journal} {Physics Letters B}\ }\textbf
  {\bibinfo {volume} {757}},\ \bibinfo {pages} {257} (\bibinfo {year}
  {2016})}\BibitemShut {NoStop}%
\bibitem [{\citenamefont {Dirac}(1973)}]{D1}%
  \BibitemOpen
  \bibfield  {author} {\bibinfo {author} {\bibfnamefont {P.~A.~M.}\
  \bibnamefont {Dirac}},\ }\bibfield  {title} {\bibinfo {title} {Long range
  forces and broken symmetries},\ }\href
  {https://doi.org/10.1098/rspa.1973.0070} {\bibfield  {journal} {\bibinfo
  {journal} {Proceedings of the Royal Society of London. A. Mathematical and
  Physical Sciences}\ }\textbf {\bibinfo {volume} {333}},\ \bibinfo {pages}
  {403} (\bibinfo {year} {1973})}\BibitemShut {NoStop}%
\bibitem [{\citenamefont {Dirac}(1974)}]{D2}%
  \BibitemOpen
  \bibfield  {author} {\bibinfo {author} {\bibfnamefont {P.~A.~M.}\
  \bibnamefont {Dirac}},\ }\bibfield  {title} {\bibinfo {title} {Cosmological
  models and the large numbers hypothesis},\ }\href
  {https://doi.org/10.1098/rspa.1974.0095} {\bibfield  {journal} {\bibinfo
  {journal} {Proceedings of the Royal Society of London. A. Mathematical and
  Physical Sciences}\ }\textbf {\bibinfo {volume} {338}},\ \bibinfo {pages}
  {439} (\bibinfo {year} {1974})}\BibitemShut {NoStop}%
\bibitem [{\citenamefont {Rosen}(1982)}]{R1}%
  \BibitemOpen
  \bibfield  {author} {\bibinfo {author} {\bibfnamefont {N.}~\bibnamefont
  {Rosen}},\ }\bibfield  {title} {\bibinfo {title} {Weyl's geometry and
  physics},\ }\href {https://doi.org/10.1007/BF00726849} {\bibfield  {journal}
  {\bibinfo  {journal} {Foundations of Physics}\ }\textbf {\bibinfo {volume}
  {12}},\ \bibinfo {pages} {213} (\bibinfo {year} {1982})}\BibitemShut
  {NoStop}%
\bibitem [{\citenamefont {Israelit}(2011)}]{R2}%
  \BibitemOpen
  \bibfield  {author} {\bibinfo {author} {\bibfnamefont {M.}~\bibnamefont
  {Israelit}},\ }\bibfield  {title} {\bibinfo {title} {{A Weyl-Dirac
  cosmological model with DM and DE}},\ }\href
  {https://doi.org/10.1007/s10714-010-1092-3} {\bibfield  {journal} {\bibinfo
  {journal} {General Relativity and Gravitation}\ }\textbf {\bibinfo {volume}
  {43}},\ \bibinfo {pages} {751} (\bibinfo {year} {2011})}\BibitemShut
  {NoStop}%
\bibitem [{\citenamefont {Utiyama}(1973)}]{U1}%
  \BibitemOpen
  \bibfield  {author} {\bibinfo {author} {\bibfnamefont {R.}~\bibnamefont
  {Utiyama}},\ }\bibfield  {title} {\bibinfo {title} {{On Weyl’s Gauge
  Field}},\ }\href {https://doi.org/10.1143/PTP.50.2080} {\bibfield  {journal}
  {\bibinfo  {journal} {Progress of Theoretical Physics}\ }\textbf {\bibinfo
  {volume} {50}},\ \bibinfo {pages} {2080} (\bibinfo {year}
  {1973})}\BibitemShut {NoStop}%
\bibitem [{\citenamefont {Utiyama}(1975{\natexlab{a}})}]{U2}%
  \BibitemOpen
  \bibfield  {author} {\bibinfo {author} {\bibfnamefont {R.}~\bibnamefont
  {Utiyama}},\ }\bibfield  {title} {\bibinfo {title} {{On Weyl's gauge
  field}},\ }\href {https://doi.org/10.1007/BF00766599} {\bibfield  {journal}
  {\bibinfo  {journal} {General Relativity and Gravitation}\ }\textbf {\bibinfo
  {volume} {6}},\ \bibinfo {pages} {41} (\bibinfo {year}
  {1975}{\natexlab{a}})}\BibitemShut {NoStop}%
\bibitem [{\citenamefont {Utiyama}(1975{\natexlab{b}})}]{U3}%
  \BibitemOpen
  \bibfield  {author} {\bibinfo {author} {\bibfnamefont {R.}~\bibnamefont
  {Utiyama}},\ }\bibfield  {title} {\bibinfo {title} {{On Weyl's Gauge Field.
  II}},\ }\href {https://doi.org/10.1143/PTP.53.565} {\bibfield  {journal}
  {\bibinfo  {journal} {Progress of Theoretical Physics}\ }\textbf {\bibinfo
  {volume} {53}},\ \bibinfo {pages} {565} (\bibinfo {year}
  {1975}{\natexlab{b}})}\BibitemShut {NoStop}%
\bibitem [{\citenamefont {{Mannheim}}\ and\ \citenamefont
  {{Kazanas}}(1989)}]{M1}%
  \BibitemOpen
  \bibfield  {author} {\bibinfo {author} {\bibfnamefont {P.~D.}\ \bibnamefont
  {{Mannheim}}}\ and\ \bibinfo {author} {\bibfnamefont {D.}~\bibnamefont
  {{Kazanas}}},\ }\bibfield  {title} {\bibinfo {title} {{Exact Vacuum Solution
  to Conformal Weyl Gravity and Galactic Rotation Curves}},\ }\href
  {https://doi.org/10.1086/167623} {\bibfield  {journal} {\bibinfo  {journal}
  {The Astrophysical Journal}\ }\textbf {\bibinfo {volume} {342}},\ \bibinfo
  {pages} {635} (\bibinfo {year} {1989})}\BibitemShut {NoStop}%
\bibitem [{\citenamefont {Mannheim}(1994)}]{M2}%
  \BibitemOpen
  \bibfield  {author} {\bibinfo {author} {\bibfnamefont {P.~D.}\ \bibnamefont
  {Mannheim}},\ }\bibfield  {title} {\bibinfo {title} {Open questions in
  classical gravity},\ }\href {https://doi.org/10.1007/BF02058060} {\bibfield
  {journal} {\bibinfo  {journal} {Foundations of Physics}\ }\textbf {\bibinfo
  {volume} {24}},\ \bibinfo {pages} {487} (\bibinfo {year} {1994})}\BibitemShut
  {NoStop}%
\bibitem [{\citenamefont {Mannheim}(1996)}]{M3}%
  \BibitemOpen
  \bibfield  {author} {\bibinfo {author} {\bibfnamefont {P.~D.}\ \bibnamefont
  {Mannheim}},\ }\bibfield  {title} {\bibinfo {title} {Local and global
  gravity},\ }\href {https://doi.org/10.1007/BF02282129} {\bibfield  {journal}
  {\bibinfo  {journal} {Foundations of Physics}\ }\textbf {\bibinfo {volume}
  {26}},\ \bibinfo {pages} {1683} (\bibinfo {year} {1996})}\BibitemShut
  {NoStop}%
\bibitem [{\citenamefont {Mannheim}(2000)}]{M4}%
  \BibitemOpen
  \bibfield  {author} {\bibinfo {author} {\bibfnamefont {P.~D.}\ \bibnamefont
  {Mannheim}},\ }\bibfield  {title} {\bibinfo {title} {Attractive and repulsive
  gravity},\ }\href {https://doi.org/10.1023/A:1003737011054} {\bibfield
  {journal} {\bibinfo  {journal} {Foundations of Physics}\ }\textbf {\bibinfo
  {volume} {30}},\ \bibinfo {pages} {709} (\bibinfo {year} {2000})}\BibitemShut
  {NoStop}%
\bibitem [{\citenamefont {Mannheim}(2007)}]{M5}%
  \BibitemOpen
  \bibfield  {author} {\bibinfo {author} {\bibfnamefont {P.~D.}\ \bibnamefont
  {Mannheim}},\ }\bibfield  {title} {\bibinfo {title} {Solution to the ghost
  problem in fourth order derivative theories},\ }\href
  {https://doi.org/10.1007/s10701-007-9119-7} {\bibfield  {journal} {\bibinfo
  {journal} {Foundations of Physics}\ }\textbf {\bibinfo {volume} {37}},\
  \bibinfo {pages} {532} (\bibinfo {year} {2007})}\BibitemShut {NoStop}%
\bibitem [{\citenamefont {Mannheim}(2012)}]{M6}%
  \BibitemOpen
  \bibfield  {author} {\bibinfo {author} {\bibfnamefont {P.~D.}\ \bibnamefont
  {Mannheim}},\ }\bibfield  {title} {\bibinfo {title} {Making the case for
  conformal gravity},\ }\href {https://doi.org/10.1007/s10701-011-9608-6}
  {\bibfield  {journal} {\bibinfo  {journal} {Foundations of Physics}\ }\textbf
  {\bibinfo {volume} {42}},\ \bibinfo {pages} {388} (\bibinfo {year}
  {2012})}\BibitemShut {NoStop}%
\bibitem [{\citenamefont {Ghilencea}(2019{\natexlab{a}})}]{Gh1}%
  \BibitemOpen
  \bibfield  {author} {\bibinfo {author} {\bibfnamefont {D.~M.}\ \bibnamefont
  {Ghilencea}},\ }\bibfield  {title} {\bibinfo {title} {{Spontaneous breaking
  of Weyl quadratic gravity to Einstein action and Higgs potential}},\ }\href
  {https://doi.org/10.1007/JHEP03(2019)049} {\bibfield  {journal} {\bibinfo
  {journal} {Journal of High Energy Physics}\ }\textbf {\bibinfo {volume}
  {2019}},\ \bibinfo {pages} {49} (\bibinfo {year}
  {2019}{\natexlab{a}})}\BibitemShut {NoStop}%
\bibitem [{\citenamefont {Ghilencea}\ and\ \citenamefont {Lee}(2019)}]{Gh2}%
  \BibitemOpen
  \bibfield  {author} {\bibinfo {author} {\bibfnamefont {D.~M.}\ \bibnamefont
  {Ghilencea}}\ and\ \bibinfo {author} {\bibfnamefont {H.~M.}\ \bibnamefont
  {Lee}},\ }\bibfield  {title} {\bibinfo {title} {Weyl gauge symmetry and its
  spontaneous breaking in the standard model and inflation},\ }\href
  {https://doi.org/10.1103/PhysRevD.99.115007} {\bibfield  {journal} {\bibinfo
  {journal} {Physical Review D}\ }\textbf {\bibinfo {volume} {99}},\ \bibinfo
  {pages} {115007} (\bibinfo {year} {2019})}\BibitemShut {NoStop}%
\bibitem [{\citenamefont {Ghilencea}(2019{\natexlab{b}})}]{Gh3}%
  \BibitemOpen
  \bibfield  {author} {\bibinfo {author} {\bibfnamefont {D.~M.}\ \bibnamefont
  {Ghilencea}},\ }\bibfield  {title} {\bibinfo {title} {{Weyl $R^2$ inflation
  with an emergent Planck scale}},\ }\href
  {https://doi.org/10.1007/JHEP10(2019)209} {\bibfield  {journal} {\bibinfo
  {journal} {Journal of High Energy Physics}\ }\textbf {\bibinfo {volume}
  {2019}},\ \bibinfo {pages} {209} (\bibinfo {year}
  {2019}{\natexlab{b}})}\BibitemShut {NoStop}%
\bibitem [{\citenamefont {Ghilencea}(2020{\natexlab{a}})}]{Gh4}%
  \BibitemOpen
  \bibfield  {author} {\bibinfo {author} {\bibfnamefont {D.~M.}\ \bibnamefont
  {Ghilencea}},\ }\bibfield  {title} {\bibinfo {title} {{Stueckelberg breaking
  of Weyl conformal geometry and applications to gravity}},\ }\href
  {https://doi.org/10.1103/PhysRevD.101.045010} {\bibfield  {journal} {\bibinfo
   {journal} {Physical Review D}\ }\textbf {\bibinfo {volume} {101}},\ \bibinfo
  {pages} {045010} (\bibinfo {year} {2020}{\natexlab{a}})}\BibitemShut
  {NoStop}%
\bibitem [{\citenamefont {Ghilencea}(2020{\natexlab{b}})}]{Gh5}%
  \BibitemOpen
  \bibfield  {author} {\bibinfo {author} {\bibfnamefont {D.~M.}\ \bibnamefont
  {Ghilencea}},\ }\bibfield  {title} {\bibinfo {title} {Palatini quadratic
  gravity: spontaneous breaking of gauged scale symmetry and inflation},\
  }\href {https://doi.org/10.1140/epjc/s10052-020-08722-0} {\bibfield
  {journal} {\bibinfo  {journal} {The European Physical Journal C}\ }\textbf
  {\bibinfo {volume} {80}},\ \bibinfo {pages} {1147} (\bibinfo {year}
  {2020}{\natexlab{b}})}\BibitemShut {NoStop}%
\bibitem [{\citenamefont {Ghilencea}(2021)}]{Gh6}%
  \BibitemOpen
  \bibfield  {author} {\bibinfo {author} {\bibfnamefont {D.~M.}\ \bibnamefont
  {Ghilencea}},\ }\bibfield  {title} {\bibinfo {title} {{Gauging scale symmetry
  and inflation: Weyl versus Palatini gravity}},\ }\href
  {https://doi.org/10.1140/epjc/s10052-021-09226-1} {\bibfield  {journal}
  {\bibinfo  {journal} {The European Physical Journal C}\ }\textbf {\bibinfo
  {volume} {81}},\ \bibinfo {pages} {510} (\bibinfo {year} {2021})}\BibitemShut
  {NoStop}%
\bibitem [{\citenamefont {Ghilencea}(2022)}]{Gh7}%
  \BibitemOpen
  \bibfield  {author} {\bibinfo {author} {\bibfnamefont {D.~M.}\ \bibnamefont
  {Ghilencea}},\ }\bibfield  {title} {\bibinfo {title} {{Standard Model in Weyl
  conformal geometry}},\ }\href
  {https://doi.org/10.1140/epjc/s10052-021-09887-y} {\bibfield  {journal}
  {\bibinfo  {journal} {The European Physical Journal C}\ }\textbf {\bibinfo
  {volume} {82}},\ \bibinfo {pages} {23} (\bibinfo {year} {2022})}\BibitemShut
  {NoStop}%
\bibitem [{\citenamefont {Ghilencea}(2023)}]{Gh8}%
  \BibitemOpen
  \bibfield  {author} {\bibinfo {author} {\bibfnamefont {D.~M.}\ \bibnamefont
  {Ghilencea}},\ }\bibfield  {title} {\bibinfo {title} {Non-metric geometry as
  the origin of mass in gauge theories of scale invariance},\ }\href
  {https://doi.org/10.1140/epjc/s10052-023-11237-z} {\bibfield  {journal}
  {\bibinfo  {journal} {The European Physical Journal C}\ }\textbf {\bibinfo
  {volume} {83}},\ \bibinfo {pages} {176} (\bibinfo {year} {2023})}\BibitemShut
  {NoStop}%
\bibitem [{\citenamefont {Yang}\ \emph {et~al.}(2022)\citenamefont {Yang},
  \citenamefont {Shahidi},\ and\ \citenamefont {Harko}}]{Yang}%
  \BibitemOpen
  \bibfield  {author} {\bibinfo {author} {\bibfnamefont {J.-Z.}\ \bibnamefont
  {Yang}}, \bibinfo {author} {\bibfnamefont {S.}~\bibnamefont {Shahidi}},\ and\
  \bibinfo {author} {\bibfnamefont {T.}~\bibnamefont {Harko}},\ }\bibfield
  {title} {\bibinfo {title} {{Black hole solutions in the quadratic Weyl
  conformal geometric theory of gravity}},\ }\href
  {https://doi.org/10.1140/epjc/s10052-022-11131-0} {\bibfield  {journal}
  {\bibinfo  {journal} {The European Physical Journal C}\ }\textbf {\bibinfo
  {volume} {82}},\ \bibinfo {pages} {1171} (\bibinfo {year}
  {2022})}\BibitemShut {NoStop}%
\bibitem [{\citenamefont {Burikham}\ \emph {et~al.}(2023)\citenamefont
  {Burikham}, \citenamefont {Harko}, \citenamefont {Pimsamarn},\ and\
  \citenamefont {Shahidi}}]{Pi1}%
  \BibitemOpen
  \bibfield  {author} {\bibinfo {author} {\bibfnamefont {P.}~\bibnamefont
  {Burikham}}, \bibinfo {author} {\bibfnamefont {T.}~\bibnamefont {Harko}},
  \bibinfo {author} {\bibfnamefont {K.}~\bibnamefont {Pimsamarn}},\ and\
  \bibinfo {author} {\bibfnamefont {S.}~\bibnamefont {Shahidi}},\ }\bibfield
  {title} {\bibinfo {title} {{Dark matter as a Weyl geometric effect}},\ }\href
  {https://doi.org/10.1103/PhysRevD.107.064008} {\bibfield  {journal} {\bibinfo
   {journal} {Physical Review D}\ }\textbf {\bibinfo {volume} {107}},\ \bibinfo
  {pages} {064008} (\bibinfo {year} {2023})}\BibitemShut {NoStop}%
\bibitem [{\citenamefont {Sinova}\ \emph {et~al.}(2015)\citenamefont {Sinova},
  \citenamefont {Valenzuela}, \citenamefont {Wunderlich}, \citenamefont
  {Back},\ and\ \citenamefont {Jungwirth}}]{SHE_review}%
  \BibitemOpen
  \bibfield  {author} {\bibinfo {author} {\bibfnamefont {J.}~\bibnamefont
  {Sinova}}, \bibinfo {author} {\bibfnamefont {S.~O.}\ \bibnamefont
  {Valenzuela}}, \bibinfo {author} {\bibfnamefont {J.}~\bibnamefont
  {Wunderlich}}, \bibinfo {author} {\bibfnamefont {C.~H.}\ \bibnamefont
  {Back}},\ and\ \bibinfo {author} {\bibfnamefont {T.}~\bibnamefont
  {Jungwirth}},\ }\bibfield  {title} {\bibinfo {title} {{Spin Hall effects}},\
  }\href {https://doi.org/10.1103/RevModPhys.87.1213} {\bibfield  {journal}
  {\bibinfo  {journal} {Reviews of Modern Physics}\ }\textbf {\bibinfo {volume}
  {87}},\ \bibinfo {pages} {1213} (\bibinfo {year} {2015})}\BibitemShut
  {NoStop}%
\bibitem [{\citenamefont {Dyakonov}\ and\ \citenamefont
  {Khaetskii}(2008)}]{SHE_review1}%
  \BibitemOpen
  \bibfield  {author} {\bibinfo {author} {\bibfnamefont {M.~I.}\ \bibnamefont
  {Dyakonov}}\ and\ \bibinfo {author} {\bibfnamefont {A.~V.}\ \bibnamefont
  {Khaetskii}},\ }\bibinfo {title} {{Spin Hall Effect}},\ in\ \href
  {https://doi.org/10.1007/978-3-540-78820-1_8} {\emph {\bibinfo {booktitle}
  {Spin Physics in Semiconductors}}},\ \bibinfo {editor} {edited by\ \bibinfo
  {editor} {\bibfnamefont {M.~I.}\ \bibnamefont {Dyakonov}}}\ (\bibinfo
  {publisher} {Springer Berlin Heidelberg},\ \bibinfo {year} {2008})\ pp.\
  \bibinfo {pages} {211--243}\BibitemShut {NoStop}%
\bibitem [{\citenamefont {Bliokh}\ \emph
  {et~al.}(2015{\natexlab{a}})\citenamefont {Bliokh}, \citenamefont
  {Rodr\'{i}guez-Fortu\~{n}o}, \citenamefont {Nori},\ and\ \citenamefont
  {Zayats}}]{opt1}%
  \BibitemOpen
  \bibfield  {author} {\bibinfo {author} {\bibfnamefont {K.~Y.}\ \bibnamefont
  {Bliokh}}, \bibinfo {author} {\bibfnamefont {F.~J.}\ \bibnamefont
  {Rodr\'{i}guez-Fortu\~{n}o}}, \bibinfo {author} {\bibfnamefont
  {F.}~\bibnamefont {Nori}},\ and\ \bibinfo {author} {\bibfnamefont {A.~V.}\
  \bibnamefont {Zayats}},\ }\bibfield  {title} {\bibinfo {title} {{Spin-orbit
  interactions of light}},\ }\href {https://doi.org/10.1038/nphoton.2015.201}
  {\bibfield  {journal} {\bibinfo  {journal} {Nature Photonics}\ }\textbf
  {\bibinfo {volume} {9}},\ \bibinfo {pages} {796} (\bibinfo {year}
  {2015}{\natexlab{a}})}\BibitemShut {NoStop}%
\bibitem [{\citenamefont {Kim}\ \emph {et~al.}(2023)\citenamefont {Kim},
  \citenamefont {Yang}, \citenamefont {Lee}, \citenamefont {Kim}, \citenamefont
  {Kim},\ and\ \citenamefont {Rho}}]{rev}%
  \BibitemOpen
  \bibfield  {author} {\bibinfo {author} {\bibfnamefont {M.}~\bibnamefont
  {Kim}}, \bibinfo {author} {\bibfnamefont {Y.}~\bibnamefont {Yang}}, \bibinfo
  {author} {\bibfnamefont {D.}~\bibnamefont {Lee}}, \bibinfo {author}
  {\bibfnamefont {Y.}~\bibnamefont {Kim}}, \bibinfo {author} {\bibfnamefont
  {H.}~\bibnamefont {Kim}},\ and\ \bibinfo {author} {\bibfnamefont
  {J.}~\bibnamefont {Rho}},\ }\bibfield  {title} {\bibinfo {title} {{Spin Hall
  Effect of Light: From Fundamentals To Recent Advancements}},\ }\href
  {https://doi.org/https://doi.org/10.1002/lpor.202200046} {\bibfield
  {journal} {\bibinfo  {journal} {Laser \& Photonics Reviews}\ }\textbf
  {\bibinfo {volume} {17}},\ \bibinfo {pages} {2200046} (\bibinfo {year}
  {2023})}\BibitemShut {NoStop}%
\bibitem [{\citenamefont {Dyakonov}\ and\ \citenamefont
  {Perel}(1971{\natexlab{a}})}]{originalSHE1}%
  \BibitemOpen
  \bibfield  {author} {\bibinfo {author} {\bibfnamefont {M.~I.}\ \bibnamefont
  {Dyakonov}}\ and\ \bibinfo {author} {\bibfnamefont {V.~I.}\ \bibnamefont
  {Perel}},\ }\bibfield  {title} {\bibinfo {title} {{Possibility of orienting
  electron spins with current}},\ }\href
  {http://jetpletters.ru/ps/1587/article_24366.shtml} {\bibfield  {journal}
  {\bibinfo  {journal} {Soviet Journal of Experimental and Theoretical Physics
  Letters}\ }\textbf {\bibinfo {volume} {13}},\ \bibinfo {pages} {467}
  (\bibinfo {year} {1971}{\natexlab{a}})}\BibitemShut {NoStop}%
\bibitem [{\citenamefont {Dyakonov}\ and\ \citenamefont
  {Perel}(1971{\natexlab{b}})}]{originalSHE2}%
  \BibitemOpen
  \bibfield  {author} {\bibinfo {author} {\bibfnamefont {M.~I.}\ \bibnamefont
  {Dyakonov}}\ and\ \bibinfo {author} {\bibfnamefont {V.~I.}\ \bibnamefont
  {Perel}},\ }\bibfield  {title} {\bibinfo {title} {{Current-induced spin
  orientation of electrons in semiconductors}},\ }\href
  {https://doi.org/10.1016/0375-9601(71)90196-4} {\bibfield  {journal}
  {\bibinfo  {journal} {Physics Letters A}\ }\textbf {\bibinfo {volume} {35}},\
  \bibinfo {pages} {459} (\bibinfo {year} {1971}{\natexlab{b}})}\BibitemShut
  {NoStop}%
\bibitem [{\citenamefont {{Bakun}}\ \emph {et~al.}(1984)\citenamefont
  {{Bakun}}, \citenamefont {{Zakharchenya}}, \citenamefont {{Rogachev}},
  \citenamefont {{Tkachuk}},\ and\ \citenamefont {{Fle{\v
  \i}sher}}}]{originalSHE3}%
  \BibitemOpen
  \bibfield  {author} {\bibinfo {author} {\bibfnamefont {A.~A.}\ \bibnamefont
  {{Bakun}}}, \bibinfo {author} {\bibfnamefont {B.~P.}\ \bibnamefont
  {{Zakharchenya}}}, \bibinfo {author} {\bibfnamefont {A.~A.}\ \bibnamefont
  {{Rogachev}}}, \bibinfo {author} {\bibfnamefont {M.~N.}\ \bibnamefont
  {{Tkachuk}}},\ and\ \bibinfo {author} {\bibfnamefont {V.~G.}\ \bibnamefont
  {{Fle{\v \i}sher}}},\ }\bibfield  {title} {\bibinfo {title} {{Observation of
  a surface photocurrent caused by optical orientation of electrons in a
  semiconductor}},\ }\href {http://jetpletters.ru/ps/1262/article_19087.shtml}
  {\bibfield  {journal} {\bibinfo  {journal} {Soviet Journal of Experimental
  and Theoretical Physics Letters}\ }\textbf {\bibinfo {volume} {40}},\
  \bibinfo {pages} {1293} (\bibinfo {year} {1984})}\BibitemShut {NoStop}%
\bibitem [{\citenamefont {Kato}\ \emph {et~al.}(2004)\citenamefont {Kato},
  \citenamefont {Myers}, \citenamefont {Gossard},\ and\ \citenamefont
  {Awschalom}}]{originalSHE4}%
  \BibitemOpen
  \bibfield  {author} {\bibinfo {author} {\bibfnamefont {Y.~K.}\ \bibnamefont
  {Kato}}, \bibinfo {author} {\bibfnamefont {R.~C.}\ \bibnamefont {Myers}},
  \bibinfo {author} {\bibfnamefont {A.~C.}\ \bibnamefont {Gossard}},\ and\
  \bibinfo {author} {\bibfnamefont {D.~D.}\ \bibnamefont {Awschalom}},\
  }\bibfield  {title} {\bibinfo {title} {{Observation of the spin Hall effect
  in semiconductors}},\ }\href {https://doi.org/10.1126/science.1105514}
  {\bibfield  {journal} {\bibinfo  {journal} {Science}\ }\textbf {\bibinfo
  {volume} {306}},\ \bibinfo {pages} {1910} (\bibinfo {year}
  {2004})}\BibitemShut {NoStop}%
\bibitem [{\citenamefont {Bliokh}\ \emph
  {et~al.}(2015{\natexlab{b}})\citenamefont {Bliokh}, \citenamefont
  {Smirnova},\ and\ \citenamefont {Nori}}]{opt2}%
  \BibitemOpen
  \bibfield  {author} {\bibinfo {author} {\bibfnamefont {K.~Y.}\ \bibnamefont
  {Bliokh}}, \bibinfo {author} {\bibfnamefont {D.}~\bibnamefont {Smirnova}},\
  and\ \bibinfo {author} {\bibfnamefont {F.}~\bibnamefont {Nori}},\ }\bibfield
  {title} {\bibinfo {title} {{Quantum spin Hall effect of light}},\ }\href
  {https://doi.org/10.1126/science.aaa9519} {\bibfield  {journal} {\bibinfo
  {journal} {Science}\ }\textbf {\bibinfo {volume} {348}},\ \bibinfo {pages}
  {1448} (\bibinfo {year} {2015}{\natexlab{b}})}\BibitemShut {NoStop}%
\bibitem [{\citenamefont {Hosten}\ and\ \citenamefont {Kwiat}(2008)}]{opt5}%
  \BibitemOpen
  \bibfield  {author} {\bibinfo {author} {\bibfnamefont {O.}~\bibnamefont
  {Hosten}}\ and\ \bibinfo {author} {\bibfnamefont {P.}~\bibnamefont {Kwiat}},\
  }\bibfield  {title} {\bibinfo {title} {{Observation of the spin Hall effect
  of light via weak measurements}},\ }\href
  {https://doi.org/10.1126/science.1152697} {\bibfield  {journal} {\bibinfo
  {journal} {Science}\ }\textbf {\bibinfo {volume} {319}},\ \bibinfo {pages}
  {787} (\bibinfo {year} {2008})}\BibitemShut {NoStop}%
\bibitem [{\citenamefont {Bliokh}\ \emph
  {et~al.}(2008{\natexlab{a}})\citenamefont {Bliokh}, \citenamefont {Niv},
  \citenamefont {Kleiner},\ and\ \citenamefont {Hasman}}]{B1}%
  \BibitemOpen
  \bibfield  {author} {\bibinfo {author} {\bibfnamefont {K.~Y.}\ \bibnamefont
  {Bliokh}}, \bibinfo {author} {\bibfnamefont {A.}~\bibnamefont {Niv}},
  \bibinfo {author} {\bibfnamefont {V.}~\bibnamefont {Kleiner}},\ and\ \bibinfo
  {author} {\bibfnamefont {E.}~\bibnamefont {Hasman}},\ }\bibfield  {title}
  {\bibinfo {title} {{Geometrodynamics of spinning light}},\ }\href
  {https://doi.org/10.1038/nphoton.2008.229} {\bibfield  {journal} {\bibinfo
  {journal} {Nature Photonics}\ }\textbf {\bibinfo {volume} {2}},\ \bibinfo
  {pages} {748} (\bibinfo {year} {2008}{\natexlab{a}})}\BibitemShut {NoStop}%
\bibitem [{\citenamefont {Bliokh}\ and\ \citenamefont {Bliokh}(2006)}]{opt3}%
  \BibitemOpen
  \bibfield  {author} {\bibinfo {author} {\bibfnamefont {K.~Y.}\ \bibnamefont
  {Bliokh}}\ and\ \bibinfo {author} {\bibfnamefont {Y.~P.}\ \bibnamefont
  {Bliokh}},\ }\bibfield  {title} {\bibinfo {title} {{Conservation of Angular
  Momentum, Transverse Shift, and Spin Hall Effect in Reflection and Refraction
  of an Electromagnetic Wave Packet}},\ }\href
  {https://doi.org/10.1103/PhysRevLett.96.073903} {\bibfield  {journal}
  {\bibinfo  {journal} {Physical Review Letters}\ }\textbf {\bibinfo {volume}
  {96}},\ \bibinfo {pages} {073903} (\bibinfo {year} {2006})}\BibitemShut
  {NoStop}%
\bibitem [{\citenamefont {Onoda}\ \emph {et~al.}(2004)\citenamefont {Onoda},
  \citenamefont {Murakami},\ and\ \citenamefont {Nagaosa}}]{opt4}%
  \BibitemOpen
  \bibfield  {author} {\bibinfo {author} {\bibfnamefont {M.}~\bibnamefont
  {Onoda}}, \bibinfo {author} {\bibfnamefont {S.}~\bibnamefont {Murakami}},\
  and\ \bibinfo {author} {\bibfnamefont {N.}~\bibnamefont {Nagaosa}},\
  }\bibfield  {title} {\bibinfo {title} {Hall effect of light},\ }\href
  {https://doi.org/10.1103/PhysRevLett.93.083901} {\bibfield  {journal}
  {\bibinfo  {journal} {Physical Review Letters}\ }\textbf {\bibinfo {volume}
  {93}},\ \bibinfo {pages} {083901} (\bibinfo {year} {2004})}\BibitemShut
  {NoStop}%
\bibitem [{\citenamefont {Plebanski}(1960)}]{Plebansky-Maxwell}%
  \BibitemOpen
  \bibfield  {author} {\bibinfo {author} {\bibfnamefont {J.}~\bibnamefont
  {Plebanski}},\ }\bibfield  {title} {\bibinfo {title} {Electromagnetic waves
  in gravitational fields},\ }\href {https://doi.org/10.1103/PhysRev.118.1396}
  {\bibfield  {journal} {\bibinfo  {journal} {Physical Review}\ }\textbf
  {\bibinfo {volume} {118}},\ \bibinfo {pages} {1396} (\bibinfo {year}
  {1960})}\BibitemShut {NoStop}%
\bibitem [{\citenamefont {Mashhoon}(1973)}]{Mashoon1}%
  \BibitemOpen
  \bibfield  {author} {\bibinfo {author} {\bibfnamefont {B.}~\bibnamefont
  {Mashhoon}},\ }\bibfield  {title} {\bibinfo {title} {Scattering of
  electromagnetic radiation from a black hole},\ }\href
  {https://doi.org/10.1103/PhysRevD.7.2807} {\bibfield  {journal} {\bibinfo
  {journal} {Physical Review D}\ }\textbf {\bibinfo {volume} {7}},\ \bibinfo
  {pages} {2807} (\bibinfo {year} {1973})}\BibitemShut {NoStop}%
\bibitem [{\citenamefont {Mashhoon}(1974{\natexlab{a}})}]{Mashoon2}%
  \BibitemOpen
  \bibfield  {author} {\bibinfo {author} {\bibfnamefont {B.}~\bibnamefont
  {Mashhoon}},\ }\bibfield  {title} {\bibinfo {title} {{Can Einstein's theory
  of gravitation be tested beyond the geometrical optics limit?}},\ }\href
  {https://doi.org/doi.org/10.1038/250316a0} {\bibfield  {journal} {\bibinfo
  {journal} {Nature}\ }\textbf {\bibinfo {volume} {250}},\ \bibinfo {pages}
  {316} (\bibinfo {year} {1974}{\natexlab{a}})}\BibitemShut {NoStop}%
\bibitem [{\citenamefont {Mashhoon}(1974{\natexlab{b}})}]{Mashoon3}%
  \BibitemOpen
  \bibfield  {author} {\bibinfo {author} {\bibfnamefont {B.}~\bibnamefont
  {Mashhoon}},\ }\bibfield  {title} {\bibinfo {title} {Electromagnetic
  scattering from a black hole and the glory effect},\ }\href
  {https://doi.org/10.1103/PhysRevD.10.1059} {\bibfield  {journal} {\bibinfo
  {journal} {Physical Review D}\ }\textbf {\bibinfo {volume} {10}},\ \bibinfo
  {pages} {1059} (\bibinfo {year} {1974}{\natexlab{b}})}\BibitemShut {NoStop}%
\bibitem [{\citenamefont {Mashhoon}(1975)}]{Mashoon4}%
  \BibitemOpen
  \bibfield  {author} {\bibinfo {author} {\bibfnamefont {B.}~\bibnamefont
  {Mashhoon}},\ }\bibfield  {title} {\bibinfo {title} {Influence of gravitation
  on the propagation of electromagnetic radiation},\ }\href
  {https://doi.org/10.1103/PhysRevD.11.2679} {\bibfield  {journal} {\bibinfo
  {journal} {Physical Review D}\ }\textbf {\bibinfo {volume} {11}},\ \bibinfo
  {pages} {2679} (\bibinfo {year} {1975})}\BibitemShut {NoStop}%
\bibitem [{\citenamefont {Kopeikin}\ and\ \citenamefont
  {Mashhoon}(2002)}]{Mashoon5}%
  \BibitemOpen
  \bibfield  {author} {\bibinfo {author} {\bibfnamefont {S.}~\bibnamefont
  {Kopeikin}}\ and\ \bibinfo {author} {\bibfnamefont {B.}~\bibnamefont
  {Mashhoon}},\ }\bibfield  {title} {\bibinfo {title} {Gravitomagnetic effects
  in the propagation of electromagnetic waves in variable gravitational fields
  of arbitrary-moving and spinning bodies},\ }\href
  {https://doi.org/10.1103/PhysRevD.65.064025} {\bibfield  {journal} {\bibinfo
  {journal} {Physical Review D}\ }\textbf {\bibinfo {volume} {65}},\ \bibinfo
  {pages} {064025} (\bibinfo {year} {2002})}\BibitemShut {NoStop}%
\bibitem [{\citenamefont {Leite}\ \emph {et~al.}(2017)\citenamefont {Leite},
  \citenamefont {Dolan},\ and\ \citenamefont {Crispino}}]{leite2017absorption}%
  \BibitemOpen
  \bibfield  {author} {\bibinfo {author} {\bibfnamefont {L.~C.~S.}\
  \bibnamefont {Leite}}, \bibinfo {author} {\bibfnamefont {S.~R.}\ \bibnamefont
  {Dolan}},\ and\ \bibinfo {author} {\bibfnamefont {L.~C.~B.}\ \bibnamefont
  {Crispino}},\ }\bibfield  {title} {\bibinfo {title} {{Absorption of
  electromagnetic and gravitational waves by Kerr black holes}},\ }\href
  {https://doi.org/10.1016/j.physletb.2017.09.048} {\bibfield  {journal}
  {\bibinfo  {journal} {Physics Letters B}\ }\textbf {\bibinfo {volume}
  {774}},\ \bibinfo {pages} {130} (\bibinfo {year} {2017})}\BibitemShut
  {NoStop}%
\bibitem [{\citenamefont {Leite}\ \emph {et~al.}(2018)\citenamefont {Leite},
  \citenamefont {Dolan},\ and\ \citenamefont {Crispino}}]{leite2018absorption}%
  \BibitemOpen
  \bibfield  {author} {\bibinfo {author} {\bibfnamefont {L.~C.~S.}\
  \bibnamefont {Leite}}, \bibinfo {author} {\bibfnamefont {S.}~\bibnamefont
  {Dolan}},\ and\ \bibinfo {author} {\bibfnamefont {L.~C.~B.}\ \bibnamefont
  {Crispino}},\ }\bibfield  {title} {\bibinfo {title} {Absorption of
  electromagnetic plane waves by rotating black holes},\ }\href
  {https://doi.org/10.1103/PhysRevD.98.024046} {\bibfield  {journal} {\bibinfo
  {journal} {Physical Review D}\ }\textbf {\bibinfo {volume} {98}},\ \bibinfo
  {pages} {024046} (\bibinfo {year} {2018})}\BibitemShut {NoStop}%
\bibitem [{\citenamefont {Frolov}\ and\ \citenamefont {Shoom}(2011)}]{Frolov1}%
  \BibitemOpen
  \bibfield  {author} {\bibinfo {author} {\bibfnamefont {V.~P.}\ \bibnamefont
  {Frolov}}\ and\ \bibinfo {author} {\bibfnamefont {A.~A.}\ \bibnamefont
  {Shoom}},\ }\bibfield  {title} {\bibinfo {title} {Spinoptics in a stationary
  spacetime},\ }\href {https://doi.org/10.1103/PhysRevD.84.044026} {\bibfield
  {journal} {\bibinfo  {journal} {Physical Review D}\ }\textbf {\bibinfo
  {volume} {84}},\ \bibinfo {pages} {044026} (\bibinfo {year}
  {2011})}\BibitemShut {NoStop}%
\bibitem [{\citenamefont {Frolov}\ and\ \citenamefont {Shoom}(2012)}]{Frolov2}%
  \BibitemOpen
  \bibfield  {author} {\bibinfo {author} {\bibfnamefont {V.~P.}\ \bibnamefont
  {Frolov}}\ and\ \bibinfo {author} {\bibfnamefont {A.~A.}\ \bibnamefont
  {Shoom}},\ }\bibfield  {title} {\bibinfo {title} {Scattering of circularly
  polarized light by a rotating black hole},\ }\href
  {https://doi.org/10.1103/PhysRevD.86.024010} {\bibfield  {journal} {\bibinfo
  {journal} {Physical Review D}\ }\textbf {\bibinfo {volume} {86}},\ \bibinfo
  {pages} {024010} (\bibinfo {year} {2012})}\BibitemShut {NoStop}%
\bibitem [{\citenamefont {Yoo}(2012)}]{covariantSpinoptics}%
  \BibitemOpen
  \bibfield  {author} {\bibinfo {author} {\bibfnamefont {C.-M.}\ \bibnamefont
  {Yoo}},\ }\bibfield  {title} {\bibinfo {title} {Notes on spinoptics in a
  stationary spacetime},\ }\href {https://doi.org/10.1103/PhysRevD.86.084005}
  {\bibfield  {journal} {\bibinfo  {journal} {Physical Review D}\ }\textbf
  {\bibinfo {volume} {86}},\ \bibinfo {pages} {084005} (\bibinfo {year}
  {2012})}\BibitemShut {NoStop}%
\bibitem [{\citenamefont {Oancea}\ \emph {et~al.}(2020)\citenamefont {Oancea},
  \citenamefont {Joudioux}, \citenamefont {Dodin}, \citenamefont {Ruiz},
  \citenamefont {Paganini},\ and\ \citenamefont {Andersson}}]{O1}%
  \BibitemOpen
  \bibfield  {author} {\bibinfo {author} {\bibfnamefont {M.~A.}\ \bibnamefont
  {Oancea}}, \bibinfo {author} {\bibfnamefont {J.}~\bibnamefont {Joudioux}},
  \bibinfo {author} {\bibfnamefont {I.~Y.}\ \bibnamefont {Dodin}}, \bibinfo
  {author} {\bibfnamefont {D.~E.}\ \bibnamefont {Ruiz}}, \bibinfo {author}
  {\bibfnamefont {C.~F.}\ \bibnamefont {Paganini}},\ and\ \bibinfo {author}
  {\bibfnamefont {L.}~\bibnamefont {Andersson}},\ }\bibfield  {title} {\bibinfo
  {title} {{Gravitational spin Hall effect of light}},\ }\href
  {https://doi.org/10.1103/PhysRevD.102.024075} {\bibfield  {journal} {\bibinfo
   {journal} {Physical Review D}\ }\textbf {\bibinfo {volume} {102}},\ \bibinfo
  {pages} {024075} (\bibinfo {year} {2020})}\BibitemShut {NoStop}%
\bibitem [{\citenamefont {Harte}\ and\ \citenamefont {Oancea}(2022)}]{O2}%
  \BibitemOpen
  \bibfield  {author} {\bibinfo {author} {\bibfnamefont {A.~I.}\ \bibnamefont
  {Harte}}\ and\ \bibinfo {author} {\bibfnamefont {M.~A.}\ \bibnamefont
  {Oancea}},\ }\bibfield  {title} {\bibinfo {title} {{Spin Hall effects and the
  localization of massless spinning particles}},\ }\href
  {https://doi.org/10.1103/PhysRevD.105.104061} {\bibfield  {journal} {\bibinfo
   {journal} {Physical Review D}\ }\textbf {\bibinfo {volume} {105}},\ \bibinfo
  {pages} {104061} (\bibinfo {year} {2022})}\BibitemShut {NoStop}%
\bibitem [{\citenamefont {Gosselin}\ \emph {et~al.}(2007)\citenamefont
  {Gosselin}, \citenamefont {B{\'e}rard},\ and\ \citenamefont {Mohrbach}}]{G1}%
  \BibitemOpen
  \bibfield  {author} {\bibinfo {author} {\bibfnamefont {P.}~\bibnamefont
  {Gosselin}}, \bibinfo {author} {\bibfnamefont {A.}~\bibnamefont
  {B{\'e}rard}},\ and\ \bibinfo {author} {\bibfnamefont {H.}~\bibnamefont
  {Mohrbach}},\ }\bibfield  {title} {\bibinfo {title} {{Spin Hall effect of
  photons in a static gravitational field}},\ }\href
  {https://doi.org/10.1103/PhysRevD.75.084035} {\bibfield  {journal} {\bibinfo
  {journal} {Physical Review D}\ }\textbf {\bibinfo {volume} {75}},\ \bibinfo
  {pages} {084035} (\bibinfo {year} {2007})}\BibitemShut {NoStop}%
\bibitem [{\citenamefont {Frolov}(2020)}]{Frolov2020}%
  \BibitemOpen
  \bibfield  {author} {\bibinfo {author} {\bibfnamefont {V.~P.}\ \bibnamefont
  {Frolov}},\ }\bibfield  {title} {\bibinfo {title} {Maxwell equations in a
  curved spacetime: Spin optics approximation},\ }\href
  {https://doi.org/10.1103/PhysRevD.102.084013} {\bibfield  {journal} {\bibinfo
   {journal} {Physical Review D}\ }\textbf {\bibinfo {volume} {102}},\ \bibinfo
  {pages} {084013} (\bibinfo {year} {2020})}\BibitemShut {NoStop}%
\bibitem [{\citenamefont {Dolan}(2018{\natexlab{a}})}]{spinorSpinoptics2}%
  \BibitemOpen
  \bibfield  {author} {\bibinfo {author} {\bibfnamefont {S.~R.}\ \bibnamefont
  {Dolan}},\ }\bibfield  {title} {\bibinfo {title} {{Geometrical optics for
  scalar, electromagnetic and gravitational waves on curved spacetime}},\
  }\href {https://doi.org/10.1142/S0218271818430101} {\bibfield  {journal}
  {\bibinfo  {journal} {International Journal of Modern Physics D}\ }\textbf
  {\bibinfo {volume} {27}},\ \bibinfo {pages} {1843010} (\bibinfo {year}
  {2018}{\natexlab{a}})}\BibitemShut {NoStop}%
\bibitem [{\citenamefont {Dolan}(2018{\natexlab{b}})}]{Dolan}%
  \BibitemOpen
  \bibfield  {author} {\bibinfo {author} {\bibfnamefont {S.~R.}\ \bibnamefont
  {Dolan}},\ }\bibfield  {title} {\bibinfo {title} {{Higher-order geometrical
  optics for electromagnetic waves on a curved spacetime}},\ }\href
  {https://doi.org/10.48550/arXiv.1801.02273} {\bibfield  {journal} {\bibinfo
  {journal} {arXiv:1801.02273}\ } (\bibinfo {year}
  {2018}{\natexlab{b}})}\BibitemShut {NoStop}%
\bibitem [{\citenamefont {Harte}(2019{\natexlab{a}})}]{Harte2018}%
  \BibitemOpen
  \bibfield  {author} {\bibinfo {author} {\bibfnamefont {A.~I.}\ \bibnamefont
  {Harte}},\ }\bibfield  {title} {\bibinfo {title} {{Gravitational lensing
  beyond geometric optics: I. Formalism and observables}},\ }\href
  {https://doi.org/10.1007/s10714-018-2494-x} {\bibfield  {journal} {\bibinfo
  {journal} {General Relativity and Gravitation}\ }\textbf {\bibinfo {volume}
  {51}},\ \bibinfo {pages} {14} (\bibinfo {year}
  {2019}{\natexlab{a}})}\BibitemShut {NoStop}%
\bibitem [{\citenamefont {Harte}(2019{\natexlab{b}})}]{HarteOptics2}%
  \BibitemOpen
  \bibfield  {author} {\bibinfo {author} {\bibfnamefont {A.~I.}\ \bibnamefont
  {Harte}},\ }\bibfield  {title} {\bibinfo {title} {{Gravitational lensing
  beyond geometric optics: II. Metric independence}},\ }\href
  {https://doi.org/10.1007/s10714-019-2646-7} {\bibfield  {journal} {\bibinfo
  {journal} {General Relativity and Gravitation}\ }\textbf {\bibinfo {volume}
  {51}},\ \bibinfo {pages} {160} (\bibinfo {year}
  {2019}{\natexlab{b}})}\BibitemShut {NoStop}%
\bibitem [{\citenamefont {Andersson}\ \emph {et~al.}(2021)\citenamefont
  {Andersson}, \citenamefont {Joudioux}, \citenamefont {Oancea},\ and\
  \citenamefont {Raj}}]{GSHE_GW}%
  \BibitemOpen
  \bibfield  {author} {\bibinfo {author} {\bibfnamefont {L.}~\bibnamefont
  {Andersson}}, \bibinfo {author} {\bibfnamefont {J.}~\bibnamefont {Joudioux}},
  \bibinfo {author} {\bibfnamefont {M.~A.}\ \bibnamefont {Oancea}},\ and\
  \bibinfo {author} {\bibfnamefont {A.}~\bibnamefont {Raj}},\ }\bibfield
  {title} {\bibinfo {title} {Propagation of polarized gravitational waves},\
  }\href {https://doi.org/10.1103/PhysRevD.103.044053} {\bibfield  {journal}
  {\bibinfo  {journal} {Physical Review D}\ }\textbf {\bibinfo {volume}
  {103}},\ \bibinfo {pages} {044053} (\bibinfo {year} {2021})}\BibitemShut
  {NoStop}%
\bibitem [{\citenamefont {Oancea}\ \emph {et~al.}(2022)\citenamefont {Oancea},
  \citenamefont {Stiskalek},\ and\ \citenamefont
  {Zumalac\'arregui}}]{GSHE_lensing}%
  \BibitemOpen
  \bibfield  {author} {\bibinfo {author} {\bibfnamefont {M.~A.}\ \bibnamefont
  {Oancea}}, \bibinfo {author} {\bibfnamefont {R.}~\bibnamefont {Stiskalek}},\
  and\ \bibinfo {author} {\bibfnamefont {M.}~\bibnamefont {Zumalac\'arregui}},\
  }\bibfield  {title} {\bibinfo {title} {{From the gates of the abyss:
  Frequency- and polarization-dependent lensing of gravitational waves in
  strong gravitational fields}},\ }\href {https://arxiv.org/abs/2209.06459}
  {\bibfield  {journal} {\bibinfo  {journal} {arXiv:2209.06459}\ } (\bibinfo
  {year} {2022})}\BibitemShut {NoStop}%
\bibitem [{\citenamefont {Oancea}\ \emph {et~al.}(2023)\citenamefont {Oancea},
  \citenamefont {Stiskalek},\ and\ \citenamefont
  {Zumalac\'arregui}}]{GSHE_lensing_letter}%
  \BibitemOpen
  \bibfield  {author} {\bibinfo {author} {\bibfnamefont {M.~A.}\ \bibnamefont
  {Oancea}}, \bibinfo {author} {\bibfnamefont {R.}~\bibnamefont {Stiskalek}},\
  and\ \bibinfo {author} {\bibfnamefont {M.}~\bibnamefont {Zumalac\'arregui}},\
  }\bibfield  {title} {\bibinfo {title} {{Probing general relativistic
  spin-orbit coupling with gravitational waves from hierarchical triple
  systems}},\ }\href {https://arxiv.org/abs/2307.01903} {\bibfield  {journal}
  {\bibinfo  {journal} {arXiv:2307.01903}\ } (\bibinfo {year}
  {2023})}\BibitemShut {NoStop}%
\bibitem [{\citenamefont {Yamamoto}(2018)}]{SHE_GW}%
  \BibitemOpen
  \bibfield  {author} {\bibinfo {author} {\bibfnamefont {N.}~\bibnamefont
  {Yamamoto}},\ }\bibfield  {title} {\bibinfo {title} {{Spin Hall effect of
  gravitational waves}},\ }\href {https://doi.org/10.1103/PhysRevD.98.061701}
  {\bibfield  {journal} {\bibinfo  {journal} {Physical Review D}\ }\textbf
  {\bibinfo {volume} {98}},\ \bibinfo {pages} {061701} (\bibinfo {year}
  {2018})}\BibitemShut {NoStop}%
\bibitem [{\citenamefont {Audretsch}(1981)}]{audretsch}%
  \BibitemOpen
  \bibfield  {author} {\bibinfo {author} {\bibfnamefont {J.}~\bibnamefont
  {Audretsch}},\ }\bibfield  {title} {\bibinfo {title} {{Trajectories and spin
  motion of massive spin-$\frac{1}{2}$ particles in gravitational fields}},\
  }\href {https://doi.org/https://doi.org/10.1088/0305-4470/14/2/017}
  {\bibfield  {journal} {\bibinfo  {journal} {Journal of Physics A:
  Mathematical and General}\ }\textbf {\bibinfo {volume} {14}},\ \bibinfo
  {pages} {411} (\bibinfo {year} {1981})}\BibitemShut {NoStop}%
\bibitem [{\citenamefont {R{\"u}diger}(1981)}]{rudiger}%
  \BibitemOpen
  \bibfield  {author} {\bibinfo {author} {\bibfnamefont {R.}~\bibnamefont
  {R{\"u}diger}},\ }\bibfield  {title} {\bibinfo {title} {{The Dirac equation
  and spinning particles in general relativity}},\ }\href
  {https://doi.org/https://doi.org/10.1098/rspa.1981.0132} {\bibfield
  {journal} {\bibinfo  {journal} {Proceedings of the Royal Society A:
  Mathematical, Physical and Engineering Sciences}\ }\textbf {\bibinfo {volume}
  {377}},\ \bibinfo {pages} {417} (\bibinfo {year} {1981})}\BibitemShut
  {NoStop}%
\bibitem [{\citenamefont {Oancea}\ and\ \citenamefont
  {Kumar}(2023)}]{GSHE_Dirac}%
  \BibitemOpen
  \bibfield  {author} {\bibinfo {author} {\bibfnamefont {M.~A.}\ \bibnamefont
  {Oancea}}\ and\ \bibinfo {author} {\bibfnamefont {A.}~\bibnamefont {Kumar}},\
  }\bibfield  {title} {\bibinfo {title} {{Semiclassical analysis of Dirac
  fields on curved spacetime}},\ }\href
  {https://doi.org/10.1103/PhysRevD.107.044029} {\bibfield  {journal} {\bibinfo
   {journal} {Physical Review D}\ }\textbf {\bibinfo {volume} {107}},\ \bibinfo
  {pages} {044029} (\bibinfo {year} {2023})}\BibitemShut {NoStop}%
\bibitem [{\citenamefont {Marsot}\ \emph {et~al.}(2022)\citenamefont {Marsot},
  \citenamefont {Zhang},\ and\ \citenamefont {Horvathy}}]{marsot2022}%
  \BibitemOpen
  \bibfield  {author} {\bibinfo {author} {\bibfnamefont {L.}~\bibnamefont
  {Marsot}}, \bibinfo {author} {\bibfnamefont {P.-M.}\ \bibnamefont {Zhang}},\
  and\ \bibinfo {author} {\bibfnamefont {P.~A.}\ \bibnamefont {Horvathy}},\
  }\bibfield  {title} {\bibinfo {title} {{Anyonic spin-Hall effect on the black
  hole horizon}},\ }\href {https://doi.org/10.1103/PhysRevD.106.L121503}
  {\bibfield  {journal} {\bibinfo  {journal} {Physical Review D}\ }\textbf
  {\bibinfo {volume} {106}},\ \bibinfo {pages} {L121503} (\bibinfo {year}
  {2022})}\BibitemShut {NoStop}%
\bibitem [{\citenamefont {Gray}\ \emph {et~al.}(2023)\citenamefont {Gray},
  \citenamefont {Kubiz\ifmmode~\check{n}\else \v{n}\fi{}\'ak}, \citenamefont
  {Perche},\ and\ \citenamefont {Redondo-Yuste}}]{gray2022}%
  \BibitemOpen
  \bibfield  {author} {\bibinfo {author} {\bibfnamefont {F.}~\bibnamefont
  {Gray}}, \bibinfo {author} {\bibfnamefont {D.}~\bibnamefont
  {Kubiz\ifmmode~\check{n}\else \v{n}\fi{}\'ak}}, \bibinfo {author}
  {\bibfnamefont {T.~R.}\ \bibnamefont {Perche}},\ and\ \bibinfo {author}
  {\bibfnamefont {J.}~\bibnamefont {Redondo-Yuste}},\ }\bibfield  {title}
  {\bibinfo {title} {Carrollian motion in magnetized black hole horizons},\
  }\href {https://doi.org/10.1103/PhysRevD.107.064009} {\bibfield  {journal}
  {\bibinfo  {journal} {Physical Review D}\ }\textbf {\bibinfo {volume}
  {107}},\ \bibinfo {pages} {064009} (\bibinfo {year} {2023})}\BibitemShut
  {NoStop}%
\bibitem [{\citenamefont {Marsot}\ \emph {et~al.}(2023)\citenamefont {Marsot},
  \citenamefont {Zhang}, \citenamefont {Chernodub},\ and\ \citenamefont
  {Horvathy}}]{marsot2022hall}%
  \BibitemOpen
  \bibfield  {author} {\bibinfo {author} {\bibfnamefont {L.}~\bibnamefont
  {Marsot}}, \bibinfo {author} {\bibfnamefont {P.-M.}\ \bibnamefont {Zhang}},
  \bibinfo {author} {\bibfnamefont {M.}~\bibnamefont {Chernodub}},\ and\
  \bibinfo {author} {\bibfnamefont {P.}~\bibnamefont {Horvathy}},\ }\bibfield
  {title} {\bibinfo {title} {{Hall motions in Carroll dynamics}},\ }\href
  {https://doi.org/https://doi.org/10.1016/j.physrep.2023.07.007} {\bibfield
  {journal} {\bibinfo  {journal} {Physics Reports}\ }\textbf {\bibinfo {volume}
  {1028}},\ \bibinfo {pages} {1} (\bibinfo {year} {2023})}\BibitemShut
  {NoStop}%
\bibitem [{\citenamefont {{Bi{\v{c}}{\'a}k}}\ \emph {et~al.}(2023)\citenamefont
  {{Bi{\v{c}}{\'a}k}}, \citenamefont {Kubiz{\v{n}}{\'a}k},\ and\ \citenamefont
  {Perche}}]{bicak2023}%
  \BibitemOpen
  \bibfield  {author} {\bibinfo {author} {\bibfnamefont {J.}~\bibnamefont
  {{Bi{\v{c}}{\'a}k}}}, \bibinfo {author} {\bibfnamefont {D.}~\bibnamefont
  {Kubiz{\v{n}}{\'a}k}},\ and\ \bibinfo {author} {\bibfnamefont {T.~R.}\
  \bibnamefont {Perche}},\ }\bibfield  {title} {\bibinfo {title} {Migrating
  carrollian particles on magnetized black hole horizons},\ }\href
  {https://doi.org/10.1103/PhysRevD.107.104014} {\bibfield  {journal} {\bibinfo
   {journal} {Physical Review D}\ }\textbf {\bibinfo {volume} {107}},\ \bibinfo
  {pages} {104014} (\bibinfo {year} {2023})}\BibitemShut {NoStop}%
\bibitem [{\citenamefont {Andersson}\ and\ \citenamefont
  {Oancea}(2023)}]{GSHE_rev}%
  \BibitemOpen
  \bibfield  {author} {\bibinfo {author} {\bibfnamefont {L.}~\bibnamefont
  {Andersson}}\ and\ \bibinfo {author} {\bibfnamefont {M.~A.}\ \bibnamefont
  {Oancea}},\ }\bibfield  {title} {\bibinfo {title} {{Spin Hall effects in the
  sky}},\ }\href {https://doi.org/10.1088/1361-6382/ace021} {\bibfield
  {journal} {\bibinfo  {journal} {Classical and Quantum Gravity}\ }\textbf
  {\bibinfo {volume} {40}},\ \bibinfo {pages} {154002} (\bibinfo {year}
  {2023})}\BibitemShut {NoStop}%
\bibitem [{\citenamefont {Oancea}\ \emph {et~al.}(2019)\citenamefont {Oancea},
  \citenamefont {Paganini}, \citenamefont {Joudioux},\ and\ \citenamefont
  {Andersson}}]{GSHE_review}%
  \BibitemOpen
  \bibfield  {author} {\bibinfo {author} {\bibfnamefont {M.~A.}\ \bibnamefont
  {Oancea}}, \bibinfo {author} {\bibfnamefont {C.~F.}\ \bibnamefont
  {Paganini}}, \bibinfo {author} {\bibfnamefont {J.}~\bibnamefont {Joudioux}},\
  and\ \bibinfo {author} {\bibfnamefont {L.}~\bibnamefont {Andersson}},\
  }\bibfield  {title} {\bibinfo {title} {{An overview of the gravitational spin
  Hall effect}},\ }\href {https://arxiv.org/abs/1904.09963} {\bibfield
  {journal} {\bibinfo  {journal} {arXiv:1904.09963}\ } (\bibinfo {year}
  {2019})}\BibitemShut {NoStop}%
\bibitem [{\citenamefont {Romero}\ \emph {et~al.}(2012)\citenamefont {Romero},
  \citenamefont {Fonseca-Neto},\ and\ \citenamefont {Pucheu}}]{Romero_2012}%
  \BibitemOpen
  \bibfield  {author} {\bibinfo {author} {\bibfnamefont {C.}~\bibnamefont
  {Romero}}, \bibinfo {author} {\bibfnamefont {J.~B.}\ \bibnamefont
  {Fonseca-Neto}},\ and\ \bibinfo {author} {\bibfnamefont {M.~L.}\ \bibnamefont
  {Pucheu}},\ }\bibfield  {title} {\bibinfo {title} {{General relativity and
  Weyl geometry}},\ }\href {https://doi.org/10.1088/0264-9381/29/15/155015}
  {\bibfield  {journal} {\bibinfo  {journal} {Classical and Quantum Gravity}\
  }\textbf {\bibinfo {volume} {29}},\ \bibinfo {pages} {155015} (\bibinfo
  {year} {2012})}\BibitemShut {NoStop}%
\bibitem [{\citenamefont {Avalos}\ \emph {et~al.}(2018)\citenamefont {Avalos},
  \citenamefont {Dahia},\ and\ \citenamefont {Romero}}]{Avalos2018}%
  \BibitemOpen
  \bibfield  {author} {\bibinfo {author} {\bibfnamefont {R.}~\bibnamefont
  {Avalos}}, \bibinfo {author} {\bibfnamefont {F.}~\bibnamefont {Dahia}},\ and\
  \bibinfo {author} {\bibfnamefont {C.}~\bibnamefont {Romero}},\ }\bibfield
  {title} {\bibinfo {title} {A note on the problem of proper time in {W}eyl
  space--time},\ }\href {https://doi.org/10.1007/s10701-017-0134-z} {\bibfield
  {journal} {\bibinfo  {journal} {Foundations of Physics}\ }\textbf {\bibinfo
  {volume} {48}},\ \bibinfo {pages} {253} (\bibinfo {year} {2018})}\BibitemShut
  {NoStop}%
\bibitem [{\citenamefont {Panpanich}\ and\ \citenamefont
  {Burikham}(2018)}]{Pi}%
  \BibitemOpen
  \bibfield  {author} {\bibinfo {author} {\bibfnamefont {S.}~\bibnamefont
  {Panpanich}}\ and\ \bibinfo {author} {\bibfnamefont {P.}~\bibnamefont
  {Burikham}},\ }\bibfield  {title} {\bibinfo {title} {{Fitting rotation curves
  of galaxies by de Rham-Gabadadze-Tolley massive gravity}},\ }\href
  {https://doi.org/10.1103/PhysRevD.98.064008} {\bibfield  {journal} {\bibinfo
  {journal} {Physical Review D}\ }\textbf {\bibinfo {volume} {98}},\ \bibinfo
  {pages} {064008} (\bibinfo {year} {2018})}\BibitemShut {NoStop}%
\bibitem [{\citenamefont {Misner}\ \emph {et~al.}(1973)\citenamefont {Misner},
  \citenamefont {Thorne},\ and\ \citenamefont {Wheeler}}]{MTW}%
  \BibitemOpen
  \bibfield  {author} {\bibinfo {author} {\bibfnamefont {C.~W.}\ \bibnamefont
  {Misner}}, \bibinfo {author} {\bibfnamefont {K.~S.}\ \bibnamefont {Thorne}},\
  and\ \bibinfo {author} {\bibfnamefont {J.~A.}\ \bibnamefont {Wheeler}},\
  }\href@noop {} {\emph {\bibinfo {title} {Gravitation}}}\ (\bibinfo
  {publisher} {W. H. Freeman San Francisco},\ \bibinfo {year}
  {1973})\BibitemShut {NoStop}%
\bibitem [{\citenamefont {Arnold}(1989)}]{Arnold_book}%
  \BibitemOpen
  \bibfield  {author} {\bibinfo {author} {\bibfnamefont {V.~I.}\ \bibnamefont
  {Arnold}},\ }\href {https://doi.org/10.1007/978-1-4757-2063-1} {\emph
  {\bibinfo {title} {Mathematical methods of classical mechanics}}},\ \bibinfo
  {series} {Graduate Texts in Mathematics}, Vol.~\bibinfo {volume} {60}\
  (\bibinfo  {publisher} {Springer-Verlag, New York},\ \bibinfo {year} {1989})\
  pp.\ \bibinfo {pages} {xvi+516}\BibitemShut {NoStop}%
\bibitem [{\citenamefont {Emmrich}\ and\ \citenamefont
  {Weinstein}(1996)}]{Emmrich1996}%
  \BibitemOpen
  \bibfield  {author} {\bibinfo {author} {\bibfnamefont {C.}~\bibnamefont
  {Emmrich}}\ and\ \bibinfo {author} {\bibfnamefont {A.}~\bibnamefont
  {Weinstein}},\ }\bibfield  {title} {\bibinfo {title} {Geometry of the
  transport equation in multicomponent {WKB} approximations},\ }\href
  {https://doi.org/https://doi.org/10.1007/BF02099256} {\bibfield  {journal}
  {\bibinfo  {journal} {Communications in Mathematical Physics}\ }\textbf
  {\bibinfo {volume} {176}},\ \bibinfo {pages} {701} (\bibinfo {year}
  {1996})}\BibitemShut {NoStop}%
\bibitem [{\citenamefont {Bates}\ and\ \citenamefont
  {Weinstein}(1997)}]{MR1806388}%
  \BibitemOpen
  \bibfield  {author} {\bibinfo {author} {\bibfnamefont {S.}~\bibnamefont
  {Bates}}\ and\ \bibinfo {author} {\bibfnamefont {A.}~\bibnamefont
  {Weinstein}},\ }\href {https://bookstore.ams.org/bmln-8} {\emph {\bibinfo
  {title} {Lectures on the Geometry of Quantization}}},\ \bibinfo {series}
  {Berkeley Mathematics Lecture Notes}, Vol.~\bibinfo {volume} {8}\ (\bibinfo
  {publisher} {American Mathematical Society, Providence, RI; Berkeley Center
  for Pure and Applied Mathematics, Berkeley, CA},\ \bibinfo {year} {1997})\
  pp.\ \bibinfo {pages} {vi+137}\BibitemShut {NoStop}%
\bibitem [{\citenamefont {Littlejohn}\ and\ \citenamefont
  {Flynn}(1991)}]{Littlejohn}%
  \BibitemOpen
  \bibfield  {author} {\bibinfo {author} {\bibfnamefont {R.~G.}\ \bibnamefont
  {Littlejohn}}\ and\ \bibinfo {author} {\bibfnamefont {W.~G.}\ \bibnamefont
  {Flynn}},\ }\bibfield  {title} {\bibinfo {title} {Geometric phases in the
  asymptotic theory of coupled wave equations},\ }\href
  {https://doi.org/10.1103/PhysRevA.44.5239} {\bibfield  {journal} {\bibinfo
  {journal} {Physical Review A}\ }\textbf {\bibinfo {volume} {44}},\ \bibinfo
  {pages} {5239} (\bibinfo {year} {1991})}\BibitemShut {NoStop}%
\bibitem [{\citenamefont {Bliokh}\ \emph
  {et~al.}(2008{\natexlab{b}})\citenamefont {Bliokh}, \citenamefont
  {Gorodetski}, \citenamefont {Kleiner},\ and\ \citenamefont {Hasman}}]{B2}%
  \BibitemOpen
  \bibfield  {author} {\bibinfo {author} {\bibfnamefont {K.~Y.}\ \bibnamefont
  {Bliokh}}, \bibinfo {author} {\bibfnamefont {Y.}~\bibnamefont {Gorodetski}},
  \bibinfo {author} {\bibfnamefont {V.}~\bibnamefont {Kleiner}},\ and\ \bibinfo
  {author} {\bibfnamefont {E.}~\bibnamefont {Hasman}},\ }\bibfield  {title}
  {\bibinfo {title} {{Coriolis Effect in Optics: Unified Geometric Phase and
  Spin-Hall Effect}},\ }\href {https://doi.org/10.1103/PhysRevLett.101.030404}
  {\bibfield  {journal} {\bibinfo  {journal} {Physical Review Letters}\
  }\textbf {\bibinfo {volume} {101}},\ \bibinfo {pages} {030404} (\bibinfo
  {year} {2008}{\natexlab{b}})}\BibitemShut {NoStop}%
\bibitem [{\citenamefont {Bliokh}(2009)}]{B3}%
  \BibitemOpen
  \bibfield  {author} {\bibinfo {author} {\bibfnamefont {K.~Y.}\ \bibnamefont
  {Bliokh}},\ }\bibfield  {title} {\bibinfo {title} {{Geometrodynamics of
  polarized light: Berry phase and spin Hall effect in a gradient-index
  medium}},\ }\href {https://doi.org/10.1088/1464-4258/11/9/094009} {\bibfield
  {journal} {\bibinfo  {journal} {Journal of Optics A: Pure and Applied
  Optics}\ }\textbf {\bibinfo {volume} {11}},\ \bibinfo {pages} {094009}
  (\bibinfo {year} {2009})}\BibitemShut {NoStop}%
\bibitem [{\citenamefont {Duval}\ \emph {et~al.}(2006)\citenamefont {Duval},
  \citenamefont {Horv\'ath},\ and\ \citenamefont {Horv\'athy}}]{Duval2006}%
  \BibitemOpen
  \bibfield  {author} {\bibinfo {author} {\bibfnamefont {C.}~\bibnamefont
  {Duval}}, \bibinfo {author} {\bibfnamefont {Z.}~\bibnamefont {Horv\'ath}},\
  and\ \bibinfo {author} {\bibfnamefont {P.~A.}\ \bibnamefont {Horv\'athy}},\
  }\bibfield  {title} {\bibinfo {title} {Fermat principle for spinning light},\
  }\href {https://doi.org/10.1103/PhysRevD.74.021701} {\bibfield  {journal}
  {\bibinfo  {journal} {Physical Review D}\ }\textbf {\bibinfo {volume} {74}},\
  \bibinfo {pages} {021701} (\bibinfo {year} {2006})}\BibitemShut {NoStop}%
\bibitem [{\citenamefont {Duval}\ \emph {et~al.}(2007)\citenamefont {Duval},
  \citenamefont {Horv\'ath},\ and\ \citenamefont {Horv\'athy}}]{Duval2007}%
  \BibitemOpen
  \bibfield  {author} {\bibinfo {author} {\bibfnamefont {C.}~\bibnamefont
  {Duval}}, \bibinfo {author} {\bibfnamefont {Z.}~\bibnamefont {Horv\'ath}},\
  and\ \bibinfo {author} {\bibfnamefont {P.~A.}\ \bibnamefont {Horv\'athy}},\
  }\bibfield  {title} {\bibinfo {title} {{Geometrical spinoptics and the
  optical Hall effect}},\ }\href
  {https://doi.org/https://doi.org/10.1016/j.geomphys.2006.07.003} {\bibfield
  {journal} {\bibinfo  {journal} {Journal of Geometry and Physics}\ }\textbf
  {\bibinfo {volume} {57}},\ \bibinfo {pages} {925} (\bibinfo {year}
  {2007})}\BibitemShut {NoStop}%
\bibitem [{\citenamefont {Ruiz}\ and\ \citenamefont {Dodin}(2015)}]{Ruiz2015}%
  \BibitemOpen
  \bibfield  {author} {\bibinfo {author} {\bibfnamefont {D.~E.}\ \bibnamefont
  {Ruiz}}\ and\ \bibinfo {author} {\bibfnamefont {I.~Y.}\ \bibnamefont
  {Dodin}},\ }\bibfield  {title} {\bibinfo {title} {First-principles
  variational formulation of polarization effects in geometrical optics},\
  }\href {https://doi.org/10.1103/PhysRevA.92.043805} {\bibfield  {journal}
  {\bibinfo  {journal} {Physical Review A}\ }\textbf {\bibinfo {volume} {92}},\
  \bibinfo {pages} {043805} (\bibinfo {year} {2015})}\BibitemShut {NoStop}%
\bibitem [{\citenamefont {Eddington}(1920)}]{Eddington}%
  \BibitemOpen
  \bibfield  {author} {\bibinfo {author} {\bibfnamefont {A.~S.}\ \bibnamefont
  {Eddington}},\ }\href@noop {} {\emph {\bibinfo {title} {{Space, time and
  gravitation: An outline of the general relativity theory}}}}\ (\bibinfo
  {publisher} {University Press},\ \bibinfo {year} {1920})\BibitemShut
  {NoStop}%
\bibitem [{\citenamefont {Gordon}(1923)}]{Gordon}%
  \BibitemOpen
  \bibfield  {author} {\bibinfo {author} {\bibfnamefont {W.}~\bibnamefont
  {Gordon}},\ }\bibfield  {title} {\bibinfo {title} {Zur lichtfortpflanzung
  nach der relativit{\"a}tstheorie},\ }\href
  {https://doi.org/10.1002/andp.19233772202} {\bibfield  {journal} {\bibinfo
  {journal} {Annalen der Physik}\ }\textbf {\bibinfo {volume} {377}},\ \bibinfo
  {pages} {421} (\bibinfo {year} {1923})}\BibitemShut {NoStop}%
\bibitem [{\citenamefont {{Skrotskii}}(1957)}]{skrotskii}%
  \BibitemOpen
  \bibfield  {author} {\bibinfo {author} {\bibfnamefont {G.~V.}\ \bibnamefont
  {{Skrotskii}}},\ }\bibfield  {title} {\bibinfo {title} {{The Influence of
  Gravitation on the Propagation of Light}},\ }\href@noop {} {\bibfield
  {journal} {\bibinfo  {journal} {Soviet Physics Doklady}\ }\textbf {\bibinfo
  {volume} {2}},\ \bibinfo {pages} {226} (\bibinfo {year} {1957})}\BibitemShut
  {NoStop}%
\bibitem [{\citenamefont {Balazs}(1958)}]{Balazs}%
  \BibitemOpen
  \bibfield  {author} {\bibinfo {author} {\bibfnamefont {N.~L.}\ \bibnamefont
  {Balazs}},\ }\bibfield  {title} {\bibinfo {title} {Effect of a gravitational
  field, due to a rotating body, on the plane of polarization of an
  electromagnetic wave},\ }\href {https://doi.org/10.1103/PhysRev.110.236}
  {\bibfield  {journal} {\bibinfo  {journal} {Physical Review}\ }\textbf
  {\bibinfo {volume} {110}},\ \bibinfo {pages} {236} (\bibinfo {year}
  {1958})}\BibitemShut {NoStop}%
\bibitem [{\citenamefont {de~Felice}(1971)}]{deFelice}%
  \BibitemOpen
  \bibfield  {author} {\bibinfo {author} {\bibfnamefont {F.}~\bibnamefont
  {de~Felice}},\ }\bibfield  {title} {\bibinfo {title} {On the gravitational
  field acting as an optical medium},\ }\href
  {https://doi.org/10.1007/BF00758153} {\bibfield  {journal} {\bibinfo
  {journal} {General Relativity and Gravitation}\ }\textbf {\bibinfo {volume}
  {2}},\ \bibinfo {pages} {347} (\bibinfo {year} {1971})}\BibitemShut {NoStop}%
\bibitem [{\citenamefont {Gerlach}(1969)}]{Gerlach1969}%
  \BibitemOpen
  \bibfield  {author} {\bibinfo {author} {\bibfnamefont {U.~H.}\ \bibnamefont
  {Gerlach}},\ }\bibfield  {title} {\bibinfo {title} {{Derivation of the Ten
  Einstein Field Equations from the Semiclassical Approximation to Quantum
  Geometrodynamics}},\ }\href {https://doi.org/10.1103/PhysRev.177.1929}
  {\bibfield  {journal} {\bibinfo  {journal} {Physical Review}\ }\textbf
  {\bibinfo {volume} {177}},\ \bibinfo {pages} {1929} (\bibinfo {year}
  {1969})}\BibitemShut {NoStop}%
\bibitem [{\citenamefont {Puetzfeld}\ and\ \citenamefont
  {Obukhov}(2014)}]{PhysRevD.90.084034}%
  \BibitemOpen
  \bibfield  {author} {\bibinfo {author} {\bibfnamefont {D.}~\bibnamefont
  {Puetzfeld}}\ and\ \bibinfo {author} {\bibfnamefont {Y.~N.}\ \bibnamefont
  {Obukhov}},\ }\bibfield  {title} {\bibinfo {title} {Equations of motion in
  metric-affine gravity: A covariant unified framework},\ }\href
  {https://doi.org/10.1103/PhysRevD.90.084034} {\bibfield  {journal} {\bibinfo
  {journal} {Physical Review D}\ }\textbf {\bibinfo {volume} {90}},\ \bibinfo
  {pages} {084034} (\bibinfo {year} {2014})}\BibitemShut {NoStop}%
\bibitem [{\citenamefont {Poisson}\ \emph {et~al.}(2011)\citenamefont
  {Poisson}, \citenamefont {Pound},\ and\ \citenamefont {Vega}}]{Poisson2011}%
  \BibitemOpen
  \bibfield  {author} {\bibinfo {author} {\bibfnamefont {E.}~\bibnamefont
  {Poisson}}, \bibinfo {author} {\bibfnamefont {A.}~\bibnamefont {Pound}},\
  and\ \bibinfo {author} {\bibfnamefont {I.}~\bibnamefont {Vega}},\ }\bibfield
  {title} {\bibinfo {title} {The motion of point particles in curved
  spacetime},\ }\href {https://doi.org/10.12942/lrr-2011-7} {\bibfield
  {journal} {\bibinfo  {journal} {Living Reviews in Relativity}\ }\textbf
  {\bibinfo {volume} {14}},\ \bibinfo {pages} {7} (\bibinfo {year}
  {2011})}\BibitemShut {NoStop}%
\bibitem [{\citenamefont {{Wolfram Research, Inc.}}(2021)}]{Mathematica}%
  \BibitemOpen
  \bibfield  {author} {\bibinfo {author} {\bibnamefont {{Wolfram Research,
  Inc.}}},\ }\href {https://www.wolfram.com/mathematica} {\bibinfo {title}
  {Mathematica, {V}ersion 13.0.0}} (\bibinfo {year} {2021}),\ \bibinfo {note}
  {{Champaign, IL}, 2021}\BibitemShut {NoStop}%
\bibitem [{\citenamefont {Oancea}(2021)}]{Oanceathesis}%
  \BibitemOpen
  \bibfield  {author} {\bibinfo {author} {\bibfnamefont {M.~A.}\ \bibnamefont
  {Oancea}},\ }\emph {\bibinfo {title} {{Spin Hall effects in General
  Relativity}}},\ \href {https://doi.org/10.25932/publishup-50229} {\bibinfo
  {type} {{PhD thesis}}},\ \bibinfo  {school} {University of Potsdam} (\bibinfo
  {year} {2021})\BibitemShut {NoStop}%
\bibitem [{\citenamefont {Yavari}\ and\ \citenamefont {Goriely}(2012)}]{CM1}%
  \BibitemOpen
  \bibfield  {author} {\bibinfo {author} {\bibfnamefont {A.}~\bibnamefont
  {Yavari}}\ and\ \bibinfo {author} {\bibfnamefont {A.}~\bibnamefont
  {Goriely}},\ }\bibfield  {title} {\bibinfo {title} {Weyl geometry and the
  nonlinear mechanics of distributed point defects},\ }\href
  {https://doi.org/10.1098/rspa.2012.0342} {\bibfield  {journal} {\bibinfo
  {journal} {Proceedings of the Royal Society A: Mathematical, Physical and
  Engineering Sciences}\ }\textbf {\bibinfo {volume} {468}},\ \bibinfo {pages}
  {3902} (\bibinfo {year} {2012})}\BibitemShut {NoStop}%
\bibitem [{\citenamefont {Roychowdhury}\ and\ \citenamefont
  {Gupta}(2017)}]{CM2}%
  \BibitemOpen
  \bibfield  {author} {\bibinfo {author} {\bibfnamefont {A.}~\bibnamefont
  {Roychowdhury}}\ and\ \bibinfo {author} {\bibfnamefont {A.}~\bibnamefont
  {Gupta}},\ }\bibfield  {title} {\bibinfo {title} {Non-metric connection and
  metric anomalies in materially uniform elastic solids},\ }\href
  {https://doi.org/10.1007/s10659-016-9578-1} {\bibfield  {journal} {\bibinfo
  {journal} {Journal of Elasticity}\ }\textbf {\bibinfo {volume} {126}},\
  \bibinfo {pages} {1} (\bibinfo {year} {2017})}\BibitemShut {NoStop}%
\bibitem [{\citenamefont {Ling}\ \emph {et~al.}(2017)\citenamefont {Ling},
  \citenamefont {Zhou}, \citenamefont {Huang}, \citenamefont {Liu},
  \citenamefont {Qiu}, \citenamefont {Luo},\ and\ \citenamefont {Wen}}]{CM3}%
  \BibitemOpen
  \bibfield  {author} {\bibinfo {author} {\bibfnamefont {X.}~\bibnamefont
  {Ling}}, \bibinfo {author} {\bibfnamefont {X.}~\bibnamefont {Zhou}}, \bibinfo
  {author} {\bibfnamefont {K.}~\bibnamefont {Huang}}, \bibinfo {author}
  {\bibfnamefont {Y.}~\bibnamefont {Liu}}, \bibinfo {author} {\bibfnamefont
  {C.-W.}\ \bibnamefont {Qiu}}, \bibinfo {author} {\bibfnamefont
  {H.}~\bibnamefont {Luo}},\ and\ \bibinfo {author} {\bibfnamefont
  {S.}~\bibnamefont {Wen}},\ }\bibfield  {title} {\bibinfo {title} {{Recent
  advances in the spin Hall effect of light}},\ }\href
  {https://doi.org/10.1088/1361-6633/aa5397} {\bibfield  {journal} {\bibinfo
  {journal} {Reports on Progress in Physics}\ }\textbf {\bibinfo {volume}
  {80}},\ \bibinfo {pages} {066401} (\bibinfo {year} {2017})}\BibitemShut
  {NoStop}%
\bibitem [{\citenamefont {Vines}\ \emph {et~al.}(2016)\citenamefont {Vines},
  \citenamefont {Kunst}, \citenamefont {Steinhoff},\ and\ \citenamefont
  {Hinderer}}]{vines2016canonical}%
  \BibitemOpen
  \bibfield  {author} {\bibinfo {author} {\bibfnamefont {J.}~\bibnamefont
  {Vines}}, \bibinfo {author} {\bibfnamefont {D.}~\bibnamefont {Kunst}},
  \bibinfo {author} {\bibfnamefont {J.}~\bibnamefont {Steinhoff}},\ and\
  \bibinfo {author} {\bibfnamefont {T.}~\bibnamefont {Hinderer}},\ }\bibfield
  {title} {\bibinfo {title} {{Canonical Hamiltonian for an extended test body
  in curved spacetime: To quadratic order in spin}},\ }\href
  {https://doi.org/https://doi.org/10.1103/PhysRevD.93.103008} {\bibfield
  {journal} {\bibinfo  {journal} {Physical Review D}\ }\textbf {\bibinfo
  {volume} {93}},\ \bibinfo {pages} {103008} (\bibinfo {year}
  {2016})}\BibitemShut {NoStop}%
\bibitem [{\citenamefont {Bliokh}\ and\ \citenamefont
  {Nori}(2012)}]{Relativistic_Hall}%
  \BibitemOpen
  \bibfield  {author} {\bibinfo {author} {\bibfnamefont {K.~Y.}\ \bibnamefont
  {Bliokh}}\ and\ \bibinfo {author} {\bibfnamefont {F.}~\bibnamefont {Nori}},\
  }\bibfield  {title} {\bibinfo {title} {{Relativistic Hall Effect}},\ }\href
  {https://doi.org/10.1103/PhysRevLett.108.120403} {\bibfield  {journal}
  {\bibinfo  {journal} {Physical Review Letters}\ }\textbf {\bibinfo {volume}
  {108}},\ \bibinfo {pages} {120403} (\bibinfo {year} {2012})}\BibitemShut
  {NoStop}%
\bibitem [{\citenamefont {Stone}\ \emph {et~al.}(2015)\citenamefont {Stone},
  \citenamefont {Dwivedi},\ and\ \citenamefont {Zhou}}]{Stone2015}%
  \BibitemOpen
  \bibfield  {author} {\bibinfo {author} {\bibfnamefont {M.}~\bibnamefont
  {Stone}}, \bibinfo {author} {\bibfnamefont {V.}~\bibnamefont {Dwivedi}},\
  and\ \bibinfo {author} {\bibfnamefont {T.}~\bibnamefont {Zhou}},\ }\bibfield
  {title} {\bibinfo {title} {{Wigner Translations and the Observer Dependence
  of the Position of Massless Spinning Particles}},\ }\href
  {https://doi.org/10.1103/PhysRevLett.114.210402} {\bibfield  {journal}
  {\bibinfo  {journal} {Physical Review Letters}\ }\textbf {\bibinfo {volume}
  {114}},\ \bibinfo {pages} {210402} (\bibinfo {year} {2015})}\BibitemShut
  {NoStop}%
\end{thebibliography}%

\end{document}